\pdfoutput=1

\documentclass[sigplan,10pt]{acmart}

\usepackage[T1]{fontenc}
\usepackage[utf8]{inputenc}
\usepackage{xcolor}
\usepackage{forest}
\usepackage{amssymb}
\usepackage{mathtools}
\usepackage{marvosym}
\newcommand{\MVUparrow}{\,\scalebox{1}[1.4]{\rotatebox[origin=c]{90}{\MVRightarrow}}}
\newcommand{\MVDnarrow}{\,\scalebox{1}[1.4]{\rotatebox[origin=c]{-90}{\MVRightarrow}}}
\usepackage{semantic}
\usepackage[normalem]{ulem}
\usepackage{enumitem}
\setlist{leftmargin=5.5mm}
\usepackage{lipsum}
\renewcommand\footnotetextcopyrightpermission[1]{} 
\newcommand{\sub}[2]{\mathclose{\{^{#2\!\!\!}\reflectbox{$\!\smallsetminus$}_{\!#1}\hspace{-.2mm}\}}}
\makeatletter
\@ifundefined{lhs2tex.lhs2tex.sty.read}%
  {\@namedef{lhs2tex.lhs2tex.sty.read}{}%
   \newcommand\SkipToFmtEnd{}%
   \newcommand\EndFmtInput{}%
   \long\def\SkipToFmtEnd#1\EndFmtInput{}%
  }\SkipToFmtEnd

\newcommand\ReadOnlyOnce[1]{\@ifundefined{#1}{\@namedef{#1}{}}\SkipToFmtEnd}
\usepackage{amstext}
\usepackage{amssymb}
\usepackage{stmaryrd}
\DeclareFontFamily{OT1}{cmtex}{}
\DeclareFontShape{OT1}{cmtex}{m}{n}
  {<5><6><7><8>cmtex8
   <9>cmtex9
   <10><10.95><12><14.4><17.28><20.74><24.88>cmtex10}{}
\DeclareFontShape{OT1}{cmtex}{m}{it}
  {<-> ssub * cmtt/m/it}{}
\newcommand{\texfamily}{\fontfamily{cmtex}\selectfont}
\DeclareFontShape{OT1}{cmtt}{bx}{n}
  {<5><6><7><8>cmtt8
   <9>cmbtt9
   <10><10.95><12><14.4><17.28><20.74><24.88>cmbtt10}{}
\DeclareFontShape{OT1}{cmtex}{bx}{n}
  {<-> ssub * cmtt/bx/n}{}
\newcommand{\tex}[1]{\text{\texfamily#1}}	

\newcommand{\Sp}{\hskip.33334em\relax}

\newcommand{\Conid}[1]{\mathit{#1}}
\newcommand{\Varid}[1]{\mathit{#1}}
\newcommand{\anonymous}{\kern0.06em \vbox{\hrule\@width.5em}}
\newcommand{\plus}{\mathbin{+\!\!\!+}}
\newcommand{\bind}{\mathbin{>\!\!\!>\mkern-6.7mu=}}
\newcommand{\rbind}{\mathbin{=\mkern-6.7mu<\!\!\!<}}
\newcommand{\sequ}{\mathbin{>\!\!\!>}}
\renewcommand{\leq}{\leqslant}
\renewcommand{\geq}{\geqslant}
\usepackage{polytable}

\@ifundefined{mathindent}%
  {\newdimen\mathindent\mathindent\leftmargini}%
  {}%

\def\resethooks{%
  \global\let\SaveRestoreHook\empty
  \global\let\ColumnHook\empty}
\newcommand*{\savecolumns}[1][default]%
  {\g@addto@macro\SaveRestoreHook{\savecolumns[#1]}}
\newcommand*{\restorecolumns}[1][default]%
  {\g@addto@macro\SaveRestoreHook{\restorecolumns[#1]}}
\newcommand*{\aligncolumn}[2]%
  {\g@addto@macro\ColumnHook{\column{#1}{#2}}}

\resethooks

\newcommand{\onelinecommentchars}{\quad-{}- }
\newcommand{\commentbeginchars}{\enskip\{-}
\newcommand{\commentendchars}{-\}\enskip}

\newcommand{\visiblecomments}{%
  \let\onelinecomment=\onelinecommentchars
  \let\commentbegin=\commentbeginchars
  \let\commentend=\commentendchars}

\newcommand{\invisiblecomments}{%
  \let\onelinecomment=\empty
  \let\commentbegin=\empty
  \let\commentend=\empty}

\visiblecomments

\newlength{\blanklineskip}
\newcommand{\hsindent}[1]{\quad}
\let\hspre\empty
\let\hspost\empty
\newcommand{\NB}{\textbf{NB}}
\newcommand{\Todo}[1]{$\langle$\textbf{To do:}~#1$\rangle$}

\EndFmtInput
\makeatother
\ReadOnlyOnce{polycode.fmt}%
\makeatletter

\newcommand{\hsnewpar}[1]%
  {{\parskip=0pt\parindent=0pt\par\vskip #1\noindent}}

\newcommand{\hscodestyle}{}

\newcommand{\sethscode}[1]%
  {\expandafter\let\expandafter\hscode\csname #1\endcsname
   \expandafter\let\expandafter\endhscode\csname end#1\endcsname}

  {\par\noindent
   \advance\leftskip\mathindent
   \hscodestyle
   \let\\=\@normalcr
   \let\hspre\(\let\hspost\)%
   \pboxed}%
  {\endpboxed\)%
   \par\noindent
   \ignorespacesafterend}

\newcommand{\compaths}{\sethscode{compathscode}}

  {\hsnewpar\abovedisplayskip
   \advance\leftskip\mathindent
   \hscodestyle
   \let\hspre\(\let\hspost\)%
   \pboxed}%
  {\endpboxed%
   \hsnewpar\belowdisplayskip
   \ignorespacesafterend}

  {\hsnewpar\abovedisplayskip
   \advance\leftskip\mathindent
   \hscodestyle
   \let\\=\@normalcr
   \(\pboxed}%
  {\endpboxed\)%
   \hsnewpar\belowdisplayskip
   \ignorespacesafterend}

\newcommand{\plainhs}{\sethscode{plainhscode}}
\newcommand{\oldplainhs}{\sethscode{oldplainhscode}}
\plainhs

  {\hsnewpar\abovedisplayskip
   \advance\leftskip\mathindent
   \hscodestyle
   \let\\=\@normalcr
   \(\parray}%
  {\endparray\)%
   \hsnewpar\belowdisplayskip
   \ignorespacesafterend}

\newcommand{\arrayhs}{\sethscode{arrayhscode}}

  {\parray}{\endparray}

\newcommand{\mathhs}{\sethscode{mathhscode}}

  {\(\parray}{\endparray\)}

\newcommand{\texths}{\sethscode{texthscode}}

\def\codeframewidth{\arrayrulewidth}
\RequirePackage{calc}

  {\parskip=\abovedisplayskip\par\noindent
   \hscodestyle
   \arrayrulewidth=\codeframewidth
   \tabular{@{}|p{\linewidth-2\arraycolsep-2\arrayrulewidth-2pt}|@{}}%
   \hline\framedhslinecorrect\\{-1.5ex}%
   \let\endoflinesave=\\
   \let\\=\@normalcr
   \(\pboxed}%
  {\endpboxed\)%
   \framedhslinecorrect\endoflinesave{.5ex}\hline
   \endtabular
   \parskip=\belowdisplayskip\par\noindent
   \ignorespacesafterend}

\newcommand{\framedhslinecorrect}[2]%
  {#1[#2]}

\newcommand{\framedhs}{\sethscode{framedhscode}}

  {\(\def\column##1##2{}%
   \let\>\undefined\let\<\undefined\let\\\undefined
   \newcommand\>[1][]{}\newcommand\<[1][]{}\newcommand\\[1][]{}%
   \def\fromto##1##2##3{##3}%
   \def\nextline{}}{\) }%

\newcommand{\inlinehs}{\sethscode{inlinehscode}}

  {\let\orighscode=\hscode
   \let\origendhscode=\endhscode
   \def\endhscode{\def\hscode{\endgroup\def\@currenvir{hscode}\\}\begingroup}
   \orighscode\def\hscode{\endgroup\def\@currenvir{hscode}}}%
  {\origendhscode
   \global\let\hscode=\orighscode
   \global\let\endhscode=\origendhscode}%

\makeatother
\EndFmtInput
\ReadOnlyOnce{colorcode.fmt}%

\RequirePackage{colortbl}
\RequirePackage{calc}

\makeatletter
  {\hsnewpar\abovedisplayskip
   \hscodestyle
   \tabular{@{}>{\columncolor{codecolor}}p{\linewidth}@{}}%
   \let\\=\@normalcr
   \(\pboxed}%
  {\endpboxed\)%
   \endtabular
   \hsnewpar\belowdisplayskip
   \ignorespacesafterend}

  {\hsnewpar\abovedisplayskip
   \hscodestyle
   \tabular{@{}>{\columncolor{codecolor}\(}l<{\)}@{}}%
   \pmboxed}%
  {\endpmboxed%
   \endtabular
   \hsnewpar\belowdisplayskip
   \ignorespacesafterend}

  {\hsnewpar\abovedisplayskip
   \hscodestyle
   \arrayrulecolor{codecolor}%
   \arrayrulewidth=\coderulewidth
   \tabular{|p{\linewidth-\arrayrulewidth-\tabcolsep}@{}}%
   \let\\=\@normalcr
   \(\pboxed}%
  {\endpboxed\)%
   \endtabular
   \hsnewpar\belowdisplayskip
   \ignorespacesafterend}
\makeatother

\def\colorcode{\columncolor{codecolor}}
\definecolor{codecolor}{rgb}{1,1,.667}
\newlength{\coderulewidth}
\newcommand{\colorhs}{\sethscode{colorhscode}}
\newcommand{\tightcolorhs}{\sethscode{tightcolorhscode}}
\newcommand{\barhs}{\sethscode{barhscode}}

\EndFmtInput

\renewcommand{\onelinecommentchars}{\color{gray}\quad-{}- }
\renewcommand{\commentbeginchars}{\color{gray}\enskip\{- }
\renewcommand{\commentendchars}{-\}\enskip}

\renewcommand{\visiblecomments}{%
  \let\onelinecomment=\onelinecommentchars
  \let\commentbegin=\commentbeginchars
  \let\commentend=\commentendchars}

\renewcommand{\invisiblecomments}{%
  \let\onelinecomment=\empty
  \let\commentbegin=\empty
  \let\commentend=\empty}

\visiblecomments

\usepackage{booktabs}   
\usepackage{subcaption} 

\newcommand{\xone}[1]{\mathop{\raisebox{-.3ex}{$\xrightarrow{\raisebox{-.5ex}{\scriptsize{\ensuremath{#1}}}}$}}}
\newcommand{\xoneb}[1]{\mathop{\raisebox{-.3ex}{$\xrightharpoonup{\raisebox{-.5ex}{\scriptsize{\ensuremath{#1}}}}_{\!\!\!\textsc{b}}$}}}
\newcommand{\OM}[0]{\ensuremath{\mathcal{O\!M}}}

\makeatletter\if@ACM@journal\makeatother
\acmJournal{PACMPL}
\acmVolume{1}
\acmNumber{1}
\acmArticle{1}
\acmYear{2017}
\acmMonth{1}
\acmDOI{10.1145/nnnnnnn.nnnnnnn}
\startPage{1}
\else\makeatother
\acmConference[Haskell'17]{ACM SIGPLAN Haskell Symposium}{September 07--08, 2017}{Oxford, UK}
\acmYear{2017}
\acmISBN{978-x-xxxx-xxxx-x/YY/MM}
\acmDOI{10.1145/nnnnnnn.nnnnnnn}
\startPage{1}
\fi

\setcopyright{none}             

\makeatletter
\begin{document}


\title{Generating Witness of Non-Bisimilarity for the pi-Calculus}



\author{Ki Yung Ahn}
\orcid{nnnn-nnnn-nnnn-nnnn}             
\affiliation{
  \institution{Nanyang Technological University}            
  \country{Singapore}
}
\email{yaki@ntu.edu.sg}          

\author{Ross Horne}
\orcid{nnnn-nnnn-nnnn-nnnn}             
\affiliation{
  \institution{Nanyang Technological University}           
  \country{Singapore}
}
\email{rhorne@ntu.edu.sg}         

\author{Alwen Tiu}
\affiliation{
  \institution{Nanyang Technological University}           
  \country{Singapore}
}
\email{atiu@ntu.edu.sg}         


\begin{abstract}In the logic programming paradigm, it is difficult to develop
an elegant solution for generating distinguishing formulae that
witness the failure of open-bisimilarity between two pi-calculus processes;
this was unexpected because the semantics of the pi-calculus and
open bisimulation have already been elegantly specified
in higher-order logic programming systems.
Our solution using Haskell defines the formulae generation as a tree transformation from
the forest of all nondeterministic bisimulation steps to a pair of distinguishing formulae.
Thanks to laziness in Haskell, only the necessary paths demanded by the tree transformation
function are generated. Our work demonstrates that Haskell and its libraries
provide an attractive platform for symbolically analyzing equivalence properties of
labeled transition systems in an environment sensitive setting.

~\\
 \end{abstract}

\ccsdesc[500]{Theory of computation~Process calculi}
\ccsdesc[300]{Theory of computation~Modal and temporal logics}
\ccsdesc[300]{Theory of computation~Constraint and logic programming}
\ccsdesc[300]{Theory of computation~Operational semantics}
\ccsdesc[100]{Theory of computation~Program verification}
\ccsdesc[500]{Software and its engineering~Functional languages}
\ccsdesc[100]{Networks~Protocol testing and verification}
\ccsdesc[100]{Networks~Formal specifications}

\keywords{process calculus, observational equivalence,
labeled transition systems, open bisimulation, modal logic, dynamic logic,
distinguishing formula, Haskell, lazy evaluation, name binding,
constraint programming, nondeterministic programming}

\maketitle

%
%
\makeatletter
\@ifundefined{lhs2tex.lhs2tex.sty.read}%
  {\@namedef{lhs2tex.lhs2tex.sty.read}{}%
   \newcommand\SkipToFmtEnd{}%
   \newcommand\EndFmtInput{}%
   \long\def\SkipToFmtEnd#1\EndFmtInput{}%
  }\SkipToFmtEnd

\newcommand\ReadOnlyOnce[1]{\@ifundefined{#1}{\@namedef{#1}{}}\SkipToFmtEnd}
\usepackage{amstext}
\usepackage{amssymb}
\usepackage{stmaryrd}
\DeclareFontFamily{OT1}{cmtex}{}
\DeclareFontShape{OT1}{cmtex}{m}{n}
  {<5><6><7><8>cmtex8
   <9>cmtex9
   <10><10.95><12><14.4><17.28><20.74><24.88>cmtex10}{}
\DeclareFontShape{OT1}{cmtex}{m}{it}
  {<-> ssub * cmtt/m/it}{}
\newcommand{\texfamily}{\fontfamily{cmtex}\selectfont}
\DeclareFontShape{OT1}{cmtt}{bx}{n}
  {<5><6><7><8>cmtt8
   <9>cmbtt9
   <10><10.95><12><14.4><17.28><20.74><24.88>cmbtt10}{}
\DeclareFontShape{OT1}{cmtex}{bx}{n}
  {<-> ssub * cmtt/bx/n}{}
\newcommand{\tex}[1]{\text{\texfamily#1}}	

\newcommand{\Sp}{\hskip.33334em\relax}

\newcommand{\Conid}[1]{\mathit{#1}}
\newcommand{\Varid}[1]{\mathit{#1}}
\newcommand{\anonymous}{\kern0.06em \vbox{\hrule\@width.5em}}
\newcommand{\plus}{\mathbin{+\!\!\!+}}
\newcommand{\bind}{\mathbin{>\!\!\!>\mkern-6.7mu=}}
\newcommand{\rbind}{\mathbin{=\mkern-6.7mu<\!\!\!<}}
\newcommand{\sequ}{\mathbin{>\!\!\!>}}
\renewcommand{\leq}{\leqslant}
\renewcommand{\geq}{\geqslant}
\usepackage{polytable}

\@ifundefined{mathindent}%
  {\newdimen\mathindent\mathindent\leftmargini}%
  {}%

\def\resethooks{%
  \global\let\SaveRestoreHook\empty
  \global\let\ColumnHook\empty}
\newcommand*{\savecolumns}[1][default]%
  {\g@addto@macro\SaveRestoreHook{\savecolumns[#1]}}
\newcommand*{\restorecolumns}[1][default]%
  {\g@addto@macro\SaveRestoreHook{\restorecolumns[#1]}}
\newcommand*{\aligncolumn}[2]%
  {\g@addto@macro\ColumnHook{\column{#1}{#2}}}

\resethooks

\newcommand{\onelinecommentchars}{\quad-{}- }
\newcommand{\commentbeginchars}{\enskip\{-}
\newcommand{\commentendchars}{-\}\enskip}

\newcommand{\visiblecomments}{%
  \let\onelinecomment=\onelinecommentchars
  \let\commentbegin=\commentbeginchars
  \let\commentend=\commentendchars}

\newcommand{\invisiblecomments}{%
  \let\onelinecomment=\empty
  \let\commentbegin=\empty
  \let\commentend=\empty}

\visiblecomments

\newlength{\blanklineskip}
\setlength{\blanklineskip}{0.66084ex}

\newcommand{\hsindent}[1]{\quad}
\let\hspre\empty
\let\hspost\empty
\newcommand{\NB}{\textbf{NB}}
\newcommand{\Todo}[1]{$\langle$\textbf{To do:}~#1$\rangle$}

\EndFmtInput
\makeatother
%
%
%
%
%
%
%
%
%
\ReadOnlyOnce{polycode.fmt}%
\makeatletter

\newcommand{\hsnewpar}[1]%
  {{\parskip=0pt\parindent=0pt\par\vskip #1\noindent}}

\newcommand{\hscodestyle}{}


\newcommand{\sethscode}[1]%
  {\expandafter\let\expandafter\hscode\csname #1\endcsname
   \expandafter\let\expandafter\endhscode\csname end#1\endcsname}


\newenvironment{compathscode}%
  {\par\noindent
   \advance\leftskip\mathindent
   \hscodestyle
   \let\\=\@normalcr
   \let\hspre\(\let\hspost\)%
   \pboxed}%
  {\endpboxed\)%
   \par\noindent
   \ignorespacesafterend}

\newcommand{\compaths}{\sethscode{compathscode}}


\newenvironment{plainhscode}%
  {\hsnewpar\abovedisplayskip
   \advance\leftskip\mathindent
   \hscodestyle
   \let\hspre\(\let\hspost\)%
   \pboxed}%
  {\endpboxed%
   \hsnewpar\belowdisplayskip
   \ignorespacesafterend}

\newenvironment{oldplainhscode}%
  {\hsnewpar\abovedisplayskip
   \advance\leftskip\mathindent
   \hscodestyle
   \let\\=\@normalcr
   \(\pboxed}%
  {\endpboxed\)%
   \hsnewpar\belowdisplayskip
   \ignorespacesafterend}


\newcommand{\plainhs}{\sethscode{plainhscode}}
\newcommand{\oldplainhs}{\sethscode{oldplainhscode}}
\plainhs


\newenvironment{arrayhscode}%
  {\hsnewpar\abovedisplayskip
   \advance\leftskip\mathindent
   \hscodestyle
   \let\\=\@normalcr
   \(\parray}%
  {\endparray\)%
   \hsnewpar\belowdisplayskip
   \ignorespacesafterend}

\newcommand{\arrayhs}{\sethscode{arrayhscode}}


\newenvironment{mathhscode}%
  {\parray}{\endparray}

\newcommand{\mathhs}{\sethscode{mathhscode}}


\newenvironment{texthscode}%
  {\(\parray}{\endparray\)}

\newcommand{\texths}{\sethscode{texthscode}}


\def\codeframewidth{\arrayrulewidth}
\RequirePackage{calc}

\newenvironment{framedhscode}%
  {\parskip=\abovedisplayskip\par\noindent
   \hscodestyle
   \arrayrulewidth=\codeframewidth
   \tabular{@{}|p{\linewidth-2\arraycolsep-2\arrayrulewidth-2pt}|@{}}%
   \hline\framedhslinecorrect\\{-1.5ex}%
   \let\endoflinesave=\\
   \let\\=\@normalcr
   \(\pboxed}%
  {\endpboxed\)%
   \framedhslinecorrect\endoflinesave{.5ex}\hline
   \endtabular
   \parskip=\belowdisplayskip\par\noindent
   \ignorespacesafterend}

\newcommand{\framedhslinecorrect}[2]%
  {#1[#2]}

\newcommand{\framedhs}{\sethscode{framedhscode}}


\newenvironment{inlinehscode}%
  {\(\def\column##1##2{}%
   \let\>\undefined\let\<\undefined\let\\\undefined
   \newcommand\>[1][]{}\newcommand\<[1][]{}\newcommand\\[1][]{}%
   \def\fromto##1##2##3{##3}%
   \def\nextline{}}{\) }%

\newcommand{\inlinehs}{\sethscode{inlinehscode}}


\newenvironment{joincode}%
  {\let\orighscode=\hscode
   \let\origendhscode=\endhscode
   \def\endhscode{\def\hscode{\endgroup\def\@currenvir{hscode}\\}\begingroup}
   \orighscode\def\hscode{\endgroup\def\@currenvir{hscode}}}%
  {\origendhscode
   \global\let\hscode=\orighscode
   \global\let\endhscode=\origendhscode}%

\makeatother
\EndFmtInput
\ReadOnlyOnce{colorcode.fmt}%

\RequirePackage{colortbl}
\RequirePackage{calc}

\makeatletter
\newenvironment{colorhscode}%
  {\hsnewpar\abovedisplayskip
   \hscodestyle
   \tabular{@{}>{\columncolor{codecolor}}p{\linewidth}@{}}%
   \let\\=\@normalcr
   \(\pboxed}%
  {\endpboxed\)%
   \endtabular
   \hsnewpar\belowdisplayskip
   \ignorespacesafterend}

\newenvironment{tightcolorhscode}%
  {\hsnewpar\abovedisplayskip
   \hscodestyle
   \tabular{@{}>{\columncolor{codecolor}\(}l<{\)}@{}}%
   \pmboxed}%
  {\endpmboxed%
   \endtabular
   \hsnewpar\belowdisplayskip
   \ignorespacesafterend}

\newenvironment{barhscode}%
  {\hsnewpar\abovedisplayskip
   \hscodestyle
   \arrayrulecolor{codecolor}%
   \arrayrulewidth=\coderulewidth
   \tabular{|p{\linewidth-\arrayrulewidth-\tabcolsep}@{}}%
   \let\\=\@normalcr
   \(\pboxed}%
  {\endpboxed\)%
   \endtabular
   \hsnewpar\belowdisplayskip
   \ignorespacesafterend}
\makeatother

\def\colorcode{\columncolor{codecolor}}
\definecolor{codecolor}{rgb}{1,1,.667}
\newlength{\coderulewidth}
\setlength{\coderulewidth}{3pt}

\newcommand{\colorhs}{\sethscode{colorhscode}}
\newcommand{\tightcolorhs}{\sethscode{tightcolorhscode}}
\newcommand{\barhs}{\sethscode{barhscode}}

\EndFmtInput

\renewcommand{\onelinecommentchars}{\color{gray}\quad-{}- }
\renewcommand{\commentbeginchars}{\color{gray}\enskip\{- }
\renewcommand{\commentendchars}{-\}\enskip}

\renewcommand{\visiblecomments}{%
  \let\onelinecomment=\onelinecommentchars
  \let\commentbegin=\commentbeginchars
  \let\commentend=\commentendchars}

\renewcommand{\invisiblecomments}{%
  \let\onelinecomment=\empty
  \let\commentbegin=\empty
  \let\commentend=\empty}

\visiblecomments

$~$\vspace*{19ex}

\section{Introduction}
\label{sec:intro}
The main idea of this paper is that Haskell and its libraries provide
a great platform for analyzing behaviors of nondeterministic transition systems 
in a symbolic way. Our main contribution is identifying an interesting problem
from process calculus and demonstrating its solution in Haskell that supports
this idea. More specifically, we implement automatic generation of
modal logic formulae for two non-open bisimilar processes in the $\pi$-calculus,
which can be machine-checked to witness that the two processes are indeed distinct.

In this section, we give a brief background on the $\pi$-calculus, bisimulation,
and its characterizing logic; discuss the motivating example; and summarize
our contributions. 

\paragraph{The $\pi$-calculus}\hspace{-1.5ex}\cite{Milner92picalcI,Milner92picalcII}
is a formal model of concurrency meant to capture a notion of mobile processes.
The notion of {\em names} plays a central role in this formal model;
communication channels are presented by names; mobility is represented by
scoping of names and {\em scope extrusion} of names. The latter is captured in
the operational semantics via transitions that may send a restricted channel name,
and thereby enlarging its scope.
There are several bisimulation equivalences for the $\pi$-calculus,
notably, early \cite{Milner92picalcII}, 
late~\cite{Milner92picalcII}, and open~\cite{Sangiorgi96acta} bisimilarities.
Only the latter is a congruence and is of main interest in this paper.

Bisimulation equivalence can be alternatively characterized using modal logics.
A modal logic is said to characterize a bisimilarity relation if whenever
two processes are bisimilar then they satisfy the same set of assertions in that modal logic
and vice versa. Such a characterization is useful for analyzing
why bisimulation between two processes fails, since an explicit witness of
non-bisimilarity, in the form of a modal logic formula (also called
a {\em distinguishing formula}), can be constructed such that
one process satisfies the formula while the other does not.
Early and late bisimilarities can be characterized using fragments of
Milner-Parrow-Walker (MPW) logic~\cite{MilParWal93lm}, and a characterization of
open bisimilarity has been recently proposed by \citet{AhnHorTiu17corr} using
a modal logic called \OM. Our work can be seen as a companion of the latter,
showing that the construction of the distinguishing formula described there can
be effectively and naturally implemented in Haskell. 

One main complication in implementing bisimulation checking for the $\pi$-calculus
(and name passing calculi in general) is that the transition system that
a process generates can have infinitely many states, so the traditional
partition-refinement-based algorithm for computing bisimulation and
distinguishing formulae~\cite{Cleaveland90} does not work. Instead,
one needs to construct the state space `on-the-fly', similar to that done in
the Mobility Workbench~\cite{VicMol94mwb}. In our work, this on-the-fly
construction is basically encapsulated in Haskell's lazy evaluation of
the search trees for distinguishing formulae. Another complicating factor is
that in $\pi$-calculus, fresh names can be generated and extruded,
and one needs to keep track of the relative scoping of names.
This is particularly
relevant in open bisimulation, where input names are treated symbolically
(i.e., represented as variables), so care needs to be taken so that,
for example, a variable representing an input name cannot be instantiated by
a fresh name generated after the input action. For this, we rely on
the \textsf{unbound} library~\cite{unbound11}, which uses a locally nameless
representation of terms with binding structures, to represent processes with
bound names and fresh name generation.



\begin{figure}\small
$\begin{array}{rrl}
{\color{ACMRed}\ensuremath{\Conid{P}}}\triangleq \hspace{-2ex}&\hspace{-2ex}
 {\color{ACMRed}\ensuremath{\scalebox{1.3}{$\tau\hspace{-.4ex}\raisebox{.29ex}{\textbf{.}}$}\;(\scalebox{1.3}{$\tau\hspace{-.4ex}\raisebox{.29ex}{\textbf{.}}$}\;\scalebox{1.1}{\bf\texttt{0}})\mathop{.\hspace{-.6ex}{+}\hspace{-.6ex}.}\scalebox{1.3}{$\tau\hspace{-.4ex}\raisebox{.29ex}{\textbf{.}}$}\;\scalebox{1.1}{\bf\texttt{0}}}} &\hspace{-1ex}
	\models \hspace{1.25ex} \langle\tau\rangle\langle\tau\rangle\top \\
& \rotatebox[origin=c]{-90}{$\not\sim_{\!o}$}\phantom{AAA} & \\
{\color{ACMDarkBlue}\ensuremath{\Conid{Q}}}\triangleq \hspace{-2ex}&\hspace{-2ex}
 {\color{ACMDarkBlue}\ensuremath{(\Varid{x}\mathop{{\leftrightarrow}\hspace{-1.48ex}\raisebox{.1ex}{:}\;\,}\Varid{y})\;(\scalebox{1.3}{$\tau\hspace{-.4ex}\raisebox{.29ex}{\textbf{.}}$}\;(\scalebox{1.3}{$\tau\hspace{-.4ex}\raisebox{.29ex}{\textbf{.}}$}\;\scalebox{1.1}{\bf\texttt{0}}))\mathop{.\hspace{-.6ex}{+}\hspace{-.6ex}.}\scalebox{1.3}{$\tau\hspace{-.4ex}\raisebox{.29ex}{\textbf{.}}$}\;\scalebox{1.1}{\bf\texttt{0}}}} &\hspace{-1ex}
	\models \hspace{1.25ex} [\tau]\big(\uwave{\langle x=y \rangle\top} \lor [\tau]\bot\big)
\end{array}
$
\begin{forest}
for tree={edge={->,line width=.75pt},l sep=4ex}
[{}, phantom, s sep=4ex
[{(1)}
   [{\!\!\!\!{\color{ACMRed}\ensuremath{\scalebox{1.3}{$\tau\hspace{-.4ex}\raisebox{.29ex}{\textbf{.}}$}\;\scalebox{1.1}{\bf\texttt{0}}}}\,,\,{\color{ACMDarkBlue}\ensuremath{\Conid{Q}}}},edge label={node[midway,left,color=ACMRed,font=\scriptsize]{\color{ACMRed}$\begin{matrix*}[r]\ensuremath{[\mskip1.5mu \mskip1.5mu]}\\\ensuremath{\ensuremath{\scalebox{1.3}{$\tau$}}}\end{matrix*}$}}, edge+={color=ACMRed}
     [{\!\!\!\!\!\!{\color{ACMRed}\ensuremath{\scalebox{1.3}{$\tau\hspace{-.4ex}\raisebox{.29ex}{\textbf{.}}$}\;\scalebox{1.1}{\bf\texttt{0}}}}\,,\,{\color{ACMDarkBlue}\ensuremath{\scalebox{1.1}{\bf\texttt{0}}}}}, edge label={node[midway,right,color=ACMDarkBlue,color=ACMDarkBlue,font=\scriptsize]{\color{ACMDarkBlue}$\begin{matrix*}[l]\ensuremath{[\mskip1.5mu \mskip1.5mu]}\\\ensuremath{\ensuremath{\scalebox{1.3}{$\tau$}}}\end{matrix*}$}}, edge+=dashed, edge+={color=ACMDarkBlue}
       [{{\color{ACMRed}\ensuremath{\scalebox{1.1}{\bf\texttt{0}}}}\,,\,{\color{ACMDarkBlue}\ensuremath{\scalebox{1.1}{\bf\texttt{0}}}}\!\!\!\!\!\!}, edge label={node[midway,left,color=ACMRed,font=\scriptsize]{\color{ACMRed}$\begin{matrix*}[r]\ensuremath{[\mskip1.5mu \mskip1.5mu]}\\\ensuremath{\ensuremath{\scalebox{1.3}{$\tau$}}}\end{matrix*}$}}, edge+={color=ACMRed} ] ] ]
]
[{(2)}
   [{{\color{ACMRed}\ensuremath{\scalebox{1.1}{\bf\texttt{0}}}}\,,\,{\color{ACMDarkBlue}\ensuremath{\Conid{Q}}}\!}, edge label={node[midway,left,color=ACMRed,font=\scriptsize]{\color{ACMRed}$\begin{matrix*}[r]\ensuremath{[\mskip1.5mu \mskip1.5mu]}\\\ensuremath{\ensuremath{\scalebox{1.3}{$\tau$}}}\end{matrix*}$}}, edge+={color=ACMRed}
     [{{\color{ACMRed}\ensuremath{\scalebox{1.1}{\bf\texttt{0}}}}\,,\,{\color{ACMDarkBlue}\ensuremath{\scalebox{1.1}{\bf\texttt{0}}}}}, edge label={node[midway,right,color=ACMDarkBlue,font=\scriptsize]{\color{ACMDarkBlue}$\begin{matrix*}[l]\ensuremath{[\mskip1.5mu \mskip1.5mu]}\\\ensuremath{\ensuremath{\scalebox{1.3}{$\tau$}}}\end{matrix*}$}}, edge+=dashed, edge+={color=ACMDarkBlue} ] ]
]
[{(3)}
   [{{\color{ACMRed}\ensuremath{\Conid{P}}}\,,\,{\color{ACMDarkBlue}\ensuremath{\scalebox{1.3}{$\tau\hspace{-.4ex}\raisebox{.29ex}{\textbf{.}}$}\;\scalebox{1.1}{\bf\texttt{0}}}}\!\!\!\!\!\!}, edge label={node[midway,right,color=ACMDarkBlue,font=\scriptsize]{\color{ACMDarkBlue}$\begin{matrix*}[l]\ensuremath{\sigma\mathrel{=}\uwave{[\mskip1.5mu (\Varid{x},\Varid{y})\mskip1.5mu]}}\\ \vspace{-4.25ex} \\\ensuremath{\ensuremath{\scalebox{1.3}{$\tau$}}}\end{matrix*}$}}, edge+={color=ACMDarkBlue}
     [{{\color{ACMRed}\ensuremath{\scalebox{1.3}{$\tau\hspace{-.4ex}\raisebox{.29ex}{\textbf{.}}$}\;\scalebox{1.1}{\bf\texttt{0}}}}\,,\,{\color{ACMDarkBlue}\ensuremath{\scalebox{1.3}{$\tau\hspace{-.4ex}\raisebox{.29ex}{\textbf{.}}$}\;\scalebox{1.1}{\bf\texttt{0}}}}}, edge label={node[midway,left,color=ACMRed,font=\scriptsize]{\color{ACMRed}$\begin{matrix*}[r]\ensuremath{\sigma}\\\ensuremath{\ensuremath{\scalebox{1.3}{$\tau$}}}\end{matrix*}$}}, edge+=dashed, edge+={color=ACMRed}, s sep=4.5ex
       [{{\color{ACMRed}\ensuremath{\scalebox{1.1}{\bf\texttt{0}}}}\,,\,{\color{ACMDarkBlue}\ensuremath{\scalebox{1.3}{$\tau\hspace{-.4ex}\raisebox{.29ex}{\textbf{.}}$}\;\scalebox{1.1}{\bf\texttt{0}}}}\!\!\!\!\!\!}, edge label={node[midway,left,color=ACMRed,font=\scriptsize]{\color{ACMRed}$\begin{matrix*}[r]\ensuremath{[\mskip1.5mu \mskip1.5mu]}\\\ensuremath{\ensuremath{\scalebox{1.3}{$\tau$}}}\end{matrix*}$}}, edge+={color=ACMRed}
         [{{\color{ACMRed}\ensuremath{\scalebox{1.1}{\bf\texttt{0}}}}\,,\,{\color{ACMDarkBlue}\ensuremath{\scalebox{1.1}{\bf\texttt{0}}}}}, edge label={node[midway,right,color=ACMDarkBlue,font=\scriptsize]{\color{ACMDarkBlue}$\begin{matrix*}[l]\ensuremath{[\mskip1.5mu \mskip1.5mu]}\\\ensuremath{\ensuremath{\scalebox{1.3}{$\tau$}}}\end{matrix*}$}}, edge+=dashed, edge+={color=ACMDarkBlue} ] ]
       [{\!\!\!\!\!\!{\color{ACMRed}\ensuremath{\scalebox{1.3}{$\tau\hspace{-.4ex}\raisebox{.29ex}{\textbf{.}}$}\;\scalebox{1.1}{\bf\texttt{0}}}}\,,\,{\color{ACMDarkBlue}\ensuremath{\scalebox{1.1}{\bf\texttt{0}}}}}, edge label={node[midway,right,color=ACMDarkBlue,font=\scriptsize]{\color{ACMDarkBlue}$\begin{matrix*}[l]\ensuremath{[\mskip1.5mu \mskip1.5mu]}\\\ensuremath{\ensuremath{\scalebox{1.3}{$\tau$}}}\end{matrix*}$}}, edge+={color=ACMDarkBlue}
         [{{\color{ACMRed}\ensuremath{\scalebox{1.1}{\bf\texttt{0}}}}\,,\,{\color{ACMDarkBlue}\ensuremath{\scalebox{1.1}{\bf\texttt{0}}}}}
, edge label={node[midway,left,color=ACMRed,font=\scriptsize]{\color{ACMRed}$\begin{matrix*}[r]\ensuremath{[\mskip1.5mu \mskip1.5mu]}\\\ensuremath{\ensuremath{\scalebox{1.3}{$\tau$}}}\end{matrix*}$}}, edge+=dashed, edge+={color=ACMRed} ] ]
     ]
     [{{\color{ACMRed}\ensuremath{\scalebox{1.1}{\bf\texttt{0}}}}\,,\,{\color{ACMDarkBlue}\ensuremath{\scalebox{1.3}{$\tau\hspace{-.4ex}\raisebox{.29ex}{\textbf{.}}$}\;\scalebox{1.1}{\bf\texttt{0}}}}\!\!\!\!\!\!}, edge label={node[midway,left,color=ACMRed,font=\scriptsize]{\color{ACMRed}$\begin{matrix*}[r]\ensuremath{\sigma}\\\ensuremath{\ensuremath{\scalebox{1.3}{$\tau$}}}\end{matrix*}$}}, edge+=dashed, edge+={color=ACMRed}
       [{{\color{ACMRed}\ensuremath{\scalebox{1.1}{\bf\texttt{0}}}}\,,\,{\color{ACMDarkBlue}\ensuremath{\scalebox{1.1}{\bf\texttt{0}}}}}, edge label={node[midway,right,color=ACMDarkBlue,font=\scriptsize]{\color{ACMDarkBlue}$\begin{matrix*}[l]\ensuremath{[\mskip1.5mu \mskip1.5mu]}\\\ensuremath{\ensuremath{\scalebox{1.3}{$\tau$}}}\end{matrix*}$}}, edge+={color=ACMDarkBlue} ] ]
   ]
]
[{(4)}
  [{{\color{ACMRed}\ensuremath{\Conid{P}}}\,,\,{\color{ACMDarkBlue}\ensuremath{\scalebox{1.1}{\bf\texttt{0}}}}}, edge label={node[midway,right,color=ACMDarkBlue,font=\scriptsize]{\color{ACMDarkBlue}$\begin{matrix*}[l]\ensuremath{[\mskip1.5mu \mskip1.5mu]}\\\ensuremath{\ensuremath{\scalebox{1.3}{$\tau$}}}\end{matrix*}$}}, edge+={color=ACMDarkBlue}, s sep=4.5ex
    [{{\color{ACMRed}\ensuremath{\scalebox{1.3}{$\tau\hspace{-.4ex}\raisebox{.29ex}{\textbf{.}}$}\;\scalebox{1.1}{\bf\texttt{0}}}}\,,\,{\color{ACMDarkBlue}\ensuremath{\scalebox{1.1}{\bf\texttt{0}}}}\hspace{1ex}}, edge label={node[midway,left,color=ACMRed,font=\scriptsize]{\color{ACMRed}$\begin{matrix*}[r]\ensuremath{[\mskip1.5mu \mskip1.5mu]}\\\ensuremath{\ensuremath{\scalebox{1.3}{$\tau$}}}\end{matrix*}$}}, edge+=dashed, edge+={color=ACMRed}
      [{{\color{ACMRed}\ensuremath{\scalebox{1.1}{\bf\texttt{0}}}}\,,\,{\color{ACMDarkBlue}\ensuremath{\scalebox{1.1}{\bf\texttt{0}}}}}, edge label={node[midway,left,color=ACMRed,font=\scriptsize]{\color{ACMRed}$\begin{matrix*}[r]\ensuremath{[\mskip1.5mu \mskip1.5mu]}\\\ensuremath{\ensuremath{\scalebox{1.3}{$\tau$}}}\end{matrix*}$}}, edge+={color=ACMRed} ]
    ]
    [{\!\!\!\!{\color{ACMRed}\ensuremath{\scalebox{1.1}{\bf\texttt{0}}}}\,,\,{\color{ACMDarkBlue}\ensuremath{\scalebox{1.1}{\bf\texttt{0}}}}}, edge label={node[midway,left,color=ACMRed,font=\scriptsize]{\color{ACMRed}$\begin{matrix*}[r]\ensuremath{[\mskip1.5mu \mskip1.5mu]}\\\ensuremath{\ensuremath{\scalebox{1.3}{$\tau$}}}\end{matrix*}$}}, edge+=dashed, edge+={color=ACMRed} ]
  ]
]
]
\end{forest}
\\[-6.25ex] \hfill
$\begin{matrix*}[l] \rightarrow \text{\scriptsize(leading step)} \\
                    \dashrightarrow \text{\scriptsize(following step)} \end{matrix*}$
\vspace*{-1.5ex}
\caption{The forest of all possible open bisimulation steps of non-bisimilar processes
and their distinguishing formulae.}
\vspace*{-1.5ex}
\label{fig:example}
\end{figure}
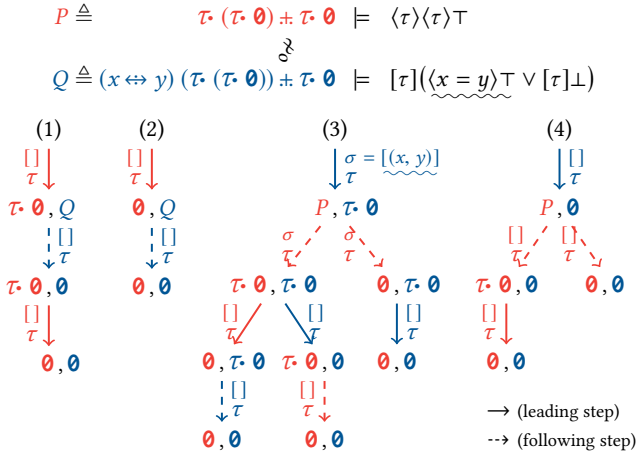

\paragraph{A motivating example\!\!}\!\! in Figure~\ref{fig:example} illustrates
two processes (left-hand side of $\models$), their distinguishing formulae
(right-hand side of $\models$), and all essential steps in an attempt to construct
a bisimulation. We give here only a high-level description of a bisimulation checking
process as search trees and postpone the detail explanation of the syntax and the operational
semantics of $\pi$-calculus to Section~\ref{sec:syntax}.

Bisimulation can be seen as a two-player game, where
every step of a player must be matched by a step by the opponent. In the figure,
the steps by the player (which we refer to as the leading steps) are
denoted by line arrows, whereas the steps by the opponent (the following steps)
are denoted by dotted arrows. 
There are initially four leading steps to consider, corresponding to the cases where
$P$ moves first ((1) and (2)) and where $Q$ moves first ((3) and (4)).


Let us visually examine whether each leading step meets the condition for bisimilarity:
(1) clearly fails the condition because no dotted arrow follows the last line arrow;
(2) clearly satisfies the condition with exactly only one dotted arrow and no more;
(3) satisfies it by taking the left branch where the subtree satisfies the condition;
and (4) also satisfies it by taking the right branch. Therefore, they are
not open bisimilar ($P \not\sim_{\!o} Q$) due to the failure in (1).

A depth first search for bisimulation, scanning from left to
right, only needs to traverse the first tree (1) to notice non-bisimilarity.
Our existing bisimulation checker (prior to this work) is
a higher-order logic program, which runs in this manner.
However, the witness we want to generate contains extra information
(\uwave{wavy underlined}), which are not found in (1) but in (3). Therefore,
simply logging all the visited steps during a run of a bisimulation check
is insufficient.

The extra information {\small\ensuremath{\sigma\mathrel{=}[\mskip1.5mu (\Varid{x},\Varid{y})\mskip1.5mu]}} represents a substitution
that unifies \ensuremath{\Varid{x}} and \ensuremath{\Varid{y}}. The third tree (3) considers the leading step
initiated by the subprocess \ensuremath{(\Varid{x}\mathop{{\leftrightarrow}\hspace{-1.48ex}\raisebox{.1ex}{:}\;\,}\Varid{y})\;(\scalebox{1.3}{$\tau\hspace{-.4ex}\raisebox{.29ex}{\textbf{.}}$}\;(\scalebox{1.3}{$\tau\hspace{-.4ex}\raisebox{.29ex}{\textbf{.}}$}\;\scalebox{1.1}{\bf\texttt{0}}))}, which
can only make a step in a world (or environment)
where \ensuremath{\Varid{x}} and \ensuremath{\Varid{y}} are equivalent. Our earlier implementation uses
a logic programming language,
relying on a representation of \ensuremath{\Varid{x}} and \ensuremath{\Varid{y}} as unifiable logic variables and 
on backtracking for nondeterminism.
However, it is difficult to access \ensuremath{\sigma} in this setting because \ensuremath{\sigma}
resides inside the system state rather than being a first-class value.
Access to logic variable substitutions since the definition of
open bisimulation and the generation of distinguishing formulae
require access to and manipulations of such substitutions.
Moreover,
the information is lost when backtracking to another branch, for instance,
from (3) to (4).

On the other hand, it is very natural in Haskell to view all possible
nondeterministic steps as tree structured data because of laziness.
Once we are able to produce the trees in Figure~\ref{fig:example}
(Section~\ref{sec:bisim}), our problem reduces to a transformation
from trees to formulae (Section~\ref{sec:df}). Thanks to laziness,
only those nodes demanded by the tree transformation function will
actually be computed. We also have constraints (i.e., substitutions)
as first-class values with an overhead of being more explicit about
substitutions compared to logic programming.

In order to produce the trees of bisimulation steps, we first
need to define the syntax (Section~\ref{sec:syntax:pi}) and
semantics (Section~\ref{sec:lts}) of the $\pi$-calculus in Haskell.
We also need to define the syntax of our modal logic formulae
(Section~\ref{sec:syntax:om}) for the return value of
the tree transformation function. However, we do not need to
implement the semantics of the logic because we can check the generated formulae
with our existing formula satisfaction ($\models$) checker.

\paragraph{Our contributions}\hspace{-1.5ex} are summarized as follows:\vspace*{-.75ex}
\begin{itemize}
\item
We identified a problem that generating certificates witnessing the failure of
process equivalence checking is non-trivial in a logic programming setting
(Figure~\ref{fig:example}), even though the equivalence property itself has been
elegantly specified as a logic program.
\vspace*{.5ex}
\item
The crux of our solution is a tree transformation from the forest of all possible
bisimulation steps to a pair of distinguishing formulae (Section~\ref{sec:df}).
The definition of tree transformation (Figure~\ref{fig:df}) is clear and easy to
understand because we are conceptually working on all possible nondeterministic steps.
Nevertheless, unnecessary computations are avoided by laziness.
\vspace*{.5ex}
\item
We demonstrate that the overhead of re-implementing the syntax
(Section~\ref{sec:syntax}), labeled transition semantics (Section~\ref{sec:lts}),
and open bisimulation checker (Section~\ref{sec:bisim}) in Haskell, which
we already had as a logic program, and then augmenting it to produce trees
is relatively small. In fact, most of the source code, omitting repetitive
symmetric cases, is laid out as figures (Figures~\ref{fig:PiCalc},\,\ref{fig:IdSubLTS},\,\ref{fig:OpenLTS},\,\ref{fig:figureOpenLTS},\;and\;\ref{fig:sim}).%
\vspace*{.5ex}
\item
Our implementation of generating distinguishing formulae is a pragmatic evidence
that reassures our recent theoretical development~\cite{AhnHorTiu17corr} of
the modal logic \OM\ being a characterizing logic for open bisimilarity
(i.e., distinguishing formulae exists iff non-open bisimilar). In this paper,
we define the syntax of \OM\ formulae in Haskell and explain their intuitive
meanings (Section~\ref{sec:syntax:om}), and provide pointers to related work
(Section~\ref{sec:relwork}).
\end{itemize}
We used \textsf{lhs2TeX} to formatt the paper from literate haskell scripts
({\small\url{https://github.com/kyagrd/hs-picalc-unbound-example}}).

%
%
\makeatletter
\@ifundefined{lhs2tex.lhs2tex.sty.read}%
  {\@namedef{lhs2tex.lhs2tex.sty.read}{}%
   \newcommand\SkipToFmtEnd{}%
   \newcommand\EndFmtInput{}%
   \long\def\SkipToFmtEnd#1\EndFmtInput{}%
  }\SkipToFmtEnd

\newcommand\ReadOnlyOnce[1]{\@ifundefined{#1}{\@namedef{#1}{}}\SkipToFmtEnd}
\usepackage{amstext}
\usepackage{amssymb}
\usepackage{stmaryrd}
\DeclareFontFamily{OT1}{cmtex}{}
\DeclareFontShape{OT1}{cmtex}{m}{n}
  {<5><6><7><8>cmtex8
   <9>cmtex9
   <10><10.95><12><14.4><17.28><20.74><24.88>cmtex10}{}
\DeclareFontShape{OT1}{cmtex}{m}{it}
  {<-> ssub * cmtt/m/it}{}
\newcommand{\texfamily}{\fontfamily{cmtex}\selectfont}
\DeclareFontShape{OT1}{cmtt}{bx}{n}
  {<5><6><7><8>cmtt8
   <9>cmbtt9
   <10><10.95><12><14.4><17.28><20.74><24.88>cmbtt10}{}
\DeclareFontShape{OT1}{cmtex}{bx}{n}
  {<-> ssub * cmtt/bx/n}{}
\newcommand{\tex}[1]{\text{\texfamily#1}}	

\newcommand{\Sp}{\hskip.33334em\relax}

\newcommand{\Conid}[1]{\mathit{#1}}
\newcommand{\Varid}[1]{\mathit{#1}}
\newcommand{\anonymous}{\kern0.06em \vbox{\hrule\@width.5em}}
\newcommand{\plus}{\mathbin{+\!\!\!+}}
\newcommand{\bind}{\mathbin{>\!\!\!>\mkern-6.7mu=}}
\newcommand{\rbind}{\mathbin{=\mkern-6.7mu<\!\!\!<}}
\newcommand{\sequ}{\mathbin{>\!\!\!>}}
\renewcommand{\leq}{\leqslant}
\renewcommand{\geq}{\geqslant}
\usepackage{polytable}

\@ifundefined{mathindent}%
  {\newdimen\mathindent\mathindent\leftmargini}%
  {}%

\def\resethooks{%
  \global\let\SaveRestoreHook\empty
  \global\let\ColumnHook\empty}
\newcommand*{\savecolumns}[1][default]%
  {\g@addto@macro\SaveRestoreHook{\savecolumns[#1]}}
\newcommand*{\restorecolumns}[1][default]%
  {\g@addto@macro\SaveRestoreHook{\restorecolumns[#1]}}
\newcommand*{\aligncolumn}[2]%
  {\g@addto@macro\ColumnHook{\column{#1}{#2}}}

\resethooks

\newcommand{\onelinecommentchars}{\quad-{}- }
\newcommand{\commentbeginchars}{\enskip\{-}
\newcommand{\commentendchars}{-\}\enskip}

\newcommand{\visiblecomments}{%
  \let\onelinecomment=\onelinecommentchars
  \let\commentbegin=\commentbeginchars
  \let\commentend=\commentendchars}

\newcommand{\invisiblecomments}{%
  \let\onelinecomment=\empty
  \let\commentbegin=\empty
  \let\commentend=\empty}

\visiblecomments

\newlength{\blanklineskip}
\setlength{\blanklineskip}{0.66084ex}

\newcommand{\hsindent}[1]{\quad}
\let\hspre\empty
\let\hspost\empty
\newcommand{\NB}{\textbf{NB}}
\newcommand{\Todo}[1]{$\langle$\textbf{To do:}~#1$\rangle$}

\EndFmtInput
\makeatother
%
%
%
%
%
%
%
%
%
\ReadOnlyOnce{polycode.fmt}%
\makeatletter

\newcommand{\hsnewpar}[1]%
  {{\parskip=0pt\parindent=0pt\par\vskip #1\noindent}}

\newcommand{\hscodestyle}{}


\newcommand{\sethscode}[1]%
  {\expandafter\let\expandafter\hscode\csname #1\endcsname
   \expandafter\let\expandafter\endhscode\csname end#1\endcsname}


\newenvironment{compathscode}%
  {\par\noindent
   \advance\leftskip\mathindent
   \hscodestyle
   \let\\=\@normalcr
   \let\hspre\(\let\hspost\)%
   \pboxed}%
  {\endpboxed\)%
   \par\noindent
   \ignorespacesafterend}

\newcommand{\compaths}{\sethscode{compathscode}}


\newenvironment{plainhscode}%
  {\hsnewpar\abovedisplayskip
   \advance\leftskip\mathindent
   \hscodestyle
   \let\hspre\(\let\hspost\)%
   \pboxed}%
  {\endpboxed%
   \hsnewpar\belowdisplayskip
   \ignorespacesafterend}

\newenvironment{oldplainhscode}%
  {\hsnewpar\abovedisplayskip
   \advance\leftskip\mathindent
   \hscodestyle
   \let\\=\@normalcr
   \(\pboxed}%
  {\endpboxed\)%
   \hsnewpar\belowdisplayskip
   \ignorespacesafterend}


\newcommand{\plainhs}{\sethscode{plainhscode}}
\newcommand{\oldplainhs}{\sethscode{oldplainhscode}}
\plainhs


\newenvironment{arrayhscode}%
  {\hsnewpar\abovedisplayskip
   \advance\leftskip\mathindent
   \hscodestyle
   \let\\=\@normalcr
   \(\parray}%
  {\endparray\)%
   \hsnewpar\belowdisplayskip
   \ignorespacesafterend}

\newcommand{\arrayhs}{\sethscode{arrayhscode}}


\newenvironment{mathhscode}%
  {\parray}{\endparray}

\newcommand{\mathhs}{\sethscode{mathhscode}}


\newenvironment{texthscode}%
  {\(\parray}{\endparray\)}

\newcommand{\texths}{\sethscode{texthscode}}


\def\codeframewidth{\arrayrulewidth}
\RequirePackage{calc}

\newenvironment{framedhscode}%
  {\parskip=\abovedisplayskip\par\noindent
   \hscodestyle
   \arrayrulewidth=\codeframewidth
   \tabular{@{}|p{\linewidth-2\arraycolsep-2\arrayrulewidth-2pt}|@{}}%
   \hline\framedhslinecorrect\\{-1.5ex}%
   \let\endoflinesave=\\
   \let\\=\@normalcr
   \(\pboxed}%
  {\endpboxed\)%
   \framedhslinecorrect\endoflinesave{.5ex}\hline
   \endtabular
   \parskip=\belowdisplayskip\par\noindent
   \ignorespacesafterend}

\newcommand{\framedhslinecorrect}[2]%
  {#1[#2]}

\newcommand{\framedhs}{\sethscode{framedhscode}}


\newenvironment{inlinehscode}%
  {\(\def\column##1##2{}%
   \let\>\undefined\let\<\undefined\let\\\undefined
   \newcommand\>[1][]{}\newcommand\<[1][]{}\newcommand\\[1][]{}%
   \def\fromto##1##2##3{##3}%
   \def\nextline{}}{\) }%

\newcommand{\inlinehs}{\sethscode{inlinehscode}}


\newenvironment{joincode}%
  {\let\orighscode=\hscode
   \let\origendhscode=\endhscode
   \def\endhscode{\def\hscode{\endgroup\def\@currenvir{hscode}\\}\begingroup}
   \orighscode\def\hscode{\endgroup\def\@currenvir{hscode}}}%
  {\origendhscode
   \global\let\hscode=\orighscode
   \global\let\endhscode=\origendhscode}%

\makeatother
\EndFmtInput
\ReadOnlyOnce{colorcode.fmt}%

\RequirePackage{colortbl}
\RequirePackage{calc}

\makeatletter
\newenvironment{colorhscode}%
  {\hsnewpar\abovedisplayskip
   \hscodestyle
   \tabular{@{}>{\columncolor{codecolor}}p{\linewidth}@{}}%
   \let\\=\@normalcr
   \(\pboxed}%
  {\endpboxed\)%
   \endtabular
   \hsnewpar\belowdisplayskip
   \ignorespacesafterend}

\newenvironment{tightcolorhscode}%
  {\hsnewpar\abovedisplayskip
   \hscodestyle
   \tabular{@{}>{\columncolor{codecolor}\(}l<{\)}@{}}%
   \pmboxed}%
  {\endpmboxed%
   \endtabular
   \hsnewpar\belowdisplayskip
   \ignorespacesafterend}

\newenvironment{barhscode}%
  {\hsnewpar\abovedisplayskip
   \hscodestyle
   \arrayrulecolor{codecolor}%
   \arrayrulewidth=\coderulewidth
   \tabular{|p{\linewidth-\arrayrulewidth-\tabcolsep}@{}}%
   \let\\=\@normalcr
   \(\pboxed}%
  {\endpboxed\)%
   \endtabular
   \hsnewpar\belowdisplayskip
   \ignorespacesafterend}
\makeatother

\def\colorcode{\columncolor{codecolor}}
\definecolor{codecolor}{rgb}{1,1,.667}
\newlength{\coderulewidth}
\setlength{\coderulewidth}{3pt}

\newcommand{\colorhs}{\sethscode{colorhscode}}
\newcommand{\tightcolorhs}{\sethscode{tightcolorhscode}}
\newcommand{\barhs}{\sethscode{barhscode}}

\EndFmtInput

\renewcommand{\onelinecommentchars}{\color{gray}\quad-{}- }
\renewcommand{\commentbeginchars}{\color{gray}\enskip\{- }
\renewcommand{\commentendchars}{-\}\enskip}

\renewcommand{\visiblecomments}{%
  \let\onelinecomment=\onelinecommentchars
  \let\commentbegin=\commentbeginchars
  \let\commentend=\commentendchars}

\renewcommand{\invisiblecomments}{%
  \let\onelinecomment=\empty
  \let\commentbegin=\empty
  \let\commentend=\empty}

\visiblecomments

\begin{figure}\small
\begin{hscode}\SaveRestoreHook
\column{B}{@{}>{\hspre}l<{\hspost}@{}}%
\column{7}{@{}>{\hspre}l<{\hspost}@{}}%
\column{10}{@{}>{\hspre}l<{\hspost}@{}}%
\column{12}{@{}>{\hspre}l<{\hspost}@{}}%
\column{19}{@{}>{\hspre}l<{\hspost}@{}}%
\column{22}{@{}>{\hspre}l<{\hspost}@{}}%
\column{24}{@{}>{\hspre}l<{\hspost}@{}}%
\column{28}{@{}>{\hspre}l<{\hspost}@{}}%
\column{29}{@{}>{\hspre}c<{\hspost}@{}}%
\column{29E}{@{}l@{}}%
\column{32}{@{}>{\hspre}l<{\hspost}@{}}%
\column{34}{@{}>{\hspre}l<{\hspost}@{}}%
\column{37}{@{}>{\hspre}l<{\hspost}@{}}%
\column{38}{@{}>{\hspre}l<{\hspost}@{}}%
\column{50}{@{}>{\hspre}l<{\hspost}@{}}%
\column{51}{@{}>{\hspre}l<{\hspost}@{}}%
\column{63}{@{}>{\hspre}l<{\hspost}@{}}%
\column{E}{@{}>{\hspre}l<{\hspost}@{}}%
\>[B]{}\mathbf{module}\;\Conid{PiCalc}\;\mathbf{where}{}\<[E]%
\\
\>[B]{}\mathbf{import}\;\Conid{\Conid{Unbound}.LocallyNameless}{}\<[E]%
\\[\blanklineskip]%
\>[B]{}\mathbf{type}\;\Conid{Nm}\mathrel{=}\Conid{Name}\;\Conid{Tm}{}\<[E]%
\\
\>[B]{}\mathbf{newtype}\;\Conid{Tm}\mathrel{=}V\!\;\Conid{Nm}\;\mathbf{deriving}\;(\Conid{Eq},\Conid{Ord},\Conid{Show}){}\<[E]%
\\[\blanklineskip]%
\>[B]{}\mathbf{data}\;\Conid{Pr}{}\<[10]%
\>[10]{}\mathrel{=}\scalebox{1.1}{\bf\texttt{0}}\mid \scalebox{1.3}{$\tau\hspace{-.4ex}\raisebox{.29ex}{\textbf{.}}$}\;\Conid{Pr}\mid \Conid{Out}\;\Conid{Tm}\;\Conid{Tm}\;\Conid{Pr}\mid \Conid{In}\;\Conid{Tm}\;\textit{Pr}_{\textsc{b}}\mid (\Conid{Tm}\mathop{{\leftrightarrow}\hspace{-1.48ex}\raisebox{.1ex}{:}\;\,}\Conid{Tm})\;\Conid{Pr}{}\<[E]%
\\
\>[10]{}\mid \Conid{Pr}\mathop{.\hspace{-.6ex}{+}\hspace{-.6ex}.}\Conid{Pr}\mid \Conid{Pr}\mathop{\|}\Conid{Pr}\mid \scalebox{1.25}{$\nu$}\!\;\textit{Pr}_{\textsc{b}}\hspace{4ex}\;\mathbf{deriving}\;(\Conid{Eq},\Conid{Ord},\Conid{Show}){}\<[E]%
\\
\>[B]{}\mathbf{type}\;\textit{Pr}_{\textsc{b}}\mathrel{=}\Conid{Bind}\;\Conid{Nm}\;\Conid{Pr}{}\<[E]%
\\
\>[B]{}\mathbf{instance}\;\Conid{Eq}\;\textit{Pr}_{\textsc{b}}\;\mathbf{where}\;(\equiv )\mathrel{=}\Varid{aeqBinders}{}\<[E]%
\\
\>[B]{}\mathbf{instance}\;\Conid{Ord}\;\textit{Pr}_{\textsc{b}}\;\mathbf{where}\;\Varid{compare}\mathrel{=}\Varid{acompare}{}\<[E]%
\\[\blanklineskip]%
\>[B]{}\mathbf{data}\;\Conid{Act}{}\<[12]%
\>[12]{}\mathrel{=}\MVUparrow\!\;\Conid{Tm}\;\Conid{Tm}{}\<[24]%
\>[24]{}\mid \ensuremath{\scalebox{1.3}{$\tau$}}\;{}\<[34]%
\>[34]{}\mathbf{deriving}\;(\Conid{Eq},\Conid{Ord},\Conid{Show}){}\<[E]%
\\
\>[B]{}\mathbf{data}\;\textit{Act}_{\textsc{b}}{}\<[12]%
\>[12]{}\mathrel{=}\MVUparrow_{\!\textsc{b}\!}\;\Conid{Tm}{}\<[24]%
\>[24]{}\mid \hspace{-.15ex}\MVDnarrow^{\hspace{-.18ex}\textsc{b}\!}\;\Conid{Tm}\;{}\<[34]%
\>[34]{}\mathbf{deriving}\;(\Conid{Eq},\Conid{Ord},\Conid{Show}){}\<[E]%
\\[\blanklineskip]%
\>[B]{}\mathbf{data}\;\Conid{Form}{}\<[12]%
\>[12]{}\mathrel{=}\bot\mid \top\mid \bigwedge\!\;[\mskip1.5mu \Conid{Form}\mskip1.5mu]\mid \bigvee\!\;[\mskip1.5mu \Conid{Form}\mskip1.5mu]{}\<[E]%
\\
\>[12]{}\mid \Diamond\!\;{}\<[19]%
\>[19]{}\Conid{Act}\;\Conid{Form}{}\<[29]%
\>[29]{}\mid {}\<[29E]%
\>[32]{}\Diamond_{^{\!}\textsc{b}}\!\;{}\<[38]%
\>[38]{}\textit{Act}_{\textsc{b}}\;\textit{Form}_{\textsc{b}}{}\<[51]%
\>[51]{}\mid \Diamond_{\!=}\!\;[\mskip1.5mu (\Conid{Tm},\Conid{Tm})\mskip1.5mu]\;\Conid{Form}{}\<[E]%
\\
\>[12]{}\mid \Box\!\;{}\<[19]%
\>[19]{}\Conid{Act}\;\Conid{Form}{}\<[29]%
\>[29]{}\mid {}\<[29E]%
\>[32]{}\Box_{^{\!}\textsc{b}}\!\;{}\<[38]%
\>[38]{}\textit{Act}_{\textsc{b}}\;\textit{Form}_{\textsc{b}}{}\<[51]%
\>[51]{}\mid \Box_{=}\!\;[\mskip1.5mu (\Conid{Tm},\Conid{Tm})\mskip1.5mu]\;\Conid{Form}{}\<[E]%
\\
\>[12]{}\mathbf{deriving}\;(\Conid{Eq},\Conid{Ord},\Conid{Show}){}\<[E]%
\\
\>[B]{}\mathbf{type}\;\textit{Form}_{\textsc{b}}\mathrel{=}\Conid{Bind}\;\Conid{Nm}\;\Conid{Form}{}\<[E]%
\\
\>[B]{}\mathbf{instance}\;\Conid{Eq}\;\textit{Form}_{\textsc{b}}\;\mathbf{where}\;(\equiv )\mathrel{=}\Varid{aeqBinders}{}\<[E]%
\\
\>[B]{}\mathbf{instance}\;\Conid{Ord}\;\textit{Form}_{\textsc{b}}\;\mathbf{where}\;\Varid{compare}\mathrel{=}\Varid{acompare}{}\<[E]%
\\[\blanklineskip]%
\>[B]{}\mathop{\texttt{\$}}(\Varid{derive}\;[\mskip1.5mu \texttt{'\!'\!\!}\;\Conid{Tm},\texttt{'\!'\!\!}\;\Conid{Act},\texttt{'\!'\!\!}\;\textit{Act}_{\textsc{b}},\texttt{'\!'\!\!}\;\Conid{Pr},\texttt{'\!'\!\!}\;\Conid{Form}\mskip1.5mu]){}\<[E]%
\\[\blanklineskip]%
\>[B]{}\mathbf{instance}\;\Conid{Alpha}\;\Conid{Tm}\hspace{.2ex};\hspace{.2ex}\;{}\<[28]%
\>[28]{}\mathbf{instance}\;\Conid{Alpha}\;\Conid{Act}\hspace{.2ex};\hspace{.2ex}\;\mathbf{instance}\;\Conid{Alpha}\;\textit{Act}_{\textsc{b}}{}\<[E]%
\\
\>[B]{}\mathbf{instance}\;\Conid{Alpha}\;\Conid{Pr}\hspace{.2ex};\hspace{.2ex}\;{}\<[28]%
\>[28]{}\mathbf{instance}\;\Conid{Alpha}\;\Conid{Form}{}\<[E]%
\\[\blanklineskip]%
\>[B]{}\mathbf{instance}\;\Conid{Subst}\;\Conid{Tm}\;\Conid{Tm}\;\mathbf{where}\;\Varid{isvar}\;(V\!\;\Varid{x})\mathrel{=}\Conid{Just}\;(\Conid{SubstName}\;\Varid{x}){}\<[E]%
\\
\>[B]{}\mathbf{instance}\;\Conid{Subst}\;\Conid{Tm}\;\Conid{Act}\hspace{.2ex};\hspace{.2ex}\;\mathbf{instance}\;\Conid{Subst}\;\Conid{Tm}\;\textit{Act}_{\textsc{b}}{}\<[E]%
\\
\>[B]{}\mathbf{instance}\;\Conid{Subst}\;\Conid{Tm}\;\Conid{Pr}\hspace{.2ex};\hspace{.2ex}\;\mathbf{instance}\;\Conid{Subst}\;\Conid{Tm}\;\Conid{Form}{}\<[E]%
\\[\blanklineskip]%
\>[B]{}\mathbf{infixr}\;\mathrm{1}\hspace{.1ex}.\hspace{-.4ex}\backslash\,\hspace{.2ex};\hspace{.2ex}\,(\hspace{.1ex}.\hspace{-.4ex}\backslash)\mathbin{::}\Conid{Alpha}\;\Varid{t}\Rightarrow \Conid{Nm}\to \Varid{t}\to \Conid{Bind}\;\Conid{Nm}\;\Varid{t}\,\hspace{.2ex};\hspace{.2ex}\,(\hspace{.1ex}.\hspace{-.4ex}\backslash)\mathrel{=}\Varid{bind}{}\<[E]%
\\[\blanklineskip]%
\>[B]{}\Varid{x}\leftrightarrow\Varid{y}\mathrel{=}(V\!\;\Varid{x}\mathop{{\leftrightarrow}\hspace{-1.48ex}\raisebox{.1ex}{:}\;\,}V\!\;\Varid{y})\,\hspace{.2ex};\hspace{.2ex}~\Varid{inp}\mathrel{=}\Conid{In}\mathbin{\circ}V\!\,\hspace{.2ex};\hspace{.2ex}~\Varid{out}\;\Varid{x}\;\Varid{y}\mathrel{=}\Conid{Out}\;(V\!\;\Varid{x})\;(V\!\;\Varid{y}){}\<[E]%
\\
\>[B]{}\tau\mathrel{=}\scalebox{1.3}{$\tau\hspace{-.4ex}\raisebox{.29ex}{\textbf{.}}$}\;\scalebox{1.1}{\bf\texttt{0}}\,\hspace{.2ex};\hspace{.2ex}\quad\tau\tau\mathrel{=}\scalebox{1.3}{$\tau\hspace{-.4ex}\raisebox{.29ex}{\textbf{.}}$}\;(\scalebox{1.3}{$\tau\hspace{-.4ex}\raisebox{.29ex}{\textbf{.}}$}\;\scalebox{1.1}{\bf\texttt{0}}){}\<[E]%
\\[\blanklineskip]%
\>[B]{}\Varid{conj}{}\<[7]%
\>[7]{}\mathrel{=}\Varid{cn}\mathbin{\circ}\Varid{filter}\;(\not\equiv \top)\;\mathbf{where}\;\Varid{cn}\;{}\<[37]%
\>[37]{}[\mskip1.5mu \mskip1.5mu]\mathrel{=}\top\hspace{.2ex};\hspace{.2ex}\Varid{cn}\;{}\<[50]%
\>[50]{}[\mskip1.5mu \Varid{f}\mskip1.5mu]\mathrel{=}\Varid{f}\hspace{.2ex};\hspace{.2ex}\Varid{cn}\;{}\<[63]%
\>[63]{}\Varid{fs}\mathrel{=}\bigwedge\!\;\Varid{fs}{}\<[E]%
\\
\>[B]{}\Varid{disj}{}\<[7]%
\>[7]{}\mathrel{=}\Varid{ds}\mathbin{\circ}\Varid{filter}\;(\not\equiv \bot)\;\mathbf{where}\;\Varid{ds}\;{}\<[37]%
\>[37]{}[\mskip1.5mu \mskip1.5mu]\mathrel{=}\bot\hspace{.2ex};\hspace{.2ex}\Varid{ds}\;{}\<[50]%
\>[50]{}[\mskip1.5mu \Varid{f}\mskip1.5mu]\mathrel{=}\Varid{f}\hspace{.2ex};\hspace{.2ex}\Varid{ds}\;{}\<[63]%
\>[63]{}\Varid{fs}\mathrel{=}\bigvee\!\;\Varid{fs}{}\<[E]%
\\[\blanklineskip]%
\>[B]{}\Varid{unbind2'}\;\Varid{b}_{1}\;\Varid{b}_{2}\mathrel{=}\mathbf{do}\;{}\<[22]%
\>[22]{}\Conid{Just}\;(\Varid{x},\Varid{p}_{1},\anonymous ,\Varid{p}_{2})\leftarrow \Varid{unbind2}\;\Varid{b}_{1}\;\Varid{b}_{2}{}\<[E]%
\\
\>[22]{}\Varid{return}\;(\Varid{x},\Varid{p}_{1},\Varid{p}_{2}){}\<[E]%
\ColumnHook
\end{hscode}\resethooks
\vspace*{-4ex}
\caption{Syntax of the $\pi$-calculus and the modal logic \OM.}
\label{fig:PiCalc}
\vspace*{-2.5ex}
\end{figure}

\section{Syntax}
\label{sec:syntax}
In this section, we define the syntax for the $\pi$-calculus and
the modal logic, which characterizes open bisimilarity.
Haskell definitions of the syntax for both are provided
in the module \ensuremath{\Conid{PiCalc}} as illustrated in Figure~\ref{fig:PiCalc}.


Since we consider only the original version of the $\pi$-calculus
with name passing, terms (\ensuremath{\Conid{Tm}}) that can be sent through channel names consist
only of names. 
Processes (\ensuremath{\Conid{Pr}}) may contain bound names due to
value passing and name restriction. In the Haskell definition,
we define these name binding constructs with the generic binding
scheme (\ensuremath{\Conid{Bind}}) from the \textsf{unbound}~\cite{unbound11} library.
We can construct a bound process (\ensuremath{\textit{Pr}_{\textsc{b}}}, i.e., \ensuremath{\Conid{Bind}\;\Conid{Nm}\;\Conid{Pr}}) by applying
the binding operator \ensuremath{(\hspace{.1ex}.\hspace{-.4ex}\backslash)} to a name (\ensuremath{\Conid{Nm}}) that may be used in
a process (\ensuremath{\Conid{Pr}}), i.e., \ensuremath{(\Varid{x}\hspace{.1ex}.\hspace{-.4ex}\backslash\Varid{p})\mathbin{::}\textit{Pr}_{\textsc{b}}} given \ensuremath{\Varid{x}\mathbin{::}\Conid{Nm}} and \ensuremath{\Varid{p}\mathbin{::}\Conid{Pr}}.
Intuitively, our Haskell expression \ensuremath{(\Varid{x}\hspace{.1ex}.\hspace{-.4ex}\backslash\Varid{p})} corresponds to
a lambda-term $(\lambda x.p)$. Similarly, we define name bindings
in the logic formulae (\ensuremath{\Conid{Form}}) with \ensuremath{\textit{Form}_{\textsc{b}}} defined as \ensuremath{\Conid{Bind}\;\Conid{Nm}\;\Conid{Form}}.
We get $\alpha$-equivalence and capture-avoiding substitutions over
processes and formulae almost for free, with a few lines of instance
declarations, thanks to the \textsf{unbound} library.

In addition to the binding operator \ensuremath{(\hspace{.1ex}.\hspace{-.4ex}\backslash)}, we define some utility functions:
(\ensuremath{\leftrightarrow}), \ensuremath{\Varid{inp}}, and \ensuremath{\Varid{out}} are wrappers to the data constructors of \ensuremath{\Conid{Pr}},
for example, \ensuremath{\Varid{out}\;\Varid{x}\;\Varid{y}\equiv \Conid{Out}\;(\Conid{V}\;\Varid{x})\;(\Conid{V}\;\Varid{y})}; \ensuremath{\tau} and \ensuremath{\tau\tau}
are shorthand names of example processes;
\ensuremath{\Varid{conj}} and \ensuremath{\Varid{disj}} are wrappers of \ensuremath{\bigwedge\!} and \ensuremath{\bigvee\!} with
obvious simplifications, for example, \ensuremath{\Varid{f}\equiv \Varid{conj}\;[\mskip1.5mu \top,\Varid{f}\mskip1.5mu]}; and
\ensuremath{\Varid{undbind2'}} is a wrapper to the library function \ensuremath{\Varid{unbind2}}, which unbinds two bound structures
by a common name, for example,\\
$\phantom{i}${\small\ensuremath{(\Varid{x},\Varid{out}\;\Varid{x}\;\Varid{x}\;\scalebox{1.1}{\bf\texttt{0}},\Varid{out}\;\Varid{x}\;\Varid{x}\;\tau)\leftarrow \Varid{unbind2}\;(\Varid{x}\hspace{.1ex}.\hspace{-.4ex}\backslash\Varid{out}\;\Varid{x}\;\Varid{x}\;\scalebox{1.1}{\bf\texttt{0}})\;(\Varid{y}\hspace{.1ex}.\hspace{-.4ex}\backslash\Varid{out}\;\Varid{y}\;\Varid{y}\;\tau)}}.
There is of course a more basic library function \textit{unbind} for a single bound structure,
which is formatted as \ensuremath{(\hspace{.1ex}.\hspace{-.4ex}\backslash)^{{\text{-}\hspace{-.2ex}1\!}}} in this paper because it acts like an inverse of \ensuremath{(\hspace{.1ex}.\hspace{-.4ex}\backslash)}.

As a convention, we use Haskell names suffixed by {\textsc{b}} to emphasize
that those definitions are related to bound structures. Naming conventions
for the values of other data types in Figure~\ref{fig:PiCalc} are:
\ensuremath{\Varid{x}}, \ensuremath{\Varid{y}}, \ensuremath{\Varid{z}}, and \ensuremath{\Varid{w}} for both terms (\ensuremath{\Conid{Tm}}) and names (\ensuremath{\Conid{Nm}});
\ensuremath{\Varid{v}} for terms (\ensuremath{\Conid{Tm}});
\ensuremath{\Varid{p}} and \ensuremath{\Varid{q}} for processes (\ensuremath{\Conid{Pr}}); \ensuremath{\Varid{b}} for bound processes (\ensuremath{\textit{Pr}_{\textsc{b}}});
\ensuremath{\Varid{a}} and \ensuremath{\Varid{l}} for both free and bound actions (\ensuremath{\Conid{Act}} and \ensuremath{\textit{Act}_{\textsc{b}}}); and
\ensuremath{\Varid{f}} for formulae (\ensuremath{\Conid{Form}}).

In the following subsections, we explain further details of
the finite $\pi$-calculus (Section~\ref{sec:syntax:pi}) and 
the modal logic (Section~\ref{sec:syntax:om}) including
the intuitive meanings of their syntax.

\subsection{Finite $\pi$-Calculus}
\label{sec:syntax:pi}
\begin{figure*}
$\displaystyle 
\begin{matrix} \phantom{a} \\ \ensuremath{\scalebox{1.3}{$\tau\hspace{-.4ex}\raisebox{.29ex}{\textbf{.}}$}\;\Varid{p}} \xone{\ensuremath{\ensuremath{\scalebox{1.3}{$\tau$}}}} \ensuremath{\Varid{p}} \end{matrix}  \qquad \quad
\begin{matrix} \phantom{a} \\ \ensuremath{\Conid{Out}\;\Varid{x}\;\Varid{y}\;\Varid{p}} \xone{\ensuremath{\MVUparrow\!\;\Varid{x}\;\Varid{y}}} \ensuremath{\Varid{p}} \end{matrix} \qquad \quad
\begin{matrix} \phantom{a} \\ \ensuremath{\Conid{In}\;\Varid{x}\;\Varid{b}} \xoneb{\ensuremath{\hspace{-.15ex}\MVDnarrow^{\hspace{-.18ex}\textsc{b}\!}\;\Varid{x}}} \ensuremath{\Varid{b}} \end{matrix} \qquad \quad
\frac{\ensuremath{\Varid{p}} \xone{a} \ensuremath{\Varid{p'}}}{\ensuremath{(\Varid{x}\mathop{{\leftrightarrow}\hspace{-1.48ex}\raisebox{.1ex}{:}\;\,}\Varid{x})\;\Varid{p}} \xone{a} \ensuremath{\Varid{p'}}} \qquad \quad
\frac{\ensuremath{\Varid{p}} \xoneb{a_\textsc{b}} b}{\ensuremath{\Varid{p}\mathop{.\hspace{-.6ex}{+}\hspace{-.6ex}.}\Varid{q}} \xoneb{a_\textsc{b}} b} \qquad 
\frac{\ensuremath{\Varid{p}} \xone{a} \ensuremath{\Varid{p'}}}{\ensuremath{\Varid{p}\mathop{.\hspace{-.6ex}{+}\hspace{-.6ex}.}\Varid{q}} \xone{a} \ensuremath{\Varid{p'}}} \qquad
$
\\[.5ex]
\[
\frac{p \xone{a} p'}{\ensuremath{\Varid{p}\mathop{\|}\Varid{q}} \xone{a} \ensuremath{\Varid{p'}\mathop{\|}\Varid{q}}}
\qquad
\frac{p \xoneb{a_\textsc{b}} \ensuremath{(\Varid{x}\hspace{.1ex}.\hspace{-.4ex}\backslash\Varid{p'})}}{\ensuremath{\Varid{p}\mathop{\|}\Varid{q}} \xoneb{a_\textsc{b}} \ensuremath{(\Varid{x}\hspace{.1ex}.\hspace{-.4ex}\backslash\Varid{p'}\mathop{\|}\Varid{q})}}
\qquad
\frac{\ensuremath{\Varid{p}} \xone{\ensuremath{\MVUparrow\!\;\Varid{x}\;\Varid{v}}} \ensuremath{\Varid{p'}} \quad \ensuremath{\Varid{q}} \xoneb{\ensuremath{\hspace{-.15ex}\MVDnarrow^{\hspace{-.18ex}\textsc{b}\!}\;\Varid{x}}} \ensuremath{(\Varid{y}\hspace{.1ex}.\hspace{-.4ex}\backslash\Varid{q'})}}{
      \ensuremath{\Varid{p}\mathop{\|}\Varid{q}} \xone{\ensuremath{\ensuremath{\scalebox{1.3}{$\tau$}}}} \ensuremath{\Varid{p'}\mathop{\|}\mathop{\sub{\Varid{y}}{\Varid{v}}\!}\!\;\Varid{q'}} }
\qquad
\frac{\ensuremath{\Varid{p}} \xoneb{\ensuremath{\MVUparrow_{\!\textsc{b}\!}\;\Varid{x}}} \ensuremath{(\Varid{y}\hspace{.1ex}.\hspace{-.4ex}\backslash\Varid{p'})} \quad \ensuremath{\Varid{q}} \xoneb{\ensuremath{\hspace{-.15ex}\MVDnarrow^{\hspace{-.18ex}\textsc{b}\!}\;\Varid{x}}} \ensuremath{(\Varid{y}\hspace{.1ex}.\hspace{-.4ex}\backslash\Varid{q'})}}{
      \ensuremath{\Varid{p}\mathop{\|}\Varid{q}} \xone{\ensuremath{\ensuremath{\scalebox{1.3}{$\tau$}}}} \ensuremath{\scalebox{1.25}{$\nu$}\!\;(\Varid{y}\hspace{.1ex}.\hspace{-.4ex}\backslash\Varid{p'}\mathop{\|}\Varid{q'})} }{\quad\text{(close scope-ext)}}
\]
\[
\quad
\frac{\ensuremath{\Varid{p}} \xone{\ensuremath{\Varid{a}}} \ensuremath{\Varid{p'}}}{\ensuremath{\scalebox{1.25}{$\nu$}\!\;(\Varid{x}\hspace{.1ex}.\hspace{-.4ex}\backslash\Varid{p})} \xone{\ensuremath{\Varid{a}}} \ensuremath{\scalebox{1.25}{$\nu$}\!\;(\Varid{x}\hspace{.1ex}.\hspace{-.4ex}\backslash\Varid{p'})}}{~~~ \ensuremath{\Varid{x}}\notin\textrm{fv}(a)}
\qquad
\frac{\ensuremath{\Varid{p}} \xoneb{a_\textsc{b}} \ensuremath{(\Varid{y}\hspace{.1ex}.\hspace{-.4ex}\backslash\Varid{p'})}}{\ensuremath{\scalebox{1.25}{$\nu$}\!\;(\Varid{x}\hspace{.1ex}.\hspace{-.4ex}\backslash\Varid{p})} \xoneb{a_\textsc{b}} \ensuremath{(\Varid{y}\hspace{.1ex}.\hspace{-.4ex}\backslash\scalebox{1.25}{$\nu$}\!\;(\Varid{x}\hspace{.1ex}.\hspace{-.4ex}\backslash\Varid{p'}))}}{~~~ \ensuremath{\Varid{x}}\notin\textrm{fv}(a)}
\qquad
\frac{\ensuremath{\Varid{p}} \xone{\ensuremath{\MVUparrow\!\;\Varid{y}\;(V\!\;\Varid{x})}} \ensuremath{\Varid{p'}}}{\ensuremath{\scalebox{1.25}{$\nu$}\!\;(\Varid{x}\hspace{.1ex}.\hspace{-.4ex}\backslash\Varid{p})} \xoneb{\ensuremath{\MVUparrow_{\!\textsc{b}\!}\;\Varid{y}}} \ensuremath{(\Varid{x}\hspace{.1ex}.\hspace{-.4ex}\backslash\Varid{p'})}}{~~~ \ensuremath{\Varid{y}\not\equiv \Varid{x}}} \quad\text{(open scope-ext)}
\]
\vspace*{-1.5ex}
\caption{Labeled transition rules of the finite $\pi$-calculus (symmetric cases for \ensuremath{\mathop{.\hspace{-.6ex}{+}\hspace{-.6ex}.}} and  \ensuremath{\mathop{\|}} are omitted).}
\label{fig:lts}
\end{figure*}

A process (\ensuremath{\Conid{Pr}}) in the finite $\pi$-calculus is either
the \ensuremath{\scalebox{1.1}{\bf\texttt{0}}} process,
a $\tau$-prefixed process \ensuremath{(\scalebox{1.3}{$\tau\hspace{-.4ex}\raisebox{.29ex}{\textbf{.}}$}\;\Varid{p})},
an input-prefixed process \ensuremath{(\Conid{In}\;\Varid{x}\;(\Varid{y}\hspace{.1ex}.\hspace{-.4ex}\backslash\Varid{p}))},
an output-prefixed process \ensuremath{(\Conid{Out}\;\Varid{x}\;\Varid{y}\;\Varid{p})},
a parallel composition of processes \ensuremath{(\Varid{p}\mathop{\|}\Varid{q})},
a nondeterministic choice between processes \ensuremath{(\Varid{p}\mathop{.\hspace{-.6ex}{+}\hspace{-.6ex}.}\Varid{q})},
a name-restricted process \ensuremath{(\scalebox{1.25}{$\nu$}\!\;(\Varid{x}\hspace{.1ex}.\hspace{-.4ex}\backslash\Varid{p}))}, or
a match-prefixed process \ensuremath{((\Varid{x}\mathop{{\leftrightarrow}\hspace{-1.48ex}\raisebox{.1ex}{:}\;\,}\Varid{y})\;\Varid{p})}.

The operational semantics of the finite $\pi$-calculus is given
in Figure~\ref{fig:lts}. Here we follow a style of
specification~\cite{McDowell96} of the $\pi$-calculus where the
continuation of an input or a bound output transition is
represented as an abstraction over processes. 

The process \ensuremath{\scalebox{1.1}{\bf\texttt{0}}} is a terminated process so that it will never make
any transitions.
\ensuremath{(\scalebox{1.3}{$\tau\hspace{-.4ex}\raisebox{.29ex}{\textbf{.}}$}\;\Varid{p})} will make a (free) transition step evolving into \ensuremath{\Varid{p}}
labeled with (free) action \ensuremath{\ensuremath{\scalebox{1.3}{$\tau$}}\mathbin{::}\Conid{Act}}, that is, $\ensuremath{\scalebox{1.3}{$\tau\hspace{-.4ex}\raisebox{.29ex}{\textbf{.}}$}\;\Varid{p}}\xone{\ensuremath{\ensuremath{\scalebox{1.3}{$\tau$}}}}\ensuremath{\Varid{p}}$.
\ensuremath{(\Conid{Out}\;\Varid{x}\;\Varid{y}\;\Varid{p})} will make a step evolving into \ensuremath{\Varid{p}} labeled with \ensuremath{\MVUparrow\!\;\Varid{x}\;\Varid{y}\mathbin{::}\Conid{Act}} and
produces a value \ensuremath{\Varid{y}} on channel \ensuremath{\Varid{x}}, which can be consumed by another process
expecting an input value on the same channel.

\ensuremath{(\Conid{In}\;\Varid{x}\;(\Varid{y}\hspace{.1ex}.\hspace{-.4ex}\backslash\Varid{p}))} can make a step evolving into \ensuremath{\Varid{p}} once an input value
is provided on channel \ensuremath{\Varid{x}}. When an input value \ensuremath{\Varid{v}\mathbin{::}\Conid{Tm}}\, is provided
on the channel, at some point in time, the process consumes the value
and evolves to \ensuremath{(\mathop{\sub{\Varid{y}}{\Varid{v}}\!}\!\;\Varid{p})}, which is a process where \ensuremath{(V\!\;\Varid{y})}
inside \ensuremath{\Varid{p}} are substituted by \ensuremath{\Varid{v}}.
This concept of a conditional step described above can be understood as if
it steps to a bound process \ensuremath{(\Varid{y}\hspace{.1ex}.\hspace{-.4ex}\backslash\Varid{p})\mathbin{::}\textit{Pr}_{\textsc{b}}}, waiting for an input value for \ensuremath{\Varid{y}}.
It is called a bound step ($\xoneb{a_\textsc{b}}$) in contrast to
the (free) step ($\xone{a}$) for the $\tau$-prefix case. Bound steps are
labeled by bound actions, which can viewed as partially applied actions.

\ensuremath{((\Varid{x}\mathop{{\leftrightarrow}\hspace{-1.48ex}\raisebox{.1ex}{:}\;\,}\Varid{y})\;\Varid{p})} behaves as \ensuremath{\Varid{p}} when \ensuremath{\Varid{x}} is same as \ensuremath{\Varid{y}}.
Otherwise, it cannot make any further steps.

\ensuremath{(\Varid{p}\mathop{.\hspace{-.6ex}{+}\hspace{-.6ex}.}\Varid{q})} nondeterministically becomes either \ensuremath{\Varid{p}} or \ensuremath{\Varid{q}}, and 
take steps thereafter. Only the rules for choosing \ensuremath{\Varid{p}} are illustrated
in Figure~\ref{fig:lts} while the rules for choosing \ensuremath{\Varid{q}} are omitted.

\ensuremath{(\Varid{p}\mathop{\|}\Varid{q})} has eight possible cases; modulo symmetry between \ensuremath{\Varid{p}} and \ensuremath{\Varid{q}}, four.
First, it may step to \ensuremath{(\Varid{p'}\mathop{\|}\Varid{q})} with action \ensuremath{\Varid{a}} when \ensuremath{\Varid{p}} steps to \ensuremath{\Varid{p'}}
with the same action. Second, there is a bound step version of the first.
Third, the two parallel processes can interact when \ensuremath{\Varid{p}} steps to \ensuremath{\Varid{p'}} with
an output action \ensuremath{\MVUparrow\!\;\Varid{x}\;\Varid{v}} and \ensuremath{\Varid{q}} steps to \ensuremath{(\Varid{y}\hspace{.1ex}.\hspace{-.4ex}\backslash\Varid{q'})} with an (bound) input
action \ensuremath{\hspace{-.15ex}\MVDnarrow^{\hspace{-.18ex}\textsc{b}\!}\;\Varid{x}} on the same channel. This interaction step is labeled with \ensuremath{\ensuremath{\scalebox{1.3}{$\tau$}}}
and the process evolves into \ensuremath{(\Varid{p}\mathop{\|}\mathop{\sub{\Varid{y}}{\Varid{v}}\!}\!\;\Varid{q'})}.
Forth {\small(close scope-ext)} is a bound interaction step similar to the third.
The differences from the third is that there is a bounded output (\ensuremath{\MVUparrow_{\!\textsc{b}\!}})
instead of a free ouput (\ensuremath{\MVUparrow\!}\;) and that the resulting process becomes
restricted with the name \ensuremath{\Varid{x}} from the output value \ensuremath{(\Conid{V}\;\Varid{x})}.
The bound output ({\small(open scope-ext)}) is driven by 
name-restricted processes, as explained next.

\ensuremath{\scalebox{1.25}{$\nu$}\!\;(\Varid{x}\hspace{.1ex}.\hspace{-.4ex}\backslash\Varid{p})} restricts actions of \ensuremath{\Varid{p}} involving the restricted name \ensuremath{\Varid{x}} from
being observed outside the scope restricted by \ensuremath{\scalebox{1.25}{$\nu$}\!}. For example, neither
\ensuremath{\scalebox{1.25}{$\nu$}\!\;\!(\Varid{x}\hspace{.1ex}.\hspace{-.4ex}\backslash\Conid{Out}\;\!(\Conid{V}\;\!\Varid{x})\;\!\Varid{v}\;\hspace{-.2ex}\Varid{p})}
nor \ensuremath{\scalebox{1.25}{$\nu$}\!\;\!(\Varid{x}\hspace{.1ex}.\hspace{-.4ex}\backslash\Conid{In}\;\!(\Conid{V}\;\!\Varid{x})\;\!(\Varid{y}\hspace{.1ex}.\hspace{-.4ex}\backslash\Varid{q}))} can make
any further steps. However, communication over the restricted channel \ensuremath{(\Conid{V}\;\!\Varid{x})}
is still possible as long as the restricted name \ensuremath{\Varid{x}} is not observable from outside.
For example, 
$ \ensuremath{\scalebox{1.25}{$\nu$}\!\;\!(\Varid{x}\hspace{.1ex}.\hspace{-.4ex}\backslash\Conid{Out}\;(\Conid{V}\;\!\Varid{x})\;\Varid{v}\;\Varid{p}\mathop{\|}\Conid{In}\;(\Conid{V}\;\!\Varid{x})\;(\Varid{y}\hspace{.1ex}.\hspace{-.4ex}\backslash\Varid{q}))}
  \xone{\ensuremath{\ensuremath{\scalebox{1.3}{$\tau$}}}} \ensuremath{\scalebox{1.25}{$\nu$}\!\;\!(\Varid{x}\hspace{.1ex}.\hspace{-.4ex}\backslash\Varid{p}\mathop{\|}\mathop{\sub{\Varid{y}}{\Varid{v}}\!}\!\;\!\Varid{q})}. $
The last rule {\small(open scope-ext)} is the source of the bounded output when
the output over a non-restricted channel happens to be the restricted value
\ensuremath{(\Conid{V}\;\!\Varid{x})}. This bound output is to be consumed by interacting with
another process waiting for an input on the same channel, according to
the rule {\small(close scope-ext)} mentioned above. For example, \vspace*{-.75ex}
{\small\setpremisesend{1pt}\setnamespace{1pt}
\[
 \inference[close]{\!\!\!\!\!\!\!\!
   \inference*[open]{\ensuremath{\Conid{Out}\;\Varid{y}\;(\Conid{V}\;\!\Varid{x})\;\Varid{p}} \xone{\ensuremath{\MVUparrow\!\;\!\Varid{y}\;\!(\Conid{V}\;\!\Varid{x})}} \ensuremath{\Varid{p}}
    }{\ensuremath{\scalebox{1.25}{$\nu$}\!\;\!(\Varid{x}\hspace{.1ex}.\hspace{-.4ex}\backslash\Conid{Out}\;\Varid{y}\;(\Conid{V}\;\!\Varid{x})\;\Varid{p})} \xoneb{\ensuremath{\MVUparrow_{\!\textsc{b}\!}\;\Varid{y}}} \ensuremath{(\Varid{x}\hspace{.1ex}.\hspace{-.4ex}\backslash\Varid{p})}}
   &
     \ensuremath{\Conid{In}\;\Varid{y}\;(\Varid{z}\hspace{.1ex}.\hspace{-.4ex}\backslash\Varid{q})} \xoneb{\ensuremath{\hspace{-.15ex}\MVDnarrow^{\hspace{-.18ex}\textsc{b}\!}\;\Varid{y}}} \ensuremath{(\Varid{z}\hspace{.1ex}.\hspace{-.4ex}\backslash\Varid{q})}
 }{
  \ensuremath{\scalebox{1.25}{$\nu$}\!\;\!(\Varid{x}\hspace{.1ex}.\hspace{-.4ex}\backslash\uwave{\Conid{Out}\;\Varid{y}\;\!(\Conid{V}\;\!\Varid{x})\;\Varid{p}})\mathop{\|}\Conid{In}\;\Varid{y}\;(\Varid{z}\hspace{.1ex}.\hspace{-.4ex}\backslash\Varid{q})} \xone{\ensuremath{\ensuremath{\scalebox{1.3}{$\tau$}}}}
  \ensuremath{\scalebox{1.25}{$\nu$}\!\;\!(\Varid{x}\hspace{.1ex}.\hspace{-.4ex}\backslash\uwave{\Varid{p}\mathop{\|}\mathop{\sub{\Varid{z}}{(\Conid{V}\;\!\Varid{x})}\!}\!\;\!\Varid{q}})} }
\]\\[-2.5ex]}
Before the interaction step, the scope of restriction (marked by \uwave{wavy underline})
did not include the input process on the right-hand side of parallel composition.
After the step, the scope of restriction includes the right-hand side,
adjusting to include the restricted output \ensuremath{(\Conid{V}\;\!\Varid{x})} extruded from
the original scope through the non-restricted channel \ensuremath{\Varid{y}}.
The rule {\small(open scope-ext)} together with
the rule {\small(close scope-ext)} descirbes
the feature known as \emph{scope extrusion} in the $\pi$-calculus.

The labeled transition rules of Figure~\ref{fig:lts} are implemented
as Haskell programs, which are to be discussed in Section~\ref{fig:lts}.

\subsection{Modal Logic \OM}
\label{sec:syntax:om}

An \OM\ formulae \ensuremath{\Varid{f}} describes a behavior pattern of processes.
$\ensuremath{\Varid{p}} \models \ensuremath{\Varid{f}}$, read as ``\ensuremath{\Varid{p}} satisfy \ensuremath{\Varid{f}}'' or ``\ensuremath{\Varid{f}} is satisfied by \ensuremath{\Varid{p}}'',
holds when the process \ensuremath{\Varid{p}} follows the behavior described by \ensuremath{\Varid{f}}.
Let $\mathcal{L}(p) = \{f \in \ensuremath{\Conid{Form}} \mid p \models f\}$,
the set of formulae satisfied by \ensuremath{\Varid{p}}.
We~\cite{AhnHorTiu17corr} recently established that $\mathcal{L}(p) \equiv \mathcal{L}(q)$
exactly coincides with $p \sim_o q$, that is, \ensuremath{\Varid{p}} and \ensuremath{\Varid{q}} are open bisimilar.
By contraposition, $\mathcal{L}(p) \not\equiv \mathcal{L}(q)$ whenever
$p \not\sim_o q$, that is, there must exists $f$ that satisfy one of
the two non-bisimilar processes but not the other. Such a formula is known as
a \emph{distinguishing formula}. This formula explains how two processes
behave differently so that it can serve as a certificate of non-bisimilarity
if we have an implementation to check satisfiability of \ensuremath{\Varid{f}} for
a given process, which we already do have \cite{AhnHorTiu17corr}.

An \OM\ formula (\ensuremath{\Conid{Form}}) is either the falsity (\ensuremath{\bot}), the truth (\ensuremath{\top}),
a conjunction (\ensuremath{\bigwedge\!\;\Varid{fs}}), a disjunction (\ensuremath{\bigvee\!\;\Varid{fs}}),
a dia-action (\ensuremath{\Diamond\!\;\Varid{a}\;\Varid{f}}),
a box-action (\ensuremath{\Box\!\;\Varid{a}\;\Varid{f}}),
a bound dia-action (\ensuremath{\Diamond_{^{\!}\textsc{b}}\!\;\Varid{a}\;(\Varid{x}\hspace{.1ex}.\hspace{-.4ex}\backslash\Varid{f})}),
a bound box-action (\ensuremath{\Box_{^{\!}\textsc{b}}\!\;\Varid{a}\;(\Varid{x}\hspace{.1ex}.\hspace{-.4ex}\backslash\Varid{f})}),
a dia-match (\ensuremath{\Diamond_{\!=}\!\;[\mskip1.5mu (\Varid{x\char95 i},\Varid{y\char95 i})\mskip1.5mu]\;\Varid{f}}), or
a box-match (\ensuremath{\Box_{=}\!\;[\mskip1.5mu (\Varid{x\char95 i},\Varid{y\char95 i})\mskip1.5mu]\;\Varid{f}}).\footnote{Standard notations
	in the literature (and also in Figure~\ref{fig:example})
	are $[a]f$ and $\langle a \rangle f$ for box- and dia-actions;
	and, $[x=y]f$ and $\langle x=y \rangle f$ for box- and dia-matches.
	The notations used for bound actions may vary between different notions of
	bisimilarities discussed in Section~\ref{sec:relwork:logic}. }
Intuitive meanings of these formulae can be best understood by 
the \emph{possible worlds} interpretation for modal logic:
\begin{itemize}
\item \ensuremath{\bot} satisfies no process;
\item \ensuremath{\top} satisfies any process, including \ensuremath{\scalebox{1.1}{\bf\texttt{0}}};
\item \ensuremath{(\bigwedge\!\;\Varid{fs})} satisfies \ensuremath{\Varid{p}} when $\ensuremath{\Varid{p}} \models \ensuremath{\Varid{f}}$ for all \ensuremath{\Varid{f}\in \Varid{fs}};
\item \ensuremath{(\bigvee\!\;\Varid{fs})} satisfies \ensuremath{\Varid{p}} when there exists \ensuremath{\Varid{f}\in \Varid{fs}} that $\ensuremath{\Varid{p}} \models \ensuremath{\Varid{f}}$;
\item \ensuremath{(\Diamond\!\;\Varid{a}\;\Varid{f})} satisfies \ensuremath{\Varid{p}} when there exists a step from \ensuremath{\Varid{p}} labeled with \ensuremath{\Varid{a}} into \ensuremath{\Varid{p'}}
in the current world such that $\ensuremath{\Varid{p'}} \models \ensuremath{\Varid{f}}$;
\item \ensuremath{(\Box\!\;\Varid{a}\;\Varid{f})} satisfies \ensuremath{\Varid{p}} when any possible step from \ensuremath{\Varid{p}} to \ensuremath{\Varid{p'}} labeled with \ensuremath{\Varid{a}}
satisfies $\ensuremath{\Varid{p'}} \models \ensuremath{\Varid{f}}$ in all possible worlds;
\item (\ensuremath{\Diamond\!\;\Varid{a}\;(\Varid{x}\hspace{.1ex}.\hspace{-.4ex}\backslash\Varid{f})}) and (\ensuremath{\Box\!\;\Varid{a}\;(\Varid{x}\hspace{.1ex}.\hspace{-.4ex}\backslash\Varid{f})}) are similar to above two items while
taking bound steps from \ensuremath{\Varid{p}} to \ensuremath{(\Varid{x}\hspace{.1ex}.\hspace{-.4ex}\backslash\Varid{p'})};\footnote{There are some subtleties on
	what values (\ensuremath{\Varid{v}}) are to be chosen to instantiate \ensuremath{\Varid{x}} for both \ensuremath{(\Varid{x}\hspace{.1ex}.\hspace{-.4ex}\backslash\Varid{p'})}
	and \ensuremath{(\Varid{x}\hspace{.1ex}.\hspace{-.4ex}\backslash\Varid{f})} in order to check $\ensuremath{\mathop{\sub{\Varid{x}}{\Varid{v}}\!}\!\;\!\Varid{p'}} \models \ensuremath{\mathop{\sub{\Varid{x}}{\Varid{v}}\!}\!\;\!\Varid{f}}$.
	The basic idea is that, for input action, all possible values should be considered
        whereas, for bound output action, \ensuremath{\Varid{x}} should be treated a fresh constant distinct from
	all the other names introduced before because \ensuremath{\Varid{x}} must have originated from
	the restricted process --- recall (open scope-ext) in Figure~\ref{fig:lts}.}
\item (\ensuremath{\Diamond_{\!=}\!\;\sigma\;\Varid{f}}) satisfies p when \ensuremath{\Varid{x\char95 i}\equiv \Varid{y\char95 i}} holds for all \ensuremath{(\Varid{x\char95 i},\Varid{y\char95 i})\in \sigma}
in the current world and $\ensuremath{\Varid{p}} \models \ensuremath{\Varid{f}}$; and
\item (\ensuremath{\Box_{=}\!\;\sigma\;\Varid{f}}) satisfies p when $\ensuremath{\Varid{p}} \models \ensuremath{\Varid{f}}$ in all possible worlds
such that \ensuremath{\Varid{x\char95 i}\equiv \Varid{y\char95 i}} holds for all \ensuremath{(\Varid{x\char95 i},\Varid{y\char95 i})\in \sigma}.
\end{itemize}

In the context of open bisimulation, possible worlds are 
determined by substitutions over the free names of processes.
For example, consider \ensuremath{\scalebox{1.3}{$\tau\hspace{-.4ex}\raisebox{.29ex}{\textbf{.}}$}\;\hspace{-.3ex}((\Varid{x}\mathop{{\leftrightarrow}\hspace{-1.48ex}\raisebox{.1ex}{:}\;\,}\Varid{y})\;\hspace{-.3ex}\tau)}.
In a world given by a substitution that unifies \ensuremath{\Varid{x}} and \ensuremath{\Varid{y}}
(i.e., both maps to same value), it can make two consecutive steps
labeled with \ensuremath{\ensuremath{\scalebox{1.3}{$\tau$}}}. On the other hand, in another world where the substitution
does not unify \ensuremath{\Varid{x}} and \ensuremath{\Varid{y}}, it can only make one \ensuremath{\ensuremath{\scalebox{1.3}{$\tau$}}}-step but no further.
Because open bisimulation is an equivalence property across all possible worlds,
\ensuremath{\scalebox{1.3}{$\tau\hspace{-.4ex}\raisebox{.29ex}{\textbf{.}}$}\;\hspace{-.3ex}((\Varid{x}\mathop{{\leftrightarrow}\hspace{-1.48ex}\raisebox{.1ex}{:}\;\,}\Varid{y})\;\hspace{-.3ex}\tau)} is bisimilar
to none of \ensuremath{\tau}, \ensuremath{\tau\tau}, and \ensuremath{\tau\mathop{.\hspace{-.6ex}{+}\hspace{-.6ex}.}\tau\tau}. In particular,
$\ensuremath{\scalebox{1.3}{$\tau\hspace{-.4ex}\raisebox{.29ex}{\textbf{.}}$}\;\hspace{-.3ex}((\Varid{x}\mathop{{\leftrightarrow}\hspace{-1.48ex}\raisebox{.1ex}{:}\;\,}\Varid{y})\;\hspace{-.3ex}\tau)} \not\sim_{\!o} \ensuremath{\tau\mathop{.\hspace{-.6ex}{+}\hspace{-.6ex}.}\tau\tau}$
exemplifies that open bisimulation distinguishes environment sensitive choices
made by match prefix from (environment insensitive) nondeterministic choices made by (\ensuremath{\mathop{.\hspace{-.6ex}{+}\hspace{-.6ex}.}}).


\section{Labeled Transition Semantics}
\label{sec:lts}
We discuss implementations of the labeled transition rules in Figure~\ref{fig:lts}.
There are two versions:
the first implements the transition step in a fixed world (Section~\ref{sec:lts:ids})
and the second implements the transition step considering all possible worlds
(Section~\ref{sec:lts:open}).
%
%
%
\makeatletter
\@ifundefined{lhs2tex.lhs2tex.sty.read}%
  {\@namedef{lhs2tex.lhs2tex.sty.read}{}%
   \newcommand\SkipToFmtEnd{}%
   \newcommand\EndFmtInput{}%
   \long\def\SkipToFmtEnd#1\EndFmtInput{}%
  }\SkipToFmtEnd

\newcommand\ReadOnlyOnce[1]{\@ifundefined{#1}{\@namedef{#1}{}}\SkipToFmtEnd}
\usepackage{amstext}
\usepackage{amssymb}
\usepackage{stmaryrd}
\DeclareFontFamily{OT1}{cmtex}{}
\DeclareFontShape{OT1}{cmtex}{m}{n}
  {<5><6><7><8>cmtex8
   <9>cmtex9
   <10><10.95><12><14.4><17.28><20.74><24.88>cmtex10}{}
\DeclareFontShape{OT1}{cmtex}{m}{it}
  {<-> ssub * cmtt/m/it}{}
\newcommand{\texfamily}{\fontfamily{cmtex}\selectfont}
\DeclareFontShape{OT1}{cmtt}{bx}{n}
  {<5><6><7><8>cmtt8
   <9>cmbtt9
   <10><10.95><12><14.4><17.28><20.74><24.88>cmbtt10}{}
\DeclareFontShape{OT1}{cmtex}{bx}{n}
  {<-> ssub * cmtt/bx/n}{}
\newcommand{\tex}[1]{\text{\texfamily#1}}	

\newcommand{\Sp}{\hskip.33334em\relax}

\newcommand{\Conid}[1]{\mathit{#1}}
\newcommand{\Varid}[1]{\mathit{#1}}
\newcommand{\anonymous}{\kern0.06em \vbox{\hrule\@width.5em}}
\newcommand{\plus}{\mathbin{+\!\!\!+}}
\newcommand{\bind}{\mathbin{>\!\!\!>\mkern-6.7mu=}}
\newcommand{\rbind}{\mathbin{=\mkern-6.7mu<\!\!\!<}}
\newcommand{\sequ}{\mathbin{>\!\!\!>}}
\renewcommand{\leq}{\leqslant}
\renewcommand{\geq}{\geqslant}
\usepackage{polytable}

\@ifundefined{mathindent}%
  {\newdimen\mathindent\mathindent\leftmargini}%
  {}%

\def\resethooks{%
  \global\let\SaveRestoreHook\empty
  \global\let\ColumnHook\empty}
\newcommand*{\savecolumns}[1][default]%
  {\g@addto@macro\SaveRestoreHook{\savecolumns[#1]}}
\newcommand*{\restorecolumns}[1][default]%
  {\g@addto@macro\SaveRestoreHook{\restorecolumns[#1]}}
\newcommand*{\aligncolumn}[2]%
  {\g@addto@macro\ColumnHook{\column{#1}{#2}}}

\resethooks

\newcommand{\onelinecommentchars}{\quad-{}- }
\newcommand{\commentbeginchars}{\enskip\{-}
\newcommand{\commentendchars}{-\}\enskip}

\newcommand{\visiblecomments}{%
  \let\onelinecomment=\onelinecommentchars
  \let\commentbegin=\commentbeginchars
  \let\commentend=\commentendchars}

\newcommand{\invisiblecomments}{%
  \let\onelinecomment=\empty
  \let\commentbegin=\empty
  \let\commentend=\empty}

\visiblecomments

\newlength{\blanklineskip}
\setlength{\blanklineskip}{0.66084ex}

\newcommand{\hsindent}[1]{\quad}
\let\hspre\empty
\let\hspost\empty
\newcommand{\NB}{\textbf{NB}}
\newcommand{\Todo}[1]{$\langle$\textbf{To do:}~#1$\rangle$}

\EndFmtInput
\makeatother
%
%
%
%
%
%
%
%
%
\ReadOnlyOnce{polycode.fmt}%
\makeatletter

\newcommand{\hsnewpar}[1]%
  {{\parskip=0pt\parindent=0pt\par\vskip #1\noindent}}

\newcommand{\hscodestyle}{}


\newcommand{\sethscode}[1]%
  {\expandafter\let\expandafter\hscode\csname #1\endcsname
   \expandafter\let\expandafter\endhscode\csname end#1\endcsname}


\newenvironment{compathscode}%
  {\par\noindent
   \advance\leftskip\mathindent
   \hscodestyle
   \let\\=\@normalcr
   \let\hspre\(\let\hspost\)%
   \pboxed}%
  {\endpboxed\)%
   \par\noindent
   \ignorespacesafterend}

\newcommand{\compaths}{\sethscode{compathscode}}


\newenvironment{plainhscode}%
  {\hsnewpar\abovedisplayskip
   \advance\leftskip\mathindent
   \hscodestyle
   \let\hspre\(\let\hspost\)%
   \pboxed}%
  {\endpboxed%
   \hsnewpar\belowdisplayskip
   \ignorespacesafterend}

\newenvironment{oldplainhscode}%
  {\hsnewpar\abovedisplayskip
   \advance\leftskip\mathindent
   \hscodestyle
   \let\\=\@normalcr
   \(\pboxed}%
  {\endpboxed\)%
   \hsnewpar\belowdisplayskip
   \ignorespacesafterend}


\newcommand{\plainhs}{\sethscode{plainhscode}}
\newcommand{\oldplainhs}{\sethscode{oldplainhscode}}
\plainhs


\newenvironment{arrayhscode}%
  {\hsnewpar\abovedisplayskip
   \advance\leftskip\mathindent
   \hscodestyle
   \let\\=\@normalcr
   \(\parray}%
  {\endparray\)%
   \hsnewpar\belowdisplayskip
   \ignorespacesafterend}

\newcommand{\arrayhs}{\sethscode{arrayhscode}}


\newenvironment{mathhscode}%
  {\parray}{\endparray}

\newcommand{\mathhs}{\sethscode{mathhscode}}


\newenvironment{texthscode}%
  {\(\parray}{\endparray\)}

\newcommand{\texths}{\sethscode{texthscode}}


\def\codeframewidth{\arrayrulewidth}
\RequirePackage{calc}

\newenvironment{framedhscode}%
  {\parskip=\abovedisplayskip\par\noindent
   \hscodestyle
   \arrayrulewidth=\codeframewidth
   \tabular{@{}|p{\linewidth-2\arraycolsep-2\arrayrulewidth-2pt}|@{}}%
   \hline\framedhslinecorrect\\{-1.5ex}%
   \let\endoflinesave=\\
   \let\\=\@normalcr
   \(\pboxed}%
  {\endpboxed\)%
   \framedhslinecorrect\endoflinesave{.5ex}\hline
   \endtabular
   \parskip=\belowdisplayskip\par\noindent
   \ignorespacesafterend}

\newcommand{\framedhslinecorrect}[2]%
  {#1[#2]}

\newcommand{\framedhs}{\sethscode{framedhscode}}


\newenvironment{inlinehscode}%
  {\(\def\column##1##2{}%
   \let\>\undefined\let\<\undefined\let\\\undefined
   \newcommand\>[1][]{}\newcommand\<[1][]{}\newcommand\\[1][]{}%
   \def\fromto##1##2##3{##3}%
   \def\nextline{}}{\) }%

\newcommand{\inlinehs}{\sethscode{inlinehscode}}


\newenvironment{joincode}%
  {\let\orighscode=\hscode
   \let\origendhscode=\endhscode
   \def\endhscode{\def\hscode{\endgroup\def\@currenvir{hscode}\\}\begingroup}
   \orighscode\def\hscode{\endgroup\def\@currenvir{hscode}}}%
  {\origendhscode
   \global\let\hscode=\orighscode
   \global\let\endhscode=\origendhscode}%

\makeatother
\EndFmtInput
\ReadOnlyOnce{colorcode.fmt}%

\RequirePackage{colortbl}
\RequirePackage{calc}

\makeatletter
\newenvironment{colorhscode}%
  {\hsnewpar\abovedisplayskip
   \hscodestyle
   \tabular{@{}>{\columncolor{codecolor}}p{\linewidth}@{}}%
   \let\\=\@normalcr
   \(\pboxed}%
  {\endpboxed\)%
   \endtabular
   \hsnewpar\belowdisplayskip
   \ignorespacesafterend}

\newenvironment{tightcolorhscode}%
  {\hsnewpar\abovedisplayskip
   \hscodestyle
   \tabular{@{}>{\columncolor{codecolor}\(}l<{\)}@{}}%
   \pmboxed}%
  {\endpmboxed%
   \endtabular
   \hsnewpar\belowdisplayskip
   \ignorespacesafterend}

\newenvironment{barhscode}%
  {\hsnewpar\abovedisplayskip
   \hscodestyle
   \arrayrulecolor{codecolor}%
   \arrayrulewidth=\coderulewidth
   \tabular{|p{\linewidth-\arrayrulewidth-\tabcolsep}@{}}%
   \let\\=\@normalcr
   \(\pboxed}%
  {\endpboxed\)%
   \endtabular
   \hsnewpar\belowdisplayskip
   \ignorespacesafterend}
\makeatother

\def\colorcode{\columncolor{codecolor}}
\definecolor{codecolor}{rgb}{1,1,.667}
\newlength{\coderulewidth}
\setlength{\coderulewidth}{3pt}

\newcommand{\colorhs}{\sethscode{colorhscode}}
\newcommand{\tightcolorhs}{\sethscode{tightcolorhscode}}
\newcommand{\barhs}{\sethscode{barhscode}}

\EndFmtInput

\renewcommand{\onelinecommentchars}{\color{gray}\quad-{}- }
\renewcommand{\commentbeginchars}{\color{gray}\enskip\{- }
\renewcommand{\commentendchars}{-\}\enskip}

\renewcommand{\visiblecomments}{%
  \let\onelinecomment=\onelinecommentchars
  \let\commentbegin=\commentbeginchars
  \let\commentend=\commentendchars}

\renewcommand{\invisiblecomments}{%
  \let\onelinecomment=\empty
  \let\commentbegin=\empty
  \let\commentend=\empty}

\visiblecomments

\subsection{Labeled Transition Steps in a Fixed World}
\label{sec:lts:ids}
Figure~\ref{fig:IdSubLTS} is a straightforward transcription of 
the transition rules (including the omitted symmetric cases)
from Figure~\ref{fig:lts} where \,\ensuremath{(\Varid{l},\Varid{p})\leftarrow \Varid{one}\;\Varid{p}}\, and \,\ensuremath{(\Varid{l},\Varid{b})\leftarrow \textit{one}_{\textsc{b}}\;\Varid{p}}\,
correspond to free and bound steps \,$p\xone{l}p'$\, and \,$p\xoneb{l}b$\,
respectively.

The type signatures of \ensuremath{\Varid{one}} and \ensuremath{\textit{one}_{\textsc{b}}} indicates that freshness of names
and nondeterminism are handled by a monadic computation that returns
a pair of a (bound) action and a (bound) process. In this paper, you may
simply consider \ensuremath{\Varid{one}} and \ensuremath{\textit{one}_{\textsc{b}}} as returning a list of all possible pairs.
For example, we can compute all the three possible next steps from the process
{\small\ensuremath{\Conid{Out}\;(V\!\;\Varid{x})\;(V\!\;\Varid{x})\;\scalebox{1.1}{\bf\texttt{0}}\mathop{\|}\Conid{Out}\;(V\!\;\Varid{w})\;(V\!\;\Varid{w})\;\scalebox{1.1}{\bf\texttt{0}}\mathop{\|}\Conid{In}\;(V\!\;\Varid{z})\;(\Varid{y}\hspace{.1ex}.\hspace{-.4ex}\backslash\scalebox{1.1}{\bf\texttt{0}})}}
using ghci as follows:\vspace*{-.5ex}\footnote{\ensuremath{\Varid{pp}\mathbin{::}\Conid{Pretty}\;\Varid{a}\Rightarrow \Varid{a}\to \Conid{IO}\;()} is
	our pretty printing utility function, which is not going to be discussed
	in this paper. It is for printing out a more readable format the default
	derived show instances provided by the unbound library.}
\begin{center}\small
\vspace*{-.5ex}
\begin{hscode}\SaveRestoreHook
\column{B}{@{}>{\hspre}l<{\hspost}@{}}%
\column{E}{@{}>{\hspre}l<{\hspost}@{}}%
\>[B]{}\texttt{*Main> :type }\Varid{runFreshMT}\mathbin{\circ}\Varid{\Conid{IdSubLTS}.one}{}\<[E]%
\\
\>[B]{}\Varid{runFreshMT}\mathbin{\circ}\Varid{\Conid{IdSubLTS}.one}\mathbin{::}\Conid{MonadPlus}\;\Varid{m}\Rightarrow \Conid{Pr}\to \Varid{m}\;(\Conid{Act},\Conid{Pr}){}\<[E]%
\\
\>[B]{}\texttt{*Main> :type }\Varid{map}\;\Varid{id}\mathbin{\circ}\Varid{runFreshMT}\mathbin{\circ}\Varid{\Conid{IdSubLTS}.one}{}\<[E]%
\\
\>[B]{}\Varid{map}\;\Varid{id}\mathbin{\circ}\Varid{runFreshMT}\mathbin{\circ}\Varid{\Conid{IdSubLTS}.one}\mathbin{::}\Conid{Pr}\to [\mskip1.5mu (\Conid{Act},\Conid{Pr})\mskip1.5mu]{}\<[E]%
\\
\>[B]{}\texttt{*Main> }\;\mathbf{let}\;\Varid{p}\mathrel{=}\Conid{Out}\;\!(V\!\;\!\Varid{x})\;(V\!\;\!\Varid{x})\;\!\scalebox{1.1}{\bf\texttt{0}}\mathop{\|}\Conid{Out}\;\!(V\!\;\!\Varid{y})\;\!(V\!\;\!\Varid{y})\;\!\scalebox{1.1}{\bf\texttt{0}}\mathop{\|}\Conid{In}\;\!(V\!\;\!\Varid{z})\;\!(\Varid{w}\hspace{.1ex}.\hspace{-.4ex}\backslash\scalebox{1.1}{\bf\texttt{0}}){}\<[E]%
\\
\>[B]{}\texttt{*Main> }\Varid{mapM}\hspace{-.2ex}\anonymous \,\;\Varid{pp}\mathbin{\circ}\Varid{runFreshMT}\mathbin{\circ}\Varid{\Conid{IdSubLTS}.one}\mathop{\texttt{\$}}\Varid{p}{}\<[E]%
\\
\>[B]{}(\MVUparrow\!\;(V\!\;\Varid{x})\;(V\!\;\Varid{x}),(\scalebox{1.1}{\bf\texttt{0}}\mathop{\|}\Conid{Out}\;(V\!\;\Varid{y})\;(V\!\;\Varid{y})\;\scalebox{1.1}{\bf\texttt{0}})\mathop{\|}\,(\Conid{In}\;(V\!\;\Varid{z})\;(\Varid{w}\hspace{.1ex}.\hspace{-.4ex}\backslash\scalebox{1.1}{\bf\texttt{0}}))){}\<[E]%
\\
\>[B]{}(\MVUparrow\!\;(V\!\;\Varid{y})\;(V\!\;\Varid{y}),(\Conid{Out}\;(V\!\;\Varid{x})\;(V\!\;\Varid{x})\;\scalebox{1.1}{\bf\texttt{0}}\mathop{\|}\scalebox{1.1}{\bf\texttt{0}})\mathop{\|}\,(\Conid{In}\;(V\!\;\Varid{z})\;(\Varid{w}\hspace{.1ex}.\hspace{-.4ex}\backslash\scalebox{1.1}{\bf\texttt{0}})))\;{}\<[E]%
\\
\>[B]{}\texttt{*Main> }\Varid{mapM}\hspace{-.2ex}\anonymous \,\;\Varid{pp}\mathbin{\circ}\Varid{runFreshMT}\mathbin{\circ}\Conid{IdSubLTS}.\textit{one}_{\textsc{b}}\mathop{\texttt{\$}}\Varid{p}{}\<[E]%
\\
\>[B]{}(\hspace{-.15ex}\MVDnarrow^{\hspace{-.18ex}\textsc{b}\!}\;(V\!\;\Varid{z}),\Varid{y}\hspace{.1ex}.\hspace{-.4ex}\backslash(((\Conid{Out}\;(V\!\;\Varid{x})\;(V\!\;\Varid{x})\;\scalebox{1.1}{\bf\texttt{0}})\mathop{\|}\,(\Conid{Out}\;(V\!\;\Varid{y})\;(V\!\;\Varid{y})\;\scalebox{1.1}{\bf\texttt{0}}))\mathop{\|}\scalebox{1.1}{\bf\texttt{0}})){}\<[E]%
\ColumnHook
\end{hscode}\resethooks
\vspace*{-3.5ex}
\end{center}

In principle, the possible worlds semantics could be implemented using
\ensuremath{\Varid{one}} and \ensuremath{\textit{one}_{\textsc{b}}} in this \ensuremath{\Conid{IdSubLTS}} module by brute force enumeration of
all substitutions over the free names in the process. For instance,
there are three free names (\ensuremath{\Varid{x}},\ensuremath{\Varid{y}},\ensuremath{\Varid{z}}) in the process (\ensuremath{\Varid{p}}) above.
Enumerating all substitutions over 3 names amounts to considering all
possible integer set partition of the 3 elements.
Let us establish a 1-to-1 mapping of \ensuremath{\Varid{x}} to 0, \ensuremath{\Varid{y}} to 1, and \ensuremath{\Varid{z}} to 2.
Then, a substitution that map \ensuremath{\Varid{x}} and \ensuremath{\Varid{z}} to the same value but \ensuremath{\Varid{y}} to
a different value corresponds to the partition [[0,2],[1]] where
0 and 2 belong to the same equivalence class. In such a world, there
is an additional possible step for \ensuremath{\Varid{p}} above, which is the interaction
between \ensuremath{\Conid{Out}\;(V\!\;\Varid{x})\;(V\!\;\Varid{x})\;\scalebox{1.1}{\bf\texttt{0}}} and \ensuremath{\Conid{In}\;(V\!\;\Varid{z})\;(\Varid{y}\hspace{.1ex}.\hspace{-.4ex}\backslash\scalebox{1.1}{\bf\texttt{0}})} due to
the unification of \ensuremath{\Varid{x}} and \ensuremath{\Varid{z}}. More generally, we can generate
all possible partitions, starting from the distinct partition [[0],[1],[2]],
by continually joining a pair of elements from different equivalence classes
until all possible joining paths reaches [[0,1,2]] where all elements are joined.
Although this brute force approach is a terminating algorithm, the number of
partition sets is exponential to the number of names \cite{Rota64bell}.


Since the original development of open bisimulation, \citet{Sangiorgi96acta}
was well aware that enumerating all possible worlds is intractable and
provided a more efficient set of transition rules, known as
the symbolic transition semantics. We implement another version of \ensuremath{\Varid{one}}
and \ensuremath{\textit{one}_{\textsc{b}}} following the style of symbolic transition in the next subsection.
Nevertheless, \ensuremath{\Varid{one}} and \ensuremath{\textit{one}_{\textsc{b}}} in this subsection are still used
in our implementation of open bisimulation, together with the symbolic version.
We will explain why we use both versions to implement open bisimulation
in Section~\ref{sec:bisim}.

\begin{figure}\small
\vspace*{2.7ex}
\begin{hscode}\SaveRestoreHook
\column{B}{@{}>{\hspre}l<{\hspost}@{}}%
\column{3}{@{}>{\hspre}c<{\hspost}@{}}%
\column{3E}{@{}l@{}}%
\column{8}{@{}>{\hspre}l<{\hspost}@{}}%
\column{12}{@{}>{\hspre}l<{\hspost}@{}}%
\column{13}{@{}>{\hspre}l<{\hspost}@{}}%
\column{18}{@{}>{\hspre}l<{\hspost}@{}}%
\column{19}{@{}>{\hspre}l<{\hspost}@{}}%
\column{20}{@{}>{\hspre}l<{\hspost}@{}}%
\column{24}{@{}>{\hspre}l<{\hspost}@{}}%
\column{28}{@{}>{\hspre}l<{\hspost}@{}}%
\column{30}{@{}>{\hspre}l<{\hspost}@{}}%
\column{39}{@{}>{\hspre}l<{\hspost}@{}}%
\column{45}{@{}>{\hspre}l<{\hspost}@{}}%
\column{49}{@{}>{\hspre}l<{\hspost}@{}}%
\column{51}{@{}>{\hspre}l<{\hspost}@{}}%
\column{52}{@{}>{\hspre}l<{\hspost}@{}}%
\column{54}{@{}>{\hspre}l<{\hspost}@{}}%
\column{62}{@{}>{\hspre}l<{\hspost}@{}}%
\column{65}{@{}>{\hspre}l<{\hspost}@{}}%
\column{E}{@{}>{\hspre}l<{\hspost}@{}}%
\>[B]{}\mathbf{module}\;\Conid{IdSubLTS}\;\mathbf{where}{}\<[E]%
\\
\>[B]{}\mathbf{import}\;\Conid{PiCalc}{}\<[E]%
\\
\>[B]{}\mathbf{import}\;\Conid{\Conid{Control}.Applicative}{}\<[E]%
\\
\>[B]{}\mathbf{import}\;\Conid{\Conid{Control}.Monad}{}\<[E]%
\\
\>[B]{}\mathbf{import}\;\Conid{\Conid{Unbound}.LocallyNameless}\;\mathbf{hiding}\;(\Varid{empty}){}\<[E]%
\\[\blanklineskip]%
\>[B]{}\Varid{one}\mathbin{::}(\Conid{Fresh}\;\Varid{m},\Conid{Alternative}\;\Varid{m})\Rightarrow \Conid{Pr}\to \Varid{m}\;(\Conid{Act},\Conid{Pr}){}\<[E]%
\\
\>[B]{}\Varid{one}\;(\Conid{Out}\;\Varid{x}\;\Varid{y}\;\Varid{p}){}\<[20]%
\>[20]{}\mathrel{=}\Varid{return}\;(\MVUparrow\!\;\Varid{x}\;\Varid{y},\Varid{p}){}\<[E]%
\\
\>[B]{}\Varid{one}\;(\scalebox{1.3}{$\tau\hspace{-.4ex}\raisebox{.29ex}{\textbf{.}}$}\;\Varid{p}){}\<[20]%
\>[20]{}\mathrel{=}\Varid{return}\;(\ensuremath{\scalebox{1.3}{$\tau$}},\Varid{p}){}\<[E]%
\\
\>[B]{}\Varid{one}\;((\Varid{x}\mathop{{\leftrightarrow}\hspace{-1.48ex}\raisebox{.1ex}{:}\;\,}\Varid{y})\;\Varid{p}){}\<[20]%
\>[20]{}\mid \Varid{x}\equiv \Varid{y}\mathrel{=}\Varid{one}\;\Varid{p}{}\<[E]%
\\
\>[B]{}\Varid{one}\;(\Varid{p}\mathop{.\hspace{-.6ex}{+}\hspace{-.6ex}.}\Varid{q})\mathrel{=}\Varid{one}\;\Varid{p}\,\mathop{\scalebox{1}[.7]{\raisebox{.3ex}{$\langle\hspace{-.1ex}\vert\hspace{-.1ex}\rangle$}}}\,\Varid{one}\;\Varid{q}{}\<[E]%
\\
\>[B]{}\Varid{one}\;(\Varid{p}\mathop{\|}\Varid{q}){}\<[E]%
\\
\>[B]{}\hsindent{3}{}\<[3]%
\>[3]{}\mathrel{=}{}\<[3E]%
\>[8]{}\mathbf{do}\;{}\<[12]%
\>[12]{}(\Varid{l},\Varid{p'})\leftarrow \Varid{one}\;\Varid{p}\hspace{.2ex};\hspace{.2ex}\;\Varid{return}\;(\Varid{l},\Varid{p'}\mathop{\|}\Varid{q}){}\<[E]%
\\
\>[B]{}\hsindent{3}{}\<[3]%
\>[3]{}\,\mathop{\scalebox{1}[.7]{\raisebox{.3ex}{$\langle\hspace{-.1ex}\vert\hspace{-.1ex}\rangle$}}}\,{}\<[3E]%
\>[8]{}\mathbf{do}\;{}\<[12]%
\>[12]{}(\Varid{l},\Varid{q'})\leftarrow \Varid{one}\;\Varid{q}\hspace{.2ex};\hspace{.2ex}\;\Varid{return}\;(\Varid{l},\Varid{p}\mathop{\|}\Varid{q'}){}\<[E]%
\\
\>[B]{}\hsindent{3}{}\<[3]%
\>[3]{}\,\mathop{\scalebox{1}[.7]{\raisebox{.3ex}{$\langle\hspace{-.1ex}\vert\hspace{-.1ex}\rangle$}}}\,{}\<[3E]%
\>[8]{}\mathbf{do}\;{}\<[12]%
\>[12]{}(l_p,b_{\!p})\leftarrow \textit{one}_{\textsc{b}}\;\Varid{p}\hspace{.2ex};\hspace{.2ex}\;(l_q,b_{\!q})\leftarrow \textit{one}_{\textsc{b}}\;\Varid{q}{}\<[E]%
\\
\>[12]{}\mathbf{case}\;(l_p,l_q)\;\mathbf{of}\;{}\<[30]%
\>[30]{}(\MVUparrow_{\!\textsc{b}\!}\;\Varid{x},\hspace{-.15ex}\MVDnarrow^{\hspace{-.18ex}\textsc{b}\!}\;\Varid{x'})\mid \Varid{x}\equiv \Varid{x'}{}\<[65]%
\>[65]{}\mbox{\onelinecomment  close}{}\<[E]%
\\
\>[30]{}\hsindent{15}{}\<[45]%
\>[45]{}\to \mathbf{do}\;{}\<[52]%
\>[52]{}(\Varid{y},\Varid{p'},\Varid{q'})\leftarrow \Varid{unbind2'}\;b_{\!p}\;b_{\!q}{}\<[E]%
\\
\>[52]{}\Varid{return}\;(\ensuremath{\scalebox{1.3}{$\tau$}},\scalebox{1.25}{$\nu$}\!\;(\Varid{y}\hspace{.1ex}.\hspace{-.4ex}\backslash\Varid{p'}\mathop{\|}\Varid{q'})){}\<[E]%
\\
\>[30]{}(\hspace{-.15ex}\MVDnarrow^{\hspace{-.18ex}\textsc{b}\!}\;\Varid{x'},\MVUparrow_{\!\textsc{b}\!}\;\Varid{x})\mid \Varid{x'}\equiv \Varid{x}{}\<[65]%
\>[65]{}\mbox{\onelinecomment  close}{}\<[E]%
\\
\>[30]{}\hsindent{15}{}\<[45]%
\>[45]{}\to \mathbf{do}\;{}\<[52]%
\>[52]{}(\Varid{y},\Varid{q'},\Varid{p'})\leftarrow \Varid{unbind2'}\;b_{\!q}\;b_{\!p}{}\<[E]%
\\
\>[52]{}\Varid{return}\;(\ensuremath{\scalebox{1.3}{$\tau$}},\scalebox{1.25}{$\nu$}\!\;(\Varid{y}\hspace{.1ex}.\hspace{-.4ex}\backslash\Varid{p'}\mathop{\|}\Varid{q'})){}\<[E]%
\\
\>[30]{}\anonymous {}\<[45]%
\>[45]{}\to \Varid{empty}{}\<[E]%
\\
\>[B]{}\hsindent{3}{}\<[3]%
\>[3]{}\,\mathop{\scalebox{1}[.7]{\raisebox{.3ex}{$\langle\hspace{-.1ex}\vert\hspace{-.1ex}\rangle$}}}\,{}\<[3E]%
\>[8]{}\mathbf{do}\;{}\<[12]%
\>[12]{}(\MVUparrow\!\;\Varid{x}\;\Varid{v},\Varid{p'})\leftarrow \Varid{one}\;\Varid{p}\hspace{.2ex};\hspace{.2ex}\;(\hspace{-.15ex}\MVDnarrow^{\hspace{-.18ex}\textsc{b}\!}\;\Varid{x'},(\Varid{y},\Varid{q'}))\leftarrow \textit{one}_{\textsc{b}}'\;\Varid{q}{}\<[E]%
\\
\>[12]{}\Varid{guard}\mathop{\texttt{\$}}\Varid{x}\equiv \Varid{x'}{}\<[E]%
\\
\>[12]{}\Varid{return}\;(\ensuremath{\scalebox{1.3}{$\tau$}},\Varid{p'}\mathop{\|}\mathop{\sub{\Varid{y}}{\Varid{v}}\!}\!\;\Varid{q'}){}\<[49]%
\>[49]{}\mbox{\onelinecomment  interaction}{}\<[E]%
\\
\>[B]{}\hsindent{3}{}\<[3]%
\>[3]{}\,\mathop{\scalebox{1}[.7]{\raisebox{.3ex}{$\langle\hspace{-.1ex}\vert\hspace{-.1ex}\rangle$}}}\,{}\<[3E]%
\>[8]{}\mathbf{do}\;{}\<[12]%
\>[12]{}(\hspace{-.15ex}\MVDnarrow^{\hspace{-.18ex}\textsc{b}\!}\;\Varid{x'},(\Varid{y},\Varid{p'}))\leftarrow \textit{one}_{\textsc{b}}'\;\Varid{p}\hspace{.2ex};\hspace{.2ex}\;(\MVUparrow\!\;\Varid{x}\;\Varid{v},\Varid{q'})\leftarrow \Varid{one}\;\Varid{q}{}\<[E]%
\\
\>[12]{}\Varid{guard}\mathop{\texttt{\$}}\Varid{x}\equiv \Varid{x'}{}\<[E]%
\\
\>[12]{}\Varid{return}\;(\ensuremath{\scalebox{1.3}{$\tau$}},\mathop{\sub{\Varid{y}}{\Varid{v}}\!}\!\;\Varid{p'}\mathop{\|}\Varid{q'}){}\<[49]%
\>[49]{}\mbox{\onelinecomment  interaction}{}\<[E]%
\\
\>[B]{}\Varid{one}\;(\scalebox{1.25}{$\nu$}\!\;\Varid{b}){}\<[13]%
\>[13]{}\mathrel{=}\mathbf{do}\;{}\<[19]%
\>[19]{}(\Varid{x},\Varid{p})\leftarrow (\hspace{.1ex}.\hspace{-.4ex}\backslash)^{{\text{-}\hspace{-.2ex}1\!}}\;\Varid{b}{}\<[E]%
\\
\>[19]{}(\Varid{l},\Varid{p'})\leftarrow \Varid{one}\;\Varid{p}{}\<[E]%
\\
\>[19]{}\mathbf{case}\;\Varid{l}\;\mathbf{of}\;{}\<[30]%
\>[30]{}\MVUparrow\!\;(V\!\;\Varid{x'})\;(V\!\;\Varid{y}){}\<[51]%
\>[51]{}\mid \Varid{x}\equiv \Varid{x'}{}\<[62]%
\>[62]{}\to \Varid{empty}{}\<[E]%
\\
\>[51]{}\mid \Varid{x}\equiv \Varid{y}{}\<[62]%
\>[62]{}\to \Varid{empty}{}\<[E]%
\\
\>[30]{}\anonymous {}\<[51]%
\>[51]{}\to \Varid{return}\;(\Varid{l},\scalebox{1.25}{$\nu$}\!\;(\Varid{x}\hspace{.1ex}.\hspace{-.4ex}\backslash\Varid{p'})){}\<[E]%
\\
\>[B]{}\Varid{one}\;\anonymous {}\<[13]%
\>[13]{}\mathrel{=}\Varid{empty}{}\<[E]%
\\[\blanklineskip]%
\>[B]{}\textit{one}_{\textsc{b}}\mathbin{::}(\Conid{Fresh}\;\Varid{m},\Conid{Alternative}\;\Varid{m})\Rightarrow \Conid{Pr}\to \Varid{m}\;(\textit{Act}_{\textsc{b}},\textit{Pr}_{\textsc{b}}){}\<[E]%
\\
\>[B]{}\textit{one}_{\textsc{b}}\;(\Conid{In}\;\Varid{x}\;\Varid{p}){}\<[20]%
\>[20]{}\mathrel{=}\Varid{return}\;(\hspace{-.15ex}\MVDnarrow^{\hspace{-.18ex}\textsc{b}\!}\;\Varid{x},\Varid{p}){}\<[E]%
\\
\>[B]{}\textit{one}_{\textsc{b}}\;((\Varid{x}\mathop{{\leftrightarrow}\hspace{-1.48ex}\raisebox{.1ex}{:}\;\,}\Varid{y})\;\Varid{p})\mid \Varid{x}\equiv \Varid{y}\mathrel{=}\textit{one}_{\textsc{b}}\;\Varid{p}{}\<[E]%
\\
\>[B]{}\textit{one}_{\textsc{b}}\;(\Varid{p}\mathop{.\hspace{-.6ex}{+}\hspace{-.6ex}.}\Varid{q}){}\<[18]%
\>[18]{}\mathrel{=}\textit{one}_{\textsc{b}}\;\Varid{p}\,\mathop{\scalebox{1}[.7]{\raisebox{.3ex}{$\langle\hspace{-.1ex}\vert\hspace{-.1ex}\rangle$}}}\,\textit{one}_{\textsc{b}}\;\Varid{q}{}\<[E]%
\\
\>[B]{}\textit{one}_{\textsc{b}}\;(\Varid{p}\mathop{\|}\Varid{q}){}\<[18]%
\>[18]{}\mathrel{=}{}\<[24]%
\>[24]{}\mathbf{do}\;{}\<[28]%
\>[28]{}(\Varid{l},(\Varid{x},\Varid{p'}))\leftarrow \textit{one}_{\textsc{b}}'\;\Varid{p}\hspace{.2ex};\hspace{.2ex}\;\Varid{return}\;(\Varid{l},\Varid{x}\hspace{.1ex}.\hspace{-.4ex}\backslash\Varid{p'}\mathop{\|}\Varid{q}){}\<[E]%
\\
\>[18]{}\,\mathop{\scalebox{1}[.7]{\raisebox{.3ex}{$\langle\hspace{-.1ex}\vert\hspace{-.1ex}\rangle$}}}\,{}\<[24]%
\>[24]{}\mathbf{do}\;{}\<[28]%
\>[28]{}(\Varid{l},(\Varid{x},\Varid{q'}))\leftarrow \textit{one}_{\textsc{b}}'\;\Varid{q}\hspace{.2ex};\hspace{.2ex}\;\Varid{return}\;(\Varid{l},\Varid{x}\hspace{.1ex}.\hspace{-.4ex}\backslash\Varid{p}\mathop{\|}\Varid{q'}){}\<[E]%
\\
\>[B]{}\textit{one}_{\textsc{b}}\;(\scalebox{1.25}{$\nu$}\!\;\Varid{b}){}\<[18]%
\>[18]{}\mathrel{=}{}\<[24]%
\>[24]{}\mathbf{do}\;{}\<[28]%
\>[28]{}(\Varid{x},\Varid{p})\leftarrow (\hspace{.1ex}.\hspace{-.4ex}\backslash)^{{\text{-}\hspace{-.2ex}1\!}}\;\Varid{b}{}\<[E]%
\\
\>[28]{}(\Varid{l},(\Varid{y},\Varid{p'}))\leftarrow \textit{one}_{\textsc{b}}'\;\Varid{p}{}\<[E]%
\\
\>[28]{}\mathbf{case}\;\Varid{l}\;\mathbf{of}\;{}\<[39]%
\>[39]{}\MVUparrow_{\!\textsc{b}\!}\;(V\!\;\Varid{x'}){}\<[54]%
\>[54]{}\mid \Varid{x}\equiv \Varid{x'}\to \Varid{empty}{}\<[E]%
\\
\>[39]{}\hspace{-.15ex}\MVDnarrow^{\hspace{-.18ex}\textsc{b}\!}\;(V\!\;\Varid{x'}){}\<[54]%
\>[54]{}\mid \Varid{x}\equiv \Varid{x'}\to \Varid{empty}{}\<[E]%
\\
\>[39]{}\anonymous {}\<[54]%
\>[54]{}\to \Varid{return}\;(\Varid{l},\Varid{y}\hspace{.1ex}.\hspace{-.4ex}\backslash\scalebox{1.25}{$\nu$}\!\;(\Varid{x}\hspace{.1ex}.\hspace{-.4ex}\backslash\Varid{p'})){}\<[E]%
\\
\>[18]{}\,\mathop{\scalebox{1}[.7]{\raisebox{.3ex}{$\langle\hspace{-.1ex}\vert\hspace{-.1ex}\rangle$}}}\,{}\<[24]%
\>[24]{}\mathbf{do}\;{}\<[28]%
\>[28]{}(\Varid{x},\Varid{p})\leftarrow (\hspace{.1ex}.\hspace{-.4ex}\backslash)^{{\text{-}\hspace{-.2ex}1\!}}\;\Varid{b}{}\<[E]%
\\
\>[28]{}(\MVUparrow\!\;\Varid{y}\;(V\!\;\Varid{x'}),\Varid{p'})\leftarrow \Varid{one}\;\Varid{p}{}\<[E]%
\\
\>[28]{}\Varid{guard}\mathop{\texttt{\$}}\Varid{x}\equiv \Varid{x'}\mathrel{\wedge}V\!\;\Varid{x}\not\equiv \Varid{y}{}\<[E]%
\\
\>[28]{}\Varid{return}\;(\MVUparrow_{\!\textsc{b}\!}\;\Varid{y},\Varid{x}\hspace{.1ex}.\hspace{-.4ex}\backslash\Varid{p'}){}\<[51]%
\>[51]{}\mbox{\onelinecomment  open}{}\<[E]%
\\
\>[B]{}\textit{one}_{\textsc{b}}\;\anonymous {}\<[18]%
\>[18]{}\mathrel{=}\Varid{empty}{}\<[E]%
\\[\blanklineskip]%
\>[B]{}\textit{one}_{\textsc{b}}'\;\Varid{p}\mathrel{=}\mathbf{do}\;(\Varid{l},\Varid{b})\leftarrow \textit{one}_{\textsc{b}}\;\Varid{p}\hspace{.2ex};\hspace{.2ex}\;\Varid{r}\leftarrow (\hspace{.1ex}.\hspace{-.4ex}\backslash)^{{\text{-}\hspace{-.2ex}1\!}}\;\Varid{b}\hspace{.2ex};\hspace{.2ex}\;\Varid{return}\;(\Varid{l},\Varid{r}){}\<[E]%
\ColumnHook
\end{hscode}\resethooks
\vspace*{-3.5ex}
\caption{Labeled transition semantics within a fixed world.}
\label{fig:IdSubLTS}
\end{figure}

%
%
\makeatletter
\@ifundefined{lhs2tex.lhs2tex.sty.read}%
  {\@namedef{lhs2tex.lhs2tex.sty.read}{}%
   \newcommand\SkipToFmtEnd{}%
   \newcommand\EndFmtInput{}%
   \long\def\SkipToFmtEnd#1\EndFmtInput{}%
  }\SkipToFmtEnd

\newcommand\ReadOnlyOnce[1]{\@ifundefined{#1}{\@namedef{#1}{}}\SkipToFmtEnd}
\usepackage{amstext}
\usepackage{amssymb}
\usepackage{stmaryrd}
\DeclareFontFamily{OT1}{cmtex}{}
\DeclareFontShape{OT1}{cmtex}{m}{n}
  {<5><6><7><8>cmtex8
   <9>cmtex9
   <10><10.95><12><14.4><17.28><20.74><24.88>cmtex10}{}
\DeclareFontShape{OT1}{cmtex}{m}{it}
  {<-> ssub * cmtt/m/it}{}
\newcommand{\texfamily}{\fontfamily{cmtex}\selectfont}
\DeclareFontShape{OT1}{cmtt}{bx}{n}
  {<5><6><7><8>cmtt8
   <9>cmbtt9
   <10><10.95><12><14.4><17.28><20.74><24.88>cmbtt10}{}
\DeclareFontShape{OT1}{cmtex}{bx}{n}
  {<-> ssub * cmtt/bx/n}{}
\newcommand{\tex}[1]{\text{\texfamily#1}}	

\newcommand{\Sp}{\hskip.33334em\relax}

\newcommand{\Conid}[1]{\mathit{#1}}
\newcommand{\Varid}[1]{\mathit{#1}}
\newcommand{\anonymous}{\kern0.06em \vbox{\hrule\@width.5em}}
\newcommand{\plus}{\mathbin{+\!\!\!+}}
\newcommand{\bind}{\mathbin{>\!\!\!>\mkern-6.7mu=}}
\newcommand{\rbind}{\mathbin{=\mkern-6.7mu<\!\!\!<}}
\newcommand{\sequ}{\mathbin{>\!\!\!>}}
\renewcommand{\leq}{\leqslant}
\renewcommand{\geq}{\geqslant}
\usepackage{polytable}

\@ifundefined{mathindent}%
  {\newdimen\mathindent\mathindent\leftmargini}%
  {}%

\def\resethooks{%
  \global\let\SaveRestoreHook\empty
  \global\let\ColumnHook\empty}
\newcommand*{\savecolumns}[1][default]%
  {\g@addto@macro\SaveRestoreHook{\savecolumns[#1]}}
\newcommand*{\restorecolumns}[1][default]%
  {\g@addto@macro\SaveRestoreHook{\restorecolumns[#1]}}
\newcommand*{\aligncolumn}[2]%
  {\g@addto@macro\ColumnHook{\column{#1}{#2}}}

\resethooks

\newcommand{\onelinecommentchars}{\quad-{}- }
\newcommand{\commentbeginchars}{\enskip\{-}
\newcommand{\commentendchars}{-\}\enskip}

\newcommand{\visiblecomments}{%
  \let\onelinecomment=\onelinecommentchars
  \let\commentbegin=\commentbeginchars
  \let\commentend=\commentendchars}

\newcommand{\invisiblecomments}{%
  \let\onelinecomment=\empty
  \let\commentbegin=\empty
  \let\commentend=\empty}

\visiblecomments

\newlength{\blanklineskip}
\setlength{\blanklineskip}{0.66084ex}

\newcommand{\hsindent}[1]{\quad}
\let\hspre\empty
\let\hspost\empty
\newcommand{\NB}{\textbf{NB}}
\newcommand{\Todo}[1]{$\langle$\textbf{To do:}~#1$\rangle$}

\EndFmtInput
\makeatother
%
%
%
%
%
%
%
%
%
\ReadOnlyOnce{polycode.fmt}%
\makeatletter

\newcommand{\hsnewpar}[1]%
  {{\parskip=0pt\parindent=0pt\par\vskip #1\noindent}}

\newcommand{\hscodestyle}{}


\newcommand{\sethscode}[1]%
  {\expandafter\let\expandafter\hscode\csname #1\endcsname
   \expandafter\let\expandafter\endhscode\csname end#1\endcsname}


\newenvironment{compathscode}%
  {\par\noindent
   \advance\leftskip\mathindent
   \hscodestyle
   \let\\=\@normalcr
   \let\hspre\(\let\hspost\)%
   \pboxed}%
  {\endpboxed\)%
   \par\noindent
   \ignorespacesafterend}

\newcommand{\compaths}{\sethscode{compathscode}}


\newenvironment{plainhscode}%
  {\hsnewpar\abovedisplayskip
   \advance\leftskip\mathindent
   \hscodestyle
   \let\hspre\(\let\hspost\)%
   \pboxed}%
  {\endpboxed%
   \hsnewpar\belowdisplayskip
   \ignorespacesafterend}

\newenvironment{oldplainhscode}%
  {\hsnewpar\abovedisplayskip
   \advance\leftskip\mathindent
   \hscodestyle
   \let\\=\@normalcr
   \(\pboxed}%
  {\endpboxed\)%
   \hsnewpar\belowdisplayskip
   \ignorespacesafterend}


\newcommand{\plainhs}{\sethscode{plainhscode}}
\newcommand{\oldplainhs}{\sethscode{oldplainhscode}}
\plainhs


\newenvironment{arrayhscode}%
  {\hsnewpar\abovedisplayskip
   \advance\leftskip\mathindent
   \hscodestyle
   \let\\=\@normalcr
   \(\parray}%
  {\endparray\)%
   \hsnewpar\belowdisplayskip
   \ignorespacesafterend}

\newcommand{\arrayhs}{\sethscode{arrayhscode}}


\newenvironment{mathhscode}%
  {\parray}{\endparray}

\newcommand{\mathhs}{\sethscode{mathhscode}}


\newenvironment{texthscode}%
  {\(\parray}{\endparray\)}

\newcommand{\texths}{\sethscode{texthscode}}


\def\codeframewidth{\arrayrulewidth}
\RequirePackage{calc}

\newenvironment{framedhscode}%
  {\parskip=\abovedisplayskip\par\noindent
   \hscodestyle
   \arrayrulewidth=\codeframewidth
   \tabular{@{}|p{\linewidth-2\arraycolsep-2\arrayrulewidth-2pt}|@{}}%
   \hline\framedhslinecorrect\\{-1.5ex}%
   \let\endoflinesave=\\
   \let\\=\@normalcr
   \(\pboxed}%
  {\endpboxed\)%
   \framedhslinecorrect\endoflinesave{.5ex}\hline
   \endtabular
   \parskip=\belowdisplayskip\par\noindent
   \ignorespacesafterend}

\newcommand{\framedhslinecorrect}[2]%
  {#1[#2]}

\newcommand{\framedhs}{\sethscode{framedhscode}}


\newenvironment{inlinehscode}%
  {\(\def\column##1##2{}%
   \let\>\undefined\let\<\undefined\let\\\undefined
   \newcommand\>[1][]{}\newcommand\<[1][]{}\newcommand\\[1][]{}%
   \def\fromto##1##2##3{##3}%
   \def\nextline{}}{\) }%

\newcommand{\inlinehs}{\sethscode{inlinehscode}}


\newenvironment{joincode}%
  {\let\orighscode=\hscode
   \let\origendhscode=\endhscode
   \def\endhscode{\def\hscode{\endgroup\def\@currenvir{hscode}\\}\begingroup}
   \orighscode\def\hscode{\endgroup\def\@currenvir{hscode}}}%
  {\origendhscode
   \global\let\hscode=\orighscode
   \global\let\endhscode=\origendhscode}%

\makeatother
\EndFmtInput
\ReadOnlyOnce{colorcode.fmt}%

\RequirePackage{colortbl}
\RequirePackage{calc}

\makeatletter
\newenvironment{colorhscode}%
  {\hsnewpar\abovedisplayskip
   \hscodestyle
   \tabular{@{}>{\columncolor{codecolor}}p{\linewidth}@{}}%
   \let\\=\@normalcr
   \(\pboxed}%
  {\endpboxed\)%
   \endtabular
   \hsnewpar\belowdisplayskip
   \ignorespacesafterend}

\newenvironment{tightcolorhscode}%
  {\hsnewpar\abovedisplayskip
   \hscodestyle
   \tabular{@{}>{\columncolor{codecolor}\(}l<{\)}@{}}%
   \pmboxed}%
  {\endpmboxed%
   \endtabular
   \hsnewpar\belowdisplayskip
   \ignorespacesafterend}

\newenvironment{barhscode}%
  {\hsnewpar\abovedisplayskip
   \hscodestyle
   \arrayrulecolor{codecolor}%
   \arrayrulewidth=\coderulewidth
   \tabular{|p{\linewidth-\arrayrulewidth-\tabcolsep}@{}}%
   \let\\=\@normalcr
   \(\pboxed}%
  {\endpboxed\)%
   \endtabular
   \hsnewpar\belowdisplayskip
   \ignorespacesafterend}
\makeatother

\def\colorcode{\columncolor{codecolor}}
\definecolor{codecolor}{rgb}{1,1,.667}
\newlength{\coderulewidth}
\setlength{\coderulewidth}{3pt}

\newcommand{\colorhs}{\sethscode{colorhscode}}
\newcommand{\tightcolorhs}{\sethscode{tightcolorhscode}}
\newcommand{\barhs}{\sethscode{barhscode}}

\EndFmtInput

\renewcommand{\onelinecommentchars}{\color{gray}\quad-{}- }
\renewcommand{\commentbeginchars}{\color{gray}\enskip\{- }
\renewcommand{\commentendchars}{-\}\enskip}

\renewcommand{\visiblecomments}{%
  \let\onelinecomment=\onelinecommentchars
  \let\commentbegin=\commentbeginchars
  \let\commentend=\commentendchars}

\renewcommand{\invisiblecomments}{%
  \let\onelinecomment=\empty
  \let\commentbegin=\empty
  \let\commentend=\empty}

\visiblecomments

\begin{figure}\small
\savecolumns
\begin{hscode}\SaveRestoreHook
\column{B}{@{}>{\hspre}l<{\hspost}@{}}%
\column{3}{@{}>{\hspre}c<{\hspost}@{}}%
\column{3E}{@{}l@{}}%
\column{8}{@{}>{\hspre}l<{\hspost}@{}}%
\column{11}{@{}>{\hspre}l<{\hspost}@{}}%
\column{12}{@{}>{\hspre}l<{\hspost}@{}}%
\column{14}{@{}>{\hspre}l<{\hspost}@{}}%
\column{25}{@{}>{\hspre}l<{\hspost}@{}}%
\column{38}{@{}>{\hspre}l<{\hspost}@{}}%
\column{40}{@{}>{\hspre}l<{\hspost}@{}}%
\column{44}{@{}>{\hspre}l<{\hspost}@{}}%
\column{47}{@{}>{\hspre}l<{\hspost}@{}}%
\column{48}{@{}>{\hspre}l<{\hspost}@{}}%
\column{63}{@{}>{\hspre}l<{\hspost}@{}}%
\column{65}{@{}>{\hspre}l<{\hspost}@{}}%
\column{E}{@{}>{\hspre}l<{\hspost}@{}}%
\>[B]{}\mbox{\onelinecomment  preamble of this \ensuremath{\Conid{OpenLTS}} module is on Figure~\ref{fig:figureOpenLTS}}{}\<[E]%
\\[\blanklineskip]%
\>[B]{}\Varid{one}\mathbin{::}(\Conid{Fresh}\;\Varid{m},\Conid{Alternative}\;\Varid{m})\Rightarrow \Conid{Ctx}\to \Conid{Pr}\to \Varid{m}\;(\Conid{EqC},(\Conid{Act},\Conid{Pr})){}\<[E]%
\\
\>[B]{}\Varid{one}\;\Gamma\;{}\<[11]%
\>[11]{}(\Conid{Out}\;\Varid{x}\;\Varid{y}\;\Varid{p}){}\<[25]%
\>[25]{}\mathrel{=}\Varid{return}\;([\mskip1.5mu \mskip1.5mu],(\MVUparrow\!\;\Varid{x}\;\Varid{y},\Varid{p})){}\<[E]%
\\
\>[B]{}\Varid{one}\;\Gamma\;{}\<[11]%
\>[11]{}(\scalebox{1.3}{$\tau\hspace{-.4ex}\raisebox{.29ex}{\textbf{.}}$}\;\Varid{p}){}\<[25]%
\>[25]{}\mathrel{=}\Varid{return}\;([\mskip1.5mu \mskip1.5mu],(\ensuremath{\scalebox{1.3}{$\tau$}},\Varid{p})){}\<[E]%
\\
\>[B]{}\Varid{one}\;\Gamma\;{}\<[11]%
\>[11]{}((V\!\;\Varid{x}\mathop{{\leftrightarrow}\hspace{-1.48ex}\raisebox{.1ex}{:}\;\,}V\!\;\Varid{y})\;\Varid{p}){}\<[38]%
\>[38]{}\mid \Varid{x}\equiv \Varid{y}{}\<[65]%
\>[65]{}\mathrel{=}\Varid{one}\;\Gamma\;\Varid{p}{}\<[E]%
\\
\>[38]{}\mid [\mskip1.5mu (\Varid{x},\Varid{y})\mskip1.5mu]\mathop{^{\,_{\backprime}\!}\Varid{respects}^{_\backprime}}\Gamma{}\<[65]%
\>[65]{}\mathrel{=}{}\<[E]%
\\
\>[38]{}\hsindent{6}{}\<[44]%
\>[44]{}\mathbf{do}\;{}\<[48]%
\>[48]{}(\sigma,\Varid{r})\leftarrow \Varid{one}\;\Gamma\;\Varid{p}{}\<[E]%
\\
\>[48]{}\mathbf{let}\;\sigma'\mathrel{=}(\Varid{x},\Varid{y})\mathop{\raisebox{.5ex}{$\curvearrowright$}\hspace{-1.9ex}\scalebox{.8}{$+$}\;}\sigma{}\<[E]%
\\
\>[48]{}\Varid{guard}\mathop{\texttt{\$}}\sigma'\mathop{^{\,_{\backprime}\!}\Varid{respects}^{_\backprime}}\Gamma{}\<[E]%
\\
\>[48]{}\Varid{return}\;(\sigma',\Varid{r}){}\<[E]%
\\
\>[B]{}\Varid{one}\;\Gamma\;{}\<[11]%
\>[11]{}(\Varid{p}\mathop{.\hspace{-.6ex}{+}\hspace{-.6ex}.}\Varid{q})\mathrel{=}\Varid{one}\;\Gamma\;\Varid{p}\,\mathop{\scalebox{1}[.7]{\raisebox{.3ex}{$\langle\hspace{-.1ex}\vert\hspace{-.1ex}\rangle$}}}\,\Varid{one}\;\Gamma\;\Varid{q}{}\<[E]%
\\
\>[B]{}\Varid{one}\;\Gamma\;{}\<[11]%
\>[11]{}(\Varid{p}\mathop{\|}\Varid{q}){}\<[E]%
\\
\>[B]{}\hsindent{3}{}\<[3]%
\>[3]{}\mathrel{=}{}\<[3E]%
\>[8]{}\mathbf{do}\;{}\<[12]%
\>[12]{}(\sigma,(\Varid{l},\Varid{p'}))\leftarrow \Varid{one}\;\Gamma\;\Varid{p}\hspace{.2ex};\hspace{.2ex}~$~$\Varid{return}\;(\sigma,(\Varid{l},\Varid{p'}\mathop{\|}\Varid{q})){}\<[E]%
\\
\>[B]{}\hsindent{3}{}\<[3]%
\>[3]{}\,\mathop{\scalebox{1}[.7]{\raisebox{.3ex}{$\langle\hspace{-.1ex}\vert\hspace{-.1ex}\rangle$}}}\,{}\<[3E]%
\>[8]{}\mathbf{do}\;{}\<[12]%
\>[12]{}(\sigma,(\Varid{l},\Varid{q'}))\leftarrow \Varid{one}\;\Gamma\;\Varid{q}\hspace{.2ex};\hspace{.2ex}~$~$\Varid{return}\;(\sigma,(\Varid{l},\Varid{p}\mathop{\|}\Varid{q'})){}\<[E]%
\\
\>[B]{}\hsindent{3}{}\<[3]%
\>[3]{}\,\mathop{\scalebox{1}[.7]{\raisebox{.3ex}{$\langle\hspace{-.1ex}\vert\hspace{-.1ex}\rangle$}}}\,{}\<[3E]%
\>[8]{}\mathbf{do}\;{}\<[12]%
\>[12]{}(\sigma_{\!p},(l_p,b_{\!p}))\leftarrow \textit{one}_{\textsc{b}}\;\Gamma\;\Varid{p}\hspace{.2ex};\hspace{.2ex}~$~$(\sigma_{\!q},(l_q,b_{\!q}))\leftarrow \textit{one}_{\textsc{b}}\;\Gamma\;\Varid{q}{}\<[E]%
\\
\>[12]{}\mathbf{case}\;(l_p,l_q)\;\mathbf{of}\;{}\<[40]%
\>[40]{}\quad\!\!\mbox{\onelinecomment  close}{}\<[E]%
\\
\>[12]{}\hsindent{2}{}\<[14]%
\>[14]{}(\hspace{-.15ex}\MVDnarrow^{\hspace{-.18ex}\textsc{b}\!}\;(V\!\;\Varid{x}),\MVUparrow_{\!\textsc{b}\!}\;(V\!\;\Varid{x'})){}\<[40]%
\>[40]{}\to \mathbf{do}\;{}\<[47]%
\>[47]{}(\Varid{y},\Varid{q'},\Varid{p'})\leftarrow {}\<[63]%
\>[63]{}\Varid{unbind2'}\;b_{\!q}\;b_{\!p}{}\<[E]%
\\
\>[47]{}\mathbf{let}\;\sigma'\mathrel{=}(\Varid{x},\Varid{x'})\mathop{\raisebox{.5ex}{$\curvearrowright$}\hspace{-1.9ex}\scalebox{.8}{$+$}\;}\sigma_{\!p}\cup\sigma_{\!q}{}\<[E]%
\\
\>[47]{}\Varid{guard}\mathop{\texttt{\$}}\sigma'\mathop{^{\,_{\backprime}\!}\Varid{respects}^{_\backprime}}\Gamma{}\<[E]%
\\
\>[47]{}\Varid{return}\;(\sigma',(\ensuremath{\scalebox{1.3}{$\tau$}},\scalebox{1.25}{$\nu$}\!\;(\Varid{y}\hspace{.1ex}.\hspace{-.4ex}\backslash\Varid{p'}\mathop{\|}\Varid{q'}))){}\<[E]%
\\
\>[12]{}\hsindent{2}{}\<[14]%
\>[14]{}(\MVUparrow_{\!\textsc{b}\!}\;(V\!\;\Varid{x'}),\hspace{-.15ex}\MVDnarrow^{\hspace{-.18ex}\textsc{b}\!}\;(V\!\;\Varid{x})){}\<[40]%
\>[40]{}\to {}\<[44]%
\>[44]{}\ldots\mbox{\onelinecomment  omitted (close)}{}\<[E]%
\ColumnHook
\end{hscode}\resethooks
\vspace*{-4.7ex}
\restorecolumns
\savecolumns
\begin{hscode}\SaveRestoreHook
\column{B}{@{}>{\hspre}l<{\hspost}@{}}%
\column{3}{@{}>{\hspre}c<{\hspost}@{}}%
\column{3E}{@{}l@{}}%
\column{8}{@{}>{\hspre}l<{\hspost}@{}}%
\column{12}{@{}>{\hspre}l<{\hspost}@{}}%
\column{14}{@{}>{\hspre}l<{\hspost}@{}}%
\column{40}{@{}>{\hspre}l<{\hspost}@{}}%
\column{E}{@{}>{\hspre}l<{\hspost}@{}}%
\>[14]{}\anonymous {}\<[40]%
\>[40]{}\to \Varid{empty}{}\<[E]%
\\
\>[3]{}\,\mathop{\scalebox{1}[.7]{\raisebox{.3ex}{$\langle\hspace{-.1ex}\vert\hspace{-.1ex}\rangle$}}}\,{}\<[3E]%
\>[8]{}\mathbf{do}\;{}\<[12]%
\>[12]{}(\sigma_{\!p},(\MVUparrow\!\;(V\!\;\Varid{x})\;\Varid{v},\Varid{p'}))\leftarrow \Varid{one}\;\Gamma\;\Varid{p}{}\<[E]%
\\
\>[12]{}(\sigma_{\!q},(\hspace{-.15ex}\MVDnarrow^{\hspace{-.18ex}\textsc{b}\!}\;(V\!\;\Varid{x'}),b_{\!q}))\leftarrow \textit{one}_{\textsc{b}}\;\Gamma\;\Varid{q}\hspace{.2ex};\hspace{.2ex}(\Varid{y},\Varid{q'})\leftarrow (\hspace{.1ex}.\hspace{-.4ex}\backslash)^{{\text{-}\hspace{-.2ex}1\!}}\;b_{\!q}{}\<[E]%
\\
\>[12]{}\mathbf{let}\;\sigma'\mathrel{=}(\Varid{x},\Varid{x'})\mathop{\raisebox{.5ex}{$\curvearrowright$}\hspace{-1.9ex}\scalebox{.8}{$+$}\;}\sigma_{\!p}\cup\sigma_{\!q}{}\<[E]%
\\
\>[12]{}\Varid{guard}\mathop{\texttt{\$}}\sigma'\mathop{^{\,_{\backprime}\!}\Varid{respects}^{_\backprime}}\Gamma{}\<[E]%
\\
\>[12]{}\Varid{return}\;(\sigma',(\ensuremath{\scalebox{1.3}{$\tau$}},\Varid{p'}\mathop{\|}\mathop{\sub{\Varid{y}}{\Varid{v}}\!}\!\;\Varid{q'}))\mbox{\onelinecomment  interaction}{}\<[E]%
\\
\>[3]{}\,\mathop{\scalebox{1}[.7]{\raisebox{.3ex}{$\langle\hspace{-.1ex}\vert\hspace{-.1ex}\rangle$}}}\,{}\<[3E]%
\>[8]{}\ldots\mbox{\onelinecomment  do block symmetric to above omitted (interaction)}{}\<[E]%
\ColumnHook
\end{hscode}\resethooks
\vspace*{-4.7ex}
\restorecolumns
\savecolumns
\begin{hscode}\SaveRestoreHook
\column{B}{@{}>{\hspre}l<{\hspost}@{}}%
\column{10}{@{}>{\hspre}l<{\hspost}@{}}%
\column{17}{@{}>{\hspre}l<{\hspost}@{}}%
\column{23}{@{}>{\hspre}l<{\hspost}@{}}%
\column{34}{@{}>{\hspre}l<{\hspost}@{}}%
\column{38}{@{}>{\hspre}l<{\hspost}@{}}%
\column{55}{@{}>{\hspre}l<{\hspost}@{}}%
\column{65}{@{}>{\hspre}l<{\hspost}@{}}%
\column{76}{@{}>{\hspre}l<{\hspost}@{}}%
\column{E}{@{}>{\hspre}l<{\hspost}@{}}%
\>[B]{}\Varid{one}\;\Gamma\;(\scalebox{1.25}{$\nu$}\!\;\Varid{b})\mathrel{=}\mathbf{do}\;{}\<[23]%
\>[23]{}(\Varid{x},\Varid{p})\leftarrow (\hspace{.1ex}.\hspace{-.4ex}\backslash)^{{\text{-}\hspace{-.2ex}1\!}}\;\Varid{b}\hspace{.2ex};\hspace{.2ex}{}\<[55]%
\>[55]{}\mathbf{let}\;\Gamma'\mathrel{=}\nabla\hspace{-.83ex}\raisebox{.1ex}{\scalebox{1.1}[.9]{$/$}}\!\;\Varid{x}\mathbin{:}\Gamma{}\<[E]%
\\
\>[23]{}(\sigma,(\Varid{l},\Varid{p'}))\leftarrow \Varid{one}\;\Gamma'\;\Varid{p}\hspace{.2ex};\hspace{.2ex}{}\<[55]%
\>[55]{}\mathbf{let}\;{\hat{\sigma}}\mathrel{=}\Varid{subs}\;\Gamma'\;\sigma{}\<[E]%
\\
\>[23]{}\mathbf{case}\;\Varid{l}\;\mathbf{of}\;{}\<[34]%
\>[34]{}\MVUparrow\!\;(V\!\;\Varid{x'})\;(V\!\;\Varid{y}){}\<[55]%
\>[55]{}\mid \Varid{x}\equiv {\hat{\sigma}}\;\Varid{x'}{}\<[76]%
\>[76]{}\to \Varid{empty}{}\<[E]%
\\
\>[55]{}\mid \Varid{x}\equiv {\hat{\sigma}}\;\Varid{y}{}\<[76]%
\>[76]{}\to \Varid{empty}{}\<[E]%
\\
\>[34]{}\anonymous {}\<[55]%
\>[55]{}\to \Varid{return}\;(\sigma,(\Varid{l},\scalebox{1.25}{$\nu$}\!\;(\Varid{x}\hspace{.1ex}.\hspace{-.4ex}\backslash\Varid{p'}))){}\<[E]%
\\
\>[B]{}\Varid{one}\;\anonymous \;{}\<[10]%
\>[10]{}\anonymous {}\<[17]%
\>[17]{}\mathrel{=}\Varid{empty}{}\<[E]%
\\[\blanklineskip]%
\>[B]{}\textit{one}_{\textsc{b}}\mathbin{::}(\Conid{Fresh}\;\Varid{m},\Conid{Alternative}\;\Varid{m})\Rightarrow \Conid{Ctx}\to \Conid{Pr}\to \Varid{m}\;(\Conid{EqC},(\textit{Act}_{\textsc{b}},\textit{Pr}_{\textsc{b}})){}\<[E]%
\\
\>[B]{}\textit{one}_{\textsc{b}}\;\Gamma\;(\Conid{In}\;\Varid{x}\;\Varid{p})\mathrel{=}\Varid{return}\;([\mskip1.5mu \mskip1.5mu],(\hspace{-.15ex}\MVDnarrow^{\hspace{-.18ex}\textsc{b}\!}\;\Varid{x},\Varid{p})){}\<[E]%
\\
\>[B]{}\textit{one}_{\textsc{b}}\;\Gamma\;((V\!\;\Varid{x}\mathop{{\leftrightarrow}\hspace{-1.48ex}\raisebox{.1ex}{:}\;\,}V\!\;\Varid{y})\;\Varid{p}){}\<[38]%
\>[38]{}\mid \Varid{x}\equiv \Varid{y}{}\<[65]%
\>[65]{}\mathrel{=}\textit{one}_{\textsc{b}}\;\Gamma\;\Varid{p}{}\<[E]%
\\
\>[38]{}\mid [\mskip1.5mu (\Varid{x},\Varid{y})\mskip1.5mu]\mathop{^{\,_{\backprime}\!}\Varid{respects}^{_\backprime}}\Gamma{}\<[65]%
\>[65]{}\mathrel{=}\ldots\mbox{\onelinecomment  omitted}{}\<[E]%
\ColumnHook
\end{hscode}\resethooks
\vspace*{-4.7ex}
\restorecolumns
\savecolumns
\begin{hscode}\SaveRestoreHook
\column{B}{@{}>{\hspre}l<{\hspost}@{}}%
\column{E}{@{}>{\hspre}l<{\hspost}@{}}%
\>[B]{}\textit{one}_{\textsc{b}}\;\Gamma\;(\Varid{p}\mathop{.\hspace{-.6ex}{+}\hspace{-.6ex}.}\Varid{q})\mathrel{=}\textit{one}_{\textsc{b}}\;\Gamma\;\Varid{p}\,\mathop{\scalebox{1}[.7]{\raisebox{.3ex}{$\langle\hspace{-.1ex}\vert\hspace{-.1ex}\rangle$}}}\,\textit{one}_{\textsc{b}}\;\Gamma\;\Varid{q}{}\<[E]%
\\
\>[B]{}\textit{one}_{\textsc{b}}\;\Gamma\;(\Varid{p}\mathop{\|}\Varid{q})\mathrel{=}\ldots\mbox{\onelinecomment  omitted }{}\<[E]%
\ColumnHook
\end{hscode}\resethooks
\vspace*{-4.7ex}
\restorecolumns
\begin{hscode}\SaveRestoreHook
\column{B}{@{}>{\hspre}l<{\hspost}@{}}%
\column{11}{@{}>{\hspre}l<{\hspost}@{}}%
\column{19}{@{}>{\hspre}c<{\hspost}@{}}%
\column{19E}{@{}l@{}}%
\column{24}{@{}>{\hspre}l<{\hspost}@{}}%
\column{28}{@{}>{\hspre}l<{\hspost}@{}}%
\column{39}{@{}>{\hspre}l<{\hspost}@{}}%
\column{53}{@{}>{\hspre}l<{\hspost}@{}}%
\column{66}{@{}>{\hspre}l<{\hspost}@{}}%
\column{72}{@{}>{\hspre}l<{\hspost}@{}}%
\column{E}{@{}>{\hspre}l<{\hspost}@{}}%
\>[B]{}\textit{one}_{\textsc{b}}\;\Gamma\;(\scalebox{1.25}{$\nu$}\!\;\Varid{b}){}\<[19]%
\>[19]{}\mathrel{=}{}\<[19E]%
\>[24]{}\mathbf{do}\;{}\<[28]%
\>[28]{}(\Varid{x},\Varid{p})\leftarrow (\hspace{.1ex}.\hspace{-.4ex}\backslash)^{{\text{-}\hspace{-.2ex}1\!}}\;\Varid{b}\hspace{.2ex};\hspace{.2ex}{}\<[66]%
\>[66]{}\mathbf{let}\;\Gamma'\mathrel{=}\nabla\hspace{-.83ex}\raisebox{.1ex}{\scalebox{1.1}[.9]{$/$}}\!\;\Varid{x}\mathbin{:}\Gamma{}\<[E]%
\\
\>[28]{}(\sigma,(\Varid{l},(\Varid{y},\Varid{p'})))\leftarrow \textit{one}_{\textsc{b}}'\;\Gamma'\;\Varid{p}\hspace{.2ex};\hspace{.2ex}{}\<[66]%
\>[66]{}\mathbf{let}\;{\hat{\sigma}}\mathrel{=}\Varid{subs}\;\Gamma'\;\sigma{}\<[E]%
\\
\>[28]{}\mathbf{case}\;\Varid{l}\;\mathbf{of}\;{}\<[39]%
\>[39]{}\MVUparrow_{\!\textsc{b}\!}\;(V\!\;\Varid{x'}){}\<[53]%
\>[53]{}\mid \Varid{x}\equiv {\hat{\sigma}}\;\Varid{x'}\to \Varid{empty}{}\<[E]%
\\
\>[39]{}\hspace{-.15ex}\MVDnarrow^{\hspace{-.18ex}\textsc{b}\!}\;(V\!\;\Varid{x'}){}\<[53]%
\>[53]{}\mid \Varid{x}\equiv {\hat{\sigma}}\;\Varid{x'}\to \Varid{empty}{}\<[E]%
\\
\>[39]{}\anonymous {}\<[53]%
\>[53]{}\to \Varid{return}\;(\sigma,(\Varid{l},\Varid{y}\hspace{.1ex}.\hspace{-.4ex}\backslash\scalebox{1.25}{$\nu$}\!\;(\Varid{x}\hspace{.1ex}.\hspace{-.4ex}\backslash\Varid{p'}))){}\<[E]%
\\
\>[19]{}\,\mathop{\scalebox{1}[.7]{\raisebox{.3ex}{$\langle\hspace{-.1ex}\vert\hspace{-.1ex}\rangle$}}}\,{}\<[19E]%
\>[24]{}\mathbf{do}\;{}\<[28]%
\>[28]{}(\Varid{x},\Varid{p})\leftarrow (\hspace{.1ex}.\hspace{-.4ex}\backslash)^{{\text{-}\hspace{-.2ex}1\!}}\;\Varid{b}\hspace{.2ex};\hspace{.2ex}{}\<[72]%
\>[72]{}\mathbf{let}\;\Gamma'\mathrel{=}\nabla\hspace{-.83ex}\raisebox{.1ex}{\scalebox{1.1}[.9]{$/$}}\!\;\Varid{x}\mathbin{:}\Gamma{}\<[E]%
\\
\>[28]{}(\sigma,(\MVUparrow\!\;\Varid{y}\;(V\!\;\Varid{x'}),\Varid{p'}))\leftarrow \Varid{one}\;\Gamma'\;\Varid{p}\hspace{.2ex};\hspace{.2ex}{}\<[72]%
\>[72]{}\mathbf{let}\;{\hat{\sigma}}\mathrel{=}\Varid{subs}\;\Gamma'\;\sigma{}\<[E]%
\\
\>[28]{}\Varid{guard}\mathop{\texttt{\$}}\Varid{x}\equiv {\hat{\sigma}}\;\Varid{x'}\mathrel{\wedge}V\!\;\Varid{x}\not\equiv {\hat{\sigma}}\;\Varid{y}{}\<[E]%
\\
\>[28]{}\Varid{return}\;(\sigma,(\MVUparrow_{\!\textsc{b}\!}\;\Varid{y},\Varid{x}\hspace{.1ex}.\hspace{-.4ex}\backslash\Varid{p'}))\mbox{\onelinecomment  open}{}\<[E]%
\\
\>[B]{}\textit{one}_{\textsc{b}}\;\anonymous \;{}\<[11]%
\>[11]{}\anonymous \mathrel{=}\Varid{empty}{}\<[E]%
\\[\blanklineskip]%
\>[B]{}\textit{one}_{\textsc{b}}'\;\Gamma\;\Varid{p}\mathrel{=}\mathbf{do}\;(\sigma,(\Varid{l},\Varid{b}))\leftarrow \textit{one}_{\textsc{b}}\;\Gamma\;\Varid{p}\hspace{.2ex};\hspace{.2ex}\Varid{r}\leftarrow (\hspace{.1ex}.\hspace{-.4ex}\backslash)^{{\text{-}\hspace{-.2ex}1\!}}\;\Varid{b}\hspace{.2ex};\hspace{.2ex}\Varid{return}\;(\sigma,(\Varid{l},\Varid{r})){}\<[E]%
\ColumnHook
\end{hscode}\resethooks
\vspace*{-3.5ex}
\caption{Symbolic labeled transition semantics.}
\label{fig:OpenLTS}
\end{figure}

\subsection{Labeled Transition Steps over Possible Worlds}
\label{sec:lts:open}
The key idea behind the symbolic transition is that it is not worth considering
every single differences between worlds. For example, consider the process
\ensuremath{\Varid{p}_{1}\mathop{\|}\mathbin{...}\mathop{\|}\Varid{p\char95 n}\mathop{\|}\,(\Varid{y}\mathop{{\leftrightarrow}\hspace{-1.48ex}\raisebox{.1ex}{:}\;\,}\Varid{z})\;\!\tau}
where \ensuremath{\Varid{p\char95 i}\mathrel{=}\Conid{Out}\;(\Conid{V}\;\!\Varid{x\char95 i})\;(\Conid{V}\;\!\Varid{x\char95 i})\;\scalebox{1.1}{\bf\texttt{0}}} for each \ensuremath{\Varid{i}\in [\mskip1.5mu \mathrm{1}\mathinner{\ldotp\ldotp}\Varid{n}\mskip1.5mu]}.
The only difference that matters is whether \ensuremath{\Varid{y}} and \ensuremath{\Varid{z}} are unified in another world
so that it can make a \ensuremath{\ensuremath{\scalebox{1.3}{$\tau$}}}-step, which were not possible in the current world.
Other details such as whether \ensuremath{\Varid{x\char95 i}} and \ensuremath{\Varid{y}}, \ensuremath{\Varid{x\char95 i}} and \ensuremath{\Varid{z}}, or \ensuremath{\Varid{x\char95 j}} and \ensuremath{\Varid{x\char95 k}}
unifies are irrelevant.

A symbolic transition step collects necessary conditions, which are equality
constraints over names in our case, for making further steps in possible worlds
and keeps track of those constraints. Here is a run of a symbolic transition step
for the same example we ran with the fixed world version:\vspace*{-.5ex}
\begin{center}\small
\vspace*{-.5ex}
\begin{hscode}\SaveRestoreHook
\column{B}{@{}>{\hspre}l<{\hspost}@{}}%
\column{E}{@{}>{\hspre}l<{\hspost}@{}}%
\>[B]{}\texttt{*Main> }\;\mathbf{let}\;\Varid{p}\mathrel{=}\Conid{Out}\;\!(V\!\;\!\Varid{x})\;(V\!\;\!\Varid{x})\;\!\scalebox{1.1}{\bf\texttt{0}}\mathop{\|}\Conid{Out}\;\!(V\!\;\!\Varid{y})\;\!(V\!\;\!\Varid{y})\;\!\scalebox{1.1}{\bf\texttt{0}}\mathop{\|}\Conid{In}\;\!(V\!\;\!\Varid{z})\;\!(\Varid{w}\hspace{.1ex}.\hspace{-.4ex}\backslash\scalebox{1.1}{\bf\texttt{0}}){}\<[E]%
\\
\>[B]{}\texttt{*Main> }\Varid{mapM}\hspace{-.2ex}\anonymous \,\;\Varid{pp}\mathbin{\circ}\Varid{runFreshMT}\mathop{\texttt{\$}}\Varid{\Conid{OpenLTS}.one}\;[\mskip1.5mu \forall\hspace{-.75ex}\raisebox{.1ex}{\scalebox{1.1}[.9]{$/$}}\!\;\!\Varid{z},\!\forall\hspace{-.75ex}\raisebox{.1ex}{\scalebox{1.1}[.9]{$/$}}\!\;\!\Varid{y},\!\forall\hspace{-.75ex}\raisebox{.1ex}{\scalebox{1.1}[.9]{$/$}}\!\;\!\Varid{x}\mskip1.5mu]\;\Varid{p}{}\<[E]%
\\
\>[B]{}([\mskip1.5mu \mskip1.5mu],(\MVUparrow\!\;(V\!\;\Varid{x})\;(V\!\;\Varid{x}),(\scalebox{1.1}{\bf\texttt{0}}\mathop{\|}\,(\Conid{Out}\;(V\!\;\Varid{y})\;(V\!\;\Varid{y})\;\scalebox{1.1}{\bf\texttt{0}}))\mathop{\|}\,(\Conid{In}\;(V\!\;\Varid{z})\;(\Varid{w}\hspace{.1ex}.\hspace{-.4ex}\backslash\scalebox{1.1}{\bf\texttt{0}})))){}\<[E]%
\\
\>[B]{}([\mskip1.5mu \mskip1.5mu],(\MVUparrow\!\;(V\!\;\Varid{y})\;(V\!\;\Varid{x}),((\Conid{Out}\;(V\!\;\Varid{x})\;(V\!\;\Varid{y})\;\scalebox{1.1}{\bf\texttt{0}})\mathop{\|}\scalebox{1.1}{\bf\texttt{0}})\mathop{\|}\,(\Conid{In}\;(V\!\;\Varid{z})\;(\Varid{w}\hspace{.1ex}.\hspace{-.4ex}\backslash\scalebox{1.1}{\bf\texttt{0}}))))\;{}\<[E]%
\\
\>[B]{}{\color{ACMDarkBlue}([\mskip1.5mu (\Varid{x},\Varid{z})\mskip1.5mu],(\ensuremath{\scalebox{1.3}{$\tau$}},(\scalebox{1.1}{\bf\texttt{0}}\mathop{\|}\,(\Conid{Out}\;(V\!\;\Varid{y})\;(V\!\;\Varid{y})\;\scalebox{1.1}{\bf\texttt{0}}))\mathop{\|}\scalebox{1.1}{\bf\texttt{0}}))\;}{}\<[E]%
\\
\>[B]{}{\color{ACMDarkBlue}([\mskip1.5mu (\Varid{y},\Varid{z})\mskip1.5mu],(\ensuremath{\scalebox{1.3}{$\tau$}},((\Conid{Out}\;(V\!\;\Varid{x})\;(V\!\;\Varid{x})\;\scalebox{1.1}{\bf\texttt{0}})\mathop{\|}\scalebox{1.1}{\bf\texttt{0}})\mathop{\|}\scalebox{1.1}{\bf\texttt{0}}))\;}{}\<[E]%
\\
\>[B]{}\texttt{*Main> }\Varid{mapM}\hspace{-.2ex}\anonymous \,\;\Varid{pp}\mathbin{\circ}\Varid{runFreshMT}\mathop{\texttt{\$}}\Conid{OpenLTS}.\textit{one}_{\textsc{b}}\;[\mskip1.5mu \forall\hspace{-.75ex}\raisebox{.1ex}{\scalebox{1.1}[.9]{$/$}}\!\;\!\Varid{z},\!\forall\hspace{-.75ex}\raisebox{.1ex}{\scalebox{1.1}[.9]{$/$}}\!\;\!\Varid{y},\!\forall\hspace{-.75ex}\raisebox{.1ex}{\scalebox{1.1}[.9]{$/$}}\!\;\!\Varid{x}\mskip1.5mu]\;\Varid{p}{}\<[E]%
\\
\>[B]{}([\mskip1.5mu \mskip1.5mu],(\hspace{-.15ex}\MVDnarrow^{\hspace{-.18ex}\textsc{b}\!}\;(V\!\;\Varid{z}),\Varid{y}\hspace{.1ex}.\hspace{-.4ex}\backslash(((\Conid{Out}\;(V\!\;\Varid{x})\;(V\!\;\Varid{x})\;\scalebox{1.1}{\bf\texttt{0}})\mathop{\|}\,(\Conid{Out}\;(V\!\;\Varid{y})\;(V\!\;\Varid{y})\;\scalebox{1.1}{\bf\texttt{0}}))\mathop{\|}\scalebox{1.1}{\bf\texttt{0}}))){}\<[E]%
\ColumnHook
\end{hscode}\resethooks
\vspace*{-3.5ex}
\end{center}
Two more interactions steps are possible: one where \ensuremath{\Varid{x}} and \ensuremath{\Varid{z}} are unified
and the other where \ensuremath{\Varid{y}} and \ensuremath{\Varid{z}} are unified.

The return types of \ensuremath{\Varid{one}} and \ensuremath{\textit{one}_{\textsc{b}}} in Figure~\ref{fig:OpenLTS} reflect such
characteristics of symbolic transition. For instance, \ensuremath{\Varid{one}} returns
the equality constraint (\ensuremath{\Conid{EqC}}) along with the transition label (\ensuremath{\Conid{Act}}) and
the process (\ensuremath{\Conid{Pr}}). Another difference from the fixed world version is that
there is an additional context (\ensuremath{\Conid{Ctx}}) argument. The definitions of \ensuremath{\Conid{EqC}}
and \ensuremath{\Conid{Ctx}} are provided in Figure~\ref{fig:figureOpenLTS} along with related
helper functions. As a naming convention, we use \ensuremath{\sigma} for equality constraints
and \ensuremath{\Gamma} for contexts. We follow through the definitions
in Figure~\ref{fig:figureOpenLTS} explaining how they are used
in the implementation symbolic transition steps in Figure~\ref{fig:OpenLTS}
while pointing out the differences from the fixed world version
in Figure~\ref{fig:IdSubLTS} laid out side-by-side.

%
%
\makeatletter
\@ifundefined{lhs2tex.lhs2tex.sty.read}%
  {\@namedef{lhs2tex.lhs2tex.sty.read}{}%
   \newcommand\SkipToFmtEnd{}%
   \newcommand\EndFmtInput{}%
   \long\def\SkipToFmtEnd#1\EndFmtInput{}%
  }\SkipToFmtEnd

\newcommand\ReadOnlyOnce[1]{\@ifundefined{#1}{\@namedef{#1}{}}\SkipToFmtEnd}
\usepackage{amstext}
\usepackage{amssymb}
\usepackage{stmaryrd}
\DeclareFontFamily{OT1}{cmtex}{}
\DeclareFontShape{OT1}{cmtex}{m}{n}
  {<5><6><7><8>cmtex8
   <9>cmtex9
   <10><10.95><12><14.4><17.28><20.74><24.88>cmtex10}{}
\DeclareFontShape{OT1}{cmtex}{m}{it}
  {<-> ssub * cmtt/m/it}{}
\newcommand{\texfamily}{\fontfamily{cmtex}\selectfont}
\DeclareFontShape{OT1}{cmtt}{bx}{n}
  {<5><6><7><8>cmtt8
   <9>cmbtt9
   <10><10.95><12><14.4><17.28><20.74><24.88>cmbtt10}{}
\DeclareFontShape{OT1}{cmtex}{bx}{n}
  {<-> ssub * cmtt/bx/n}{}
\newcommand{\tex}[1]{\text{\texfamily#1}}	

\newcommand{\Sp}{\hskip.33334em\relax}

\newcommand{\Conid}[1]{\mathit{#1}}
\newcommand{\Varid}[1]{\mathit{#1}}
\newcommand{\anonymous}{\kern0.06em \vbox{\hrule\@width.5em}}
\newcommand{\plus}{\mathbin{+\!\!\!+}}
\newcommand{\bind}{\mathbin{>\!\!\!>\mkern-6.7mu=}}
\newcommand{\rbind}{\mathbin{=\mkern-6.7mu<\!\!\!<}}
\newcommand{\sequ}{\mathbin{>\!\!\!>}}
\renewcommand{\leq}{\leqslant}
\renewcommand{\geq}{\geqslant}
\usepackage{polytable}

\@ifundefined{mathindent}%
  {\newdimen\mathindent\mathindent\leftmargini}%
  {}%

\def\resethooks{%
  \global\let\SaveRestoreHook\empty
  \global\let\ColumnHook\empty}
\newcommand*{\savecolumns}[1][default]%
  {\g@addto@macro\SaveRestoreHook{\savecolumns[#1]}}
\newcommand*{\restorecolumns}[1][default]%
  {\g@addto@macro\SaveRestoreHook{\restorecolumns[#1]}}
\newcommand*{\aligncolumn}[2]%
  {\g@addto@macro\ColumnHook{\column{#1}{#2}}}

\resethooks

\newcommand{\onelinecommentchars}{\quad-{}- }
\newcommand{\commentbeginchars}{\enskip\{-}
\newcommand{\commentendchars}{-\}\enskip}

\newcommand{\visiblecomments}{%
  \let\onelinecomment=\onelinecommentchars
  \let\commentbegin=\commentbeginchars
  \let\commentend=\commentendchars}

\newcommand{\invisiblecomments}{%
  \let\onelinecomment=\empty
  \let\commentbegin=\empty
  \let\commentend=\empty}

\visiblecomments

\newlength{\blanklineskip}
\setlength{\blanklineskip}{0.66084ex}

\newcommand{\hsindent}[1]{\quad}
\let\hspre\empty
\let\hspost\empty
\newcommand{\NB}{\textbf{NB}}
\newcommand{\Todo}[1]{$\langle$\textbf{To do:}~#1$\rangle$}

\EndFmtInput
\makeatother
%
%
%
%
%
%
%
%
%
\ReadOnlyOnce{polycode.fmt}%
\makeatletter

\newcommand{\hsnewpar}[1]%
  {{\parskip=0pt\parindent=0pt\par\vskip #1\noindent}}

\newcommand{\hscodestyle}{}


\newcommand{\sethscode}[1]%
  {\expandafter\let\expandafter\hscode\csname #1\endcsname
   \expandafter\let\expandafter\endhscode\csname end#1\endcsname}


\newenvironment{compathscode}%
  {\par\noindent
   \advance\leftskip\mathindent
   \hscodestyle
   \let\\=\@normalcr
   \let\hspre\(\let\hspost\)%
   \pboxed}%
  {\endpboxed\)%
   \par\noindent
   \ignorespacesafterend}

\newcommand{\compaths}{\sethscode{compathscode}}


\newenvironment{plainhscode}%
  {\hsnewpar\abovedisplayskip
   \advance\leftskip\mathindent
   \hscodestyle
   \let\hspre\(\let\hspost\)%
   \pboxed}%
  {\endpboxed%
   \hsnewpar\belowdisplayskip
   \ignorespacesafterend}

\newenvironment{oldplainhscode}%
  {\hsnewpar\abovedisplayskip
   \advance\leftskip\mathindent
   \hscodestyle
   \let\\=\@normalcr
   \(\pboxed}%
  {\endpboxed\)%
   \hsnewpar\belowdisplayskip
   \ignorespacesafterend}


\newcommand{\plainhs}{\sethscode{plainhscode}}
\newcommand{\oldplainhs}{\sethscode{oldplainhscode}}
\plainhs


\newenvironment{arrayhscode}%
  {\hsnewpar\abovedisplayskip
   \advance\leftskip\mathindent
   \hscodestyle
   \let\\=\@normalcr
   \(\parray}%
  {\endparray\)%
   \hsnewpar\belowdisplayskip
   \ignorespacesafterend}

\newcommand{\arrayhs}{\sethscode{arrayhscode}}


\newenvironment{mathhscode}%
  {\parray}{\endparray}

\newcommand{\mathhs}{\sethscode{mathhscode}}


\newenvironment{texthscode}%
  {\(\parray}{\endparray\)}

\newcommand{\texths}{\sethscode{texthscode}}


\def\codeframewidth{\arrayrulewidth}
\RequirePackage{calc}

\newenvironment{framedhscode}%
  {\parskip=\abovedisplayskip\par\noindent
   \hscodestyle
   \arrayrulewidth=\codeframewidth
   \tabular{@{}|p{\linewidth-2\arraycolsep-2\arrayrulewidth-2pt}|@{}}%
   \hline\framedhslinecorrect\\{-1.5ex}%
   \let\endoflinesave=\\
   \let\\=\@normalcr
   \(\pboxed}%
  {\endpboxed\)%
   \framedhslinecorrect\endoflinesave{.5ex}\hline
   \endtabular
   \parskip=\belowdisplayskip\par\noindent
   \ignorespacesafterend}

\newcommand{\framedhslinecorrect}[2]%
  {#1[#2]}

\newcommand{\framedhs}{\sethscode{framedhscode}}


\newenvironment{inlinehscode}%
  {\(\def\column##1##2{}%
   \let\>\undefined\let\<\undefined\let\\\undefined
   \newcommand\>[1][]{}\newcommand\<[1][]{}\newcommand\\[1][]{}%
   \def\fromto##1##2##3{##3}%
   \def\nextline{}}{\) }%

\newcommand{\inlinehs}{\sethscode{inlinehscode}}


\newenvironment{joincode}%
  {\let\orighscode=\hscode
   \let\origendhscode=\endhscode
   \def\endhscode{\def\hscode{\endgroup\def\@currenvir{hscode}\\}\begingroup}
   \orighscode\def\hscode{\endgroup\def\@currenvir{hscode}}}%
  {\origendhscode
   \global\let\hscode=\orighscode
   \global\let\endhscode=\origendhscode}%

\makeatother
\EndFmtInput
\ReadOnlyOnce{colorcode.fmt}%

\RequirePackage{colortbl}
\RequirePackage{calc}

\makeatletter
\newenvironment{colorhscode}%
  {\hsnewpar\abovedisplayskip
   \hscodestyle
   \tabular{@{}>{\columncolor{codecolor}}p{\linewidth}@{}}%
   \let\\=\@normalcr
   \(\pboxed}%
  {\endpboxed\)%
   \endtabular
   \hsnewpar\belowdisplayskip
   \ignorespacesafterend}

\newenvironment{tightcolorhscode}%
  {\hsnewpar\abovedisplayskip
   \hscodestyle
   \tabular{@{}>{\columncolor{codecolor}\(}l<{\)}@{}}%
   \pmboxed}%
  {\endpmboxed%
   \endtabular
   \hsnewpar\belowdisplayskip
   \ignorespacesafterend}

\newenvironment{barhscode}%
  {\hsnewpar\abovedisplayskip
   \hscodestyle
   \arrayrulecolor{codecolor}%
   \arrayrulewidth=\coderulewidth
   \tabular{|p{\linewidth-\arrayrulewidth-\tabcolsep}@{}}%
   \let\\=\@normalcr
   \(\pboxed}%
  {\endpboxed\)%
   \endtabular
   \hsnewpar\belowdisplayskip
   \ignorespacesafterend}
\makeatother

\def\colorcode{\columncolor{codecolor}}
\definecolor{codecolor}{rgb}{1,1,.667}
\newlength{\coderulewidth}
\setlength{\coderulewidth}{3pt}

\newcommand{\colorhs}{\sethscode{colorhscode}}
\newcommand{\tightcolorhs}{\sethscode{tightcolorhscode}}
\newcommand{\barhs}{\sethscode{barhscode}}

\EndFmtInput

\renewcommand{\onelinecommentchars}{\color{gray}\quad-{}- }
\renewcommand{\commentbeginchars}{\color{gray}\enskip\{- }
\renewcommand{\commentendchars}{-\}\enskip}

\renewcommand{\visiblecomments}{%
  \let\onelinecomment=\onelinecommentchars
  \let\commentbegin=\commentbeginchars
  \let\commentend=\commentendchars}

\renewcommand{\invisiblecomments}{%
  \let\onelinecomment=\empty
  \let\commentbegin=\empty
  \let\commentend=\empty}

\visiblecomments

\begin{figure}\small
\begin{hscode}\SaveRestoreHook
\column{B}{@{}>{\hspre}l<{\hspost}@{}}%
\column{3}{@{}>{\hspre}l<{\hspost}@{}}%
\column{5}{@{}>{\hspre}l<{\hspost}@{}}%
\column{10}{@{}>{\hspre}l<{\hspost}@{}}%
\column{13}{@{}>{\hspre}l<{\hspost}@{}}%
\column{16}{@{}>{\hspre}l<{\hspost}@{}}%
\column{24}{@{}>{\hspre}l<{\hspost}@{}}%
\column{26}{@{}>{\hspre}l<{\hspost}@{}}%
\column{39}{@{}>{\hspre}l<{\hspost}@{}}%
\column{50}{@{}>{\hspre}l<{\hspost}@{}}%
\column{81}{@{}>{\hspre}l<{\hspost}@{}}%
\column{E}{@{}>{\hspre}l<{\hspost}@{}}%
\>[B]{}\mathbf{module}\;\Conid{OpenLTS}\;\mathbf{where}{}\<[E]%
\\
\>[B]{}\mathbf{import}\;\Conid{PiCalc}\hspace{.2ex};\hspace{.2ex}\mathbf{import}\;\Conid{\Conid{Control}.Applicative}\hspace{.2ex};\hspace{.2ex}\mathbf{import}\;\Conid{\Conid{Control}.Monad}{}\<[E]%
\\
\>[B]{}\mathbf{import}\;\Conid{\Conid{Data}.Partition}\;\mathbf{hiding}\;(\Varid{empty}){}\<[E]%
\\
\>[B]{}\mathbf{import}\;\Conid{\Conid{Unbound}.LocallyNameless}\;\mathbf{hiding}\;(\Varid{empty},\Varid{rep},\Conid{GT}){}\<[E]%
\\
\>[B]{}\mathbf{import}\;\Conid{\Conid{Data}.\Conid{Map}.Strict}\;(\Varid{fromList},(\mathbin{!})){}\<[E]%
\\[\blanklineskip]%
\>[B]{}\mathbf{type}\;\Conid{EqC}\mathrel{=}[\mskip1.5mu (\Conid{Nm},\Conid{Nm})\mskip1.5mu]{}\<[E]%
\\
\>[B]{}\mathbf{infixr}\;\mathrm{5}\mathop{\raisebox{.5ex}{$\curvearrowright$}\hspace{-1.9ex}\scalebox{.8}{$+$}\;}\;\hspace{.2ex};\hspace{.2ex}\;(\mathop{\raisebox{.5ex}{$\curvearrowright$}\hspace{-1.9ex}\scalebox{.8}{$+$}\;})\mathbin{::}(\Conid{Nm},\Conid{Nm})\to \Conid{EqC}\to \Conid{EqC}{}\<[E]%
\\
\>[B]{}(\Varid{x},\Varid{y})\mathop{\raisebox{.5ex}{$\curvearrowright$}\hspace{-1.9ex}\scalebox{.8}{$+$}\;}\sigma\mathrel{=}\mathbf{case}\;\Varid{compare}\;\Varid{x}\;\Varid{y}\;\mathbf{of}\;{}\<[39]%
\>[39]{}\Conid{LT}\to [\mskip1.5mu (\Varid{x},\Varid{y})\mskip1.5mu]\cup\sigma{}\<[E]%
\\
\>[39]{}\Conid{EQ}\to \sigma{}\<[E]%
\\
\>[39]{}\Conid{GT}\to [\mskip1.5mu (\Varid{y},\Varid{x})\mskip1.5mu]\cup\sigma{}\<[E]%
\\
\>[B]{}\mathbf{infixr}\;\mathrm{5}\cup\;\hspace{.2ex};\hspace{.2ex}\;(\cup)\mathbin{::}\Conid{EqC}\to \Conid{EqC}\to \Conid{EqC}\hspace{.2ex};\hspace{.2ex}\;(\cup)\mathrel{=}\Varid{union}{}\<[E]%
\\[\blanklineskip]%
\>[B]{}\mathbf{type}\;\Conid{Ctx}\mathrel{=}[\mskip1.5mu \Conid{Quan}\mskip1.5mu]{}\<[E]%
\\
\>[B]{}\mathbf{data}\;\Conid{Quan}\mathrel{=}\forall\hspace{-.75ex}\raisebox{.1ex}{\scalebox{1.1}[.9]{$/$}}\!\;\Conid{Nm}\mid \nabla\hspace{-.83ex}\raisebox{.1ex}{\scalebox{1.1}[.9]{$/$}}\!\;\Conid{Nm}\;\mathbf{deriving}\;(\Conid{Eq},\Conid{Ord},\Conid{Show}){}\<[E]%
\\
\>[B]{}\Varid{quan2nm}\mathbin{::}\Conid{Quan}\to \Conid{Nm}\hspace{.2ex};\hspace{.2ex}\,\Varid{quan2nm}\;(\forall\hspace{-.75ex}\raisebox{.1ex}{\scalebox{1.1}[.9]{$/$}}\!\;\Varid{x}){}\<[50]%
\>[50]{}\mathrel{=}\Varid{x}\hspace{.2ex};\hspace{.2ex}\,\Varid{quan2nm}\;(\nabla\hspace{-.83ex}\raisebox{.1ex}{\scalebox{1.1}[.9]{$/$}}\!\;\Varid{x}){}\<[81]%
\>[81]{}\mathrel{=}\Varid{x}{}\<[E]%
\\[\blanklineskip]%
\>[B]{}\Varid{respects}\mathbin{::}\Conid{EqC}\to \Conid{Ctx}\to \Conid{Bool}{}\<[E]%
\\
\>[B]{}\Varid{respects}\;\sigma\;\Gamma\mathrel{=}\Varid{all}\;(\lambda \Varid{n}\to \Varid{rep}\;\Varid{part}\;\Varid{n}\equiv \Varid{n})\;[\mskip1.5mu \Varid{n2i}\;\Varid{x}\mid \nabla\hspace{-.83ex}\raisebox{.1ex}{\scalebox{1.1}[.9]{$/$}}\!\;\Varid{x}\leftarrow \Gamma\mskip1.5mu]{}\<[E]%
\\
\>[B]{}\hsindent{3}{}\<[3]%
\>[3]{}\mathbf{where}\;(\Varid{part},(\Varid{n2i},\anonymous ))\mathrel{=}\Varid{mkPartitionFromEqC}\;\Gamma\;\sigma{}\<[E]%
\\[\blanklineskip]%
\>[B]{}\Varid{subs}\mathbin{::}\Conid{Subst}\;\Conid{Tm}\;\Varid{b}\Rightarrow \Conid{Ctx}\to \Conid{EqC}\to \Varid{b}\to \Varid{b}{}\<[E]%
\\
\>[B]{}\Varid{subs}\;\Gamma\;\sigma\mathrel{=}\Varid{foldr}\;(\mathbin{\circ})\;\Varid{id}\;[\mskip1.5mu \mathop{\sub{\Varid{x}}{(V\!\;\!\Varid{y})}\!}\!\mid (\Varid{x},\Varid{y})\leftarrow \sigma'\mskip1.5mu]{}\<[E]%
\\
\>[B]{}\hsindent{3}{}\<[3]%
\>[3]{}\mathbf{where}\;{}\<[10]%
\>[10]{}\sigma'\mathrel{=}[\mskip1.5mu (\Varid{i2n}\;\Varid{i},\Varid{i2n}\mathop{\texttt{\$}}\Varid{rep}\;\Varid{part}\;\Varid{i})\mid \Varid{i}\leftarrow [\mskip1.5mu \mathrm{0}\mathinner{\ldotp\ldotp}\Varid{maxVal}\mskip1.5mu]\mskip1.5mu]{}\<[E]%
\\
\>[10]{}(\Varid{part},(\Varid{n2i},\Varid{i2n}))\mathrel{=}\Varid{mkPartitionFromEqC}\;\Gamma\;\sigma{}\<[E]%
\\
\>[10]{}\Varid{maxVal}\mathrel{=}\Varid{length}\;\Gamma\mathbin{-}\mathrm{1}{}\<[E]%
\\[\blanklineskip]%
\>[B]{}\Varid{mkPartitionFromEqC}\mathbin{::}{}\<[24]%
\>[24]{}\Conid{Ctx}\to \Conid{EqC}\to {}\<[E]%
\\
\>[24]{}\hsindent{2}{}\<[26]%
\>[26]{}(\Conid{Partition}\;\Conid{Int},(\Conid{Nm}\to \Conid{Int},\Conid{Int}\to \Conid{Nm})){}\<[E]%
\\
\>[B]{}\Varid{mkPartitionFromEqC}\;\Gamma\;\sigma\mathrel{=}(\Varid{part},(\Varid{n2i},\Varid{i2n})){}\<[E]%
\\
\>[B]{}\hsindent{3}{}\<[3]%
\>[3]{}\mathbf{where}{}\<[E]%
\\
\>[3]{}\hsindent{2}{}\<[5]%
\>[5]{}\Varid{part}\mathrel{=}{}\<[13]%
\>[13]{}\Varid{foldr}\;(\mathbin{\circ})\;\Varid{id}\;[\mskip1.5mu \Varid{joinElems}\;(\Varid{n2i}\;\Varid{x})\;(\Varid{n2i}\;\Varid{y})\mid (\Varid{x},\Varid{y})\leftarrow \sigma\mskip1.5mu]\;{}\<[E]%
\\
\>[13]{}\hsindent{3}{}\<[16]%
\>[16]{}\Varid{discrete}{}\<[E]%
\\
\>[3]{}\hsindent{2}{}\<[5]%
\>[5]{}\Varid{i2n}\;\Varid{i}\mathrel{=}\Varid{revns}\mathbin{!!}\Varid{i}{}\<[E]%
\\
\>[3]{}\hsindent{2}{}\<[5]%
\>[5]{}\Varid{n2i}\;\Varid{x}\mathrel{=}\Varid{n2iMap}\mathbin{!}\Varid{x}{}\<[E]%
\\
\>[3]{}\hsindent{2}{}\<[5]%
\>[5]{}\Varid{revns}\mathrel{=}\Varid{reverse}\;(\Varid{quan2nm}\mathop{\scalebox{1}[.7]{\raisebox{.3ex}{$\langle$}}\hspace{-.7ex}\mathop{\texttt{\$}}\hspace{-.7ex}\scalebox{1}[.7]{\raisebox{.3ex}{$\rangle$}}}\Gamma){}\<[E]%
\\
\>[3]{}\hsindent{2}{}\<[5]%
\>[5]{}\Varid{n2iMap}\mathrel{=}\Varid{fromList}\mathop{\texttt{\$}}\Varid{zip}\;\Varid{revns}\;[\mskip1.5mu \mathrm{0}\mathinner{\ldotp\ldotp}\Varid{maxVal}\mskip1.5mu]{}\<[E]%
\\
\>[3]{}\hsindent{2}{}\<[5]%
\>[5]{}\Varid{maxVal}\mathrel{=}\Varid{length}\;\Gamma\mathbin{-}\mathrm{1}{}\<[E]%
\ColumnHook
\end{hscode}\resethooks
\vspace*{-3.5ex}
\caption{Preamble of the \ensuremath{\Conid{OpenLTS}} module including type definitions and
helper functions for defining symbolic transition steps in Figure~\ref{fig:OpenLTS}.}
\label{fig:figureOpenLTS}
\vspace*{-1ex}
\end{figure}

An equality constraint (\ensuremath{\Conid{EqC}}) is conceptually a set of name pairs represented
as a list. Basic operations over \ensuremath{\Conid{EqC}} are single element insertion (\ensuremath{\mathop{\raisebox{.5ex}{$\curvearrowright$}\hspace{-1.9ex}\scalebox{.8}{$+$}\;}}) and
union (\ensuremath{\cup}). These operations are used on the necessary constraints 
for the additional steps, which were not possible in the current world.
Such additional steps may occur in match-prefixes, closing of scope extrusions,
and interaction steps.

A context (\ensuremath{\Conid{Ctx}}) is a list of either universally (\ensuremath{\forall\hspace{-.75ex}\raisebox{.1ex}{\scalebox{1.1}[.9]{$/$}}\!}\,) or nabla (\ensuremath{\nabla\hspace{-.83ex}\raisebox{.1ex}{\scalebox{1.1}[.9]{$/$}}\!}\,) 
quantified names (\ensuremath{\Conid{Quan}}). We assume that names in a context must be distinct
(i.e., no  duplicates). When using the symbolic transition step (\ensuremath{\Varid{one}\;\Gamma\;\Varid{p}}),
we assume that \ensuremath{\Varid{p}} is closed by \ensuremath{\Gamma}, that is, $\ensuremath{(\Varid{fv}\;\Varid{p})}\subset\ensuremath{(\Varid{quan2nm}\mathop{\scalebox{1}[.7]{\raisebox{.3ex}{$\langle$}}\hspace{-.7ex}\mathop{\texttt{\$}}\hspace{-.7ex}\scalebox{1}[.7]{\raisebox{.3ex}{$\rangle$}}}\Gamma)}$.
Similarly, for (\ensuremath{\textit{one}_{\textsc{b}}\;\Gamma\;\Varid{b}}), we assume that \ensuremath{\Varid{b}} is closed by \ensuremath{\Gamma}.

Quantified names in a context appear in reversed order from how we usually
write on paper as a mathematical notation. That is, $\forall x,\!\nabla y,\!\forall z,...$
would correspond to \ensuremath{[\mskip1.5mu \forall\hspace{-.75ex}\raisebox{.1ex}{\scalebox{1.1}[.9]{$/$}}\!\;\!\Varid{z},\nabla\hspace{-.83ex}\raisebox{.1ex}{\scalebox{1.1}[.9]{$/$}}\!\;\!\Varid{y},\forall\hspace{-.75ex}\raisebox{.1ex}{\scalebox{1.1}[.9]{$/$}}\!\;\!\Varid{x}\mskip1.5mu]}.
This reversal of layout is typical for list representation of contexts
where the most recently introduced name is added to the head of the list.
Nabla quantified names must be fresh from all previously known names.
Hence, \ensuremath{\Varid{y}} may be unified with \ensuremath{\Varid{z}} but never with \ensuremath{\Varid{x}}. A substitution
\ensuremath{\sigma\mathop{^{\,_{\backprime}\!}\Varid{respects}^{_\backprime}}\Gamma} when it obeys such nabla restrictions imposed by \ensuremath{\Gamma}.
Otherwise, i.e., \ensuremath{\neg \;(\sigma\mathop{^{\,_{\backprime}\!}\Varid{respects}^{_\backprime}}\Gamma)}, it is an impossible world,
therefore, discarded by the \ensuremath{\Varid{guard}}s involving the \ensuremath{\Varid{respect}} predicate
in Figure~\ref{fig:OpenLTS}. These are additional \ensuremath{\Varid{guard}}s that were not
present in the fixed world setting.

We use the helper function \ensuremath{\Varid{subs}} to build a substitution function
(\ensuremath{{\hat{\sigma}}\mathbin{::}\Conid{Subst}\;\Conid{Tm}\;\Varid{a}\Rightarrow \Varid{a}\to \Varid{a}}) from the context (\ensuremath{\Gamma}) and
equality constraints (\ensuremath{\sigma}). The substitution function (\ensuremath{{\hat{\sigma}}})
is used for testing name equivalence under the possible world given by \ensuremath{\sigma}
in the transition steps for the restricted process (\ensuremath{\scalebox{1.25}{$\nu$}\!\;(\Varid{x}\hspace{.1ex}.\hspace{-.4ex}\backslash\Varid{p})}).
The name (in)equality test for the restricted process in Figure~\ref{fig:IdSubLTS}
are now tested as (in)equality modulo substitution in Figure~\ref{fig:OpenLTS}.
For instance, the equality tests against the restricted name (\ensuremath{\Varid{x}}) such as
\ensuremath{\Varid{x}\equiv \Varid{x'}} and \ensuremath{\Conid{V}\;\!\Varid{x}\not\equiv \Varid{y}} for the restricted process in Figure~\ref{fig:IdSubLTS}
are replaced by \ensuremath{\Varid{x}\equiv {\hat{\sigma}}\;\!\Varid{x'}} and \ensuremath{\Conid{V}\;\!\Varid{x}\not\equiv {\hat{\sigma}}\;\!\Varid{y}}
in Figure~\ref{fig:OpenLTS}.
We need not apply \ensuremath{{\hat{\sigma}}} to the restricted name \ensuremath{\Varid{x}}, although
it would be harmless, because of our particular scheme for computing substitutions
using the helper function \ensuremath{\Varid{mkPartitionFromEqC}}, which is also used in
the definition of the \ensuremath{\Varid{respects}} predicate discussed earlier.

\subsection{Substitution modeled as Set Partitions}
\label{sec:lts:partition}
In \ensuremath{\Varid{mkPartitionFromEqC}}, we map names in \ensuremath{\Gamma} to inegers in decreasing order
so that more recently introduced names maps to larger values. For example,
consider \ensuremath{\Gamma\mathrel{=}[\mskip1.5mu \forall\hspace{-.75ex}\raisebox{.1ex}{\scalebox{1.1}[.9]{$/$}}\!\;\!\Varid{z},\nabla\hspace{-.83ex}\raisebox{.1ex}{\scalebox{1.1}[.9]{$/$}}\!\;\!\Varid{y},\forall\hspace{-.75ex}\raisebox{.1ex}{\scalebox{1.1}[.9]{$/$}}\!\;\!\Varid{x}\mskip1.5mu]},
which represents the context $\forall x,\!\nabla y,\!\forall z,...$
where \ensuremath{\Varid{x}} is mapped to 0, \ensuremath{\Varid{y}} to 1, and \ensuremath{\Varid{z}} to 2.
We model substitutions as integer set partitions using the \textsf{data-partition} library
and unification by its join operation (\ensuremath{\Varid{joinElems}}), which merges equivalence classes of
the joining elements (a.k.a., union-find algorithm). 
Consider the substitution described by \ensuremath{[\mskip1.5mu (\Varid{y},\Varid{z})\mskip1.5mu]}, which respects \ensuremath{\Gamma},
modeled by the partition $\ensuremath{\Varid{part}_1} = [[0],[1,2]]$. Also, consider \ensuremath{[\mskip1.5mu (\Varid{x},\Varid{y})\mskip1.5mu]},
which does not respect \ensuremath{\Gamma}, modeled by $\ensuremath{\Varid{part}_2} = [[0,1],[2]]$. 
The representative of an equivalence class defined to be the minimal value.
Then, we can decide whether a partition models a respectful substitution by
examining \ensuremath{(\Varid{rep}\;\Varid{part}\;\Varid{n})\mathbin{::}\Conid{Int}} for every \ensuremath{\Varid{n}} that is mapped from a nabla name.
For instance, 1 from \ensuremath{\Varid{y}} in our example. In the first partition,
\ensuremath{\Varid{rep}\;\Varid{part}_1\;\mathrm{1}\equiv \mathrm{1}} is the same as the nabla mapped value.
In the second partition, on the other hand, \ensuremath{\Varid{rep}\;\Varid{part}_2\;\mathrm{1}\equiv \mathrm{0}}
is different from the nabla mapped value.
This exactly captures the idea that a nabla quantified name only unifies with
the names introduced later (larger values)
but not with names introduced earlier (smaller values).


%
%
\makeatletter
\@ifundefined{lhs2tex.lhs2tex.sty.read}%
  {\@namedef{lhs2tex.lhs2tex.sty.read}{}%
   \newcommand\SkipToFmtEnd{}%
   \newcommand\EndFmtInput{}%
   \long\def\SkipToFmtEnd#1\EndFmtInput{}%
  }\SkipToFmtEnd

\newcommand\ReadOnlyOnce[1]{\@ifundefined{#1}{\@namedef{#1}{}}\SkipToFmtEnd}
\usepackage{amstext}
\usepackage{amssymb}
\usepackage{stmaryrd}
\DeclareFontFamily{OT1}{cmtex}{}
\DeclareFontShape{OT1}{cmtex}{m}{n}
  {<5><6><7><8>cmtex8
   <9>cmtex9
   <10><10.95><12><14.4><17.28><20.74><24.88>cmtex10}{}
\DeclareFontShape{OT1}{cmtex}{m}{it}
  {<-> ssub * cmtt/m/it}{}
\newcommand{\texfamily}{\fontfamily{cmtex}\selectfont}
\DeclareFontShape{OT1}{cmtt}{bx}{n}
  {<5><6><7><8>cmtt8
   <9>cmbtt9
   <10><10.95><12><14.4><17.28><20.74><24.88>cmbtt10}{}
\DeclareFontShape{OT1}{cmtex}{bx}{n}
  {<-> ssub * cmtt/bx/n}{}
\newcommand{\tex}[1]{\text{\texfamily#1}}	

\newcommand{\Sp}{\hskip.33334em\relax}

\newcommand{\Conid}[1]{\mathit{#1}}
\newcommand{\Varid}[1]{\mathit{#1}}
\newcommand{\anonymous}{\kern0.06em \vbox{\hrule\@width.5em}}
\newcommand{\plus}{\mathbin{+\!\!\!+}}
\newcommand{\bind}{\mathbin{>\!\!\!>\mkern-6.7mu=}}
\newcommand{\rbind}{\mathbin{=\mkern-6.7mu<\!\!\!<}}
\newcommand{\sequ}{\mathbin{>\!\!\!>}}
\renewcommand{\leq}{\leqslant}
\renewcommand{\geq}{\geqslant}
\usepackage{polytable}

\@ifundefined{mathindent}%
  {\newdimen\mathindent\mathindent\leftmargini}%
  {}%

\def\resethooks{%
  \global\let\SaveRestoreHook\empty
  \global\let\ColumnHook\empty}
\newcommand*{\savecolumns}[1][default]%
  {\g@addto@macro\SaveRestoreHook{\savecolumns[#1]}}
\newcommand*{\restorecolumns}[1][default]%
  {\g@addto@macro\SaveRestoreHook{\restorecolumns[#1]}}
\newcommand*{\aligncolumn}[2]%
  {\g@addto@macro\ColumnHook{\column{#1}{#2}}}

\resethooks

\newcommand{\onelinecommentchars}{\quad-{}- }
\newcommand{\commentbeginchars}{\enskip\{-}
\newcommand{\commentendchars}{-\}\enskip}

\newcommand{\visiblecomments}{%
  \let\onelinecomment=\onelinecommentchars
  \let\commentbegin=\commentbeginchars
  \let\commentend=\commentendchars}

\newcommand{\invisiblecomments}{%
  \let\onelinecomment=\empty
  \let\commentbegin=\empty
  \let\commentend=\empty}

\visiblecomments

\newlength{\blanklineskip}
\setlength{\blanklineskip}{0.66084ex}

\newcommand{\hsindent}[1]{\quad}
\let\hspre\empty
\let\hspost\empty
\newcommand{\NB}{\textbf{NB}}
\newcommand{\Todo}[1]{$\langle$\textbf{To do:}~#1$\rangle$}

\EndFmtInput
\makeatother
%
%
%
%
%
%
%
%
%
\ReadOnlyOnce{polycode.fmt}%
\makeatletter

\newcommand{\hsnewpar}[1]%
  {{\parskip=0pt\parindent=0pt\par\vskip #1\noindent}}

\newcommand{\hscodestyle}{}


\newcommand{\sethscode}[1]%
  {\expandafter\let\expandafter\hscode\csname #1\endcsname
   \expandafter\let\expandafter\endhscode\csname end#1\endcsname}


\newenvironment{compathscode}%
  {\par\noindent
   \advance\leftskip\mathindent
   \hscodestyle
   \let\\=\@normalcr
   \let\hspre\(\let\hspost\)%
   \pboxed}%
  {\endpboxed\)%
   \par\noindent
   \ignorespacesafterend}

\newcommand{\compaths}{\sethscode{compathscode}}


\newenvironment{plainhscode}%
  {\hsnewpar\abovedisplayskip
   \advance\leftskip\mathindent
   \hscodestyle
   \let\hspre\(\let\hspost\)%
   \pboxed}%
  {\endpboxed%
   \hsnewpar\belowdisplayskip
   \ignorespacesafterend}

\newenvironment{oldplainhscode}%
  {\hsnewpar\abovedisplayskip
   \advance\leftskip\mathindent
   \hscodestyle
   \let\\=\@normalcr
   \(\pboxed}%
  {\endpboxed\)%
   \hsnewpar\belowdisplayskip
   \ignorespacesafterend}


\newcommand{\plainhs}{\sethscode{plainhscode}}
\newcommand{\oldplainhs}{\sethscode{oldplainhscode}}
\plainhs


\newenvironment{arrayhscode}%
  {\hsnewpar\abovedisplayskip
   \advance\leftskip\mathindent
   \hscodestyle
   \let\\=\@normalcr
   \(\parray}%
  {\endparray\)%
   \hsnewpar\belowdisplayskip
   \ignorespacesafterend}

\newcommand{\arrayhs}{\sethscode{arrayhscode}}


\newenvironment{mathhscode}%
  {\parray}{\endparray}

\newcommand{\mathhs}{\sethscode{mathhscode}}


\newenvironment{texthscode}%
  {\(\parray}{\endparray\)}

\newcommand{\texths}{\sethscode{texthscode}}


\def\codeframewidth{\arrayrulewidth}
\RequirePackage{calc}

\newenvironment{framedhscode}%
  {\parskip=\abovedisplayskip\par\noindent
   \hscodestyle
   \arrayrulewidth=\codeframewidth
   \tabular{@{}|p{\linewidth-2\arraycolsep-2\arrayrulewidth-2pt}|@{}}%
   \hline\framedhslinecorrect\\{-1.5ex}%
   \let\endoflinesave=\\
   \let\\=\@normalcr
   \(\pboxed}%
  {\endpboxed\)%
   \framedhslinecorrect\endoflinesave{.5ex}\hline
   \endtabular
   \parskip=\belowdisplayskip\par\noindent
   \ignorespacesafterend}

\newcommand{\framedhslinecorrect}[2]%
  {#1[#2]}

\newcommand{\framedhs}{\sethscode{framedhscode}}


\newenvironment{inlinehscode}%
  {\(\def\column##1##2{}%
   \let\>\undefined\let\<\undefined\let\\\undefined
   \newcommand\>[1][]{}\newcommand\<[1][]{}\newcommand\\[1][]{}%
   \def\fromto##1##2##3{##3}%
   \def\nextline{}}{\) }%

\newcommand{\inlinehs}{\sethscode{inlinehscode}}


\newenvironment{joincode}%
  {\let\orighscode=\hscode
   \let\origendhscode=\endhscode
   \def\endhscode{\def\hscode{\endgroup\def\@currenvir{hscode}\\}\begingroup}
   \orighscode\def\hscode{\endgroup\def\@currenvir{hscode}}}%
  {\origendhscode
   \global\let\hscode=\orighscode
   \global\let\endhscode=\origendhscode}%

\makeatother
\EndFmtInput
\ReadOnlyOnce{colorcode.fmt}%

\RequirePackage{colortbl}
\RequirePackage{calc}

\makeatletter
\newenvironment{colorhscode}%
  {\hsnewpar\abovedisplayskip
   \hscodestyle
   \tabular{@{}>{\columncolor{codecolor}}p{\linewidth}@{}}%
   \let\\=\@normalcr
   \(\pboxed}%
  {\endpboxed\)%
   \endtabular
   \hsnewpar\belowdisplayskip
   \ignorespacesafterend}

\newenvironment{tightcolorhscode}%
  {\hsnewpar\abovedisplayskip
   \hscodestyle
   \tabular{@{}>{\columncolor{codecolor}\(}l<{\)}@{}}%
   \pmboxed}%
  {\endpmboxed%
   \endtabular
   \hsnewpar\belowdisplayskip
   \ignorespacesafterend}

\newenvironment{barhscode}%
  {\hsnewpar\abovedisplayskip
   \hscodestyle
   \arrayrulecolor{codecolor}%
   \arrayrulewidth=\coderulewidth
   \tabular{|p{\linewidth-\arrayrulewidth-\tabcolsep}@{}}%
   \let\\=\@normalcr
   \(\pboxed}%
  {\endpboxed\)%
   \endtabular
   \hsnewpar\belowdisplayskip
   \ignorespacesafterend}
\makeatother

\def\colorcode{\columncolor{codecolor}}
\definecolor{codecolor}{rgb}{1,1,.667}
\newlength{\coderulewidth}
\setlength{\coderulewidth}{3pt}

\newcommand{\colorhs}{\sethscode{colorhscode}}
\newcommand{\tightcolorhs}{\sethscode{tightcolorhscode}}
\newcommand{\barhs}{\sethscode{barhscode}}

\EndFmtInput

\renewcommand{\onelinecommentchars}{\color{gray}\quad-{}- }
\renewcommand{\commentbeginchars}{\color{gray}\enskip\{- }
\renewcommand{\commentendchars}{-\}\enskip}

\renewcommand{\visiblecomments}{%
  \let\onelinecomment=\onelinecommentchars
  \let\commentbegin=\commentbeginchars
  \let\commentend=\commentendchars}

\renewcommand{\invisiblecomments}{%
  \let\onelinecomment=\empty
  \let\commentbegin=\empty
  \let\commentend=\empty}

\visiblecomments

\section{Open Bisimulation}
\label{sec:bisim}
In this section, we discuss the definition of simulation in Haskell
to provide an understanding for the definition of bisimulation, which
shares a similar structure but twice in length.
Figure~\ref{fig:sim} illustrates two versions of the simulation definition.
The first version
\ensuremath{\Varid{sim}\mathbin{::}\Conid{Ctx}\!\to \!\Conid{Pr}\!\to \!\Conid{Pr}\!\to \!\Conid{Bool}}
is the usual simulation checker that returns a boolean value,
defined as a conjunction of the results from \ensuremath{\Varid{sim}\hspace{-.2ex}\anonymous \,}.
The second version \ensuremath{\Varid{sim'}} is almost identical to \ensuremath{\Varid{sim}\hspace{-.2ex}\anonymous \,} except
that it returns a forest that contains information about each simulation step.
Similarly, we have two versions for bisimulation,
\ensuremath{\Varid{bisim}} defined in terms of \ensuremath{\Varid{bisim}\hspace{-.2ex}\anonymous \,} and \ensuremath{\Varid{bisim'}} that returns a forest.

A process \ensuremath{\Varid{p}} is (openly) simulated by another process \ensuremath{\Varid{q}}, that is
(\ensuremath{\Varid{sim}\;\Gamma\;\Varid{p}\;\Varid{q}}) where \ensuremath{\Gamma\mathrel{=}[\mskip1.5mu \forall\hspace{-.75ex}\raisebox{.1ex}{\scalebox{1.1}[.9]{$/$}}\!\;\!\Varid{x}\mid \Varid{x}\leftarrow \Varid{fv}\;\!(\Varid{p},\Varid{q})\mskip1.5mu]},
when for every step from \ensuremath{\Varid{p}} to \ensuremath{\Varid{p'}} there exists a step from \ensuremath{\Varid{q}} to \ensuremath{\Varid{q'}}
labeled with the same action in the same word such that (\ensuremath{\Varid{sim}\;\Gamma\;\Varid{p'}\;\Varid{q'}});%
\footnote{The function (\ensuremath{\Varid{and}\mathbin{::}[\mskip1.5mu \Conid{Bool}\mskip1.5mu]\to \Conid{Bool}}) implements "for every step"
	and the function (\ensuremath{\Varid{or}\mathbin{::}[\mskip1.5mu \Conid{Bool}\mskip1.5mu]\to \Conid{Bool}}) implements "there exists a step".}
also, similarly for every bound step lead by \ensuremath{\Varid{p}} to \ensuremath{(\Varid{x}\hspace{.1ex}.\hspace{-.4ex}\backslash\Varid{p'})} is
followed by \ensuremath{\Varid{q}} to \ensuremath{(\Varid{x}\hspace{.1ex}.\hspace{-.4ex}\backslash\Varid{q'})} such that (\ensuremath{\Varid{sim}\;\Gamma'\;\Varid{p'}\;\Varid{q'}}) where
\ensuremath{\Gamma'} is a context extended from \ensuremath{\Gamma} with \ensuremath{\Varid{x}}.
In the definition of \ensuremath{\Varid{sim}\hspace{-.2ex}\anonymous \,} consists of \ensuremath{\mathbf{do}}-blocks combined by
the alternative operator (\!\ensuremath{\,\mathop{\scalebox{1}[.7]{\raisebox{.3ex}{$\langle\hspace{-.1ex}\vert\hspace{-.1ex}\rangle$}}}\,}\!). The first \ensuremath{\mathbf{do}}-block is for
the free step and the second is for the bound step. In the bound step case,
we make sure that the context (\ensuremath{\Gamma'}) used in the recursive calls after
following bound steps from \ensuremath{\Varid{q}} is extended by the same fresh variable (\ensuremath{\Varid{x'}}).%
\footnote{Having bound step children share the same fresh variable makes it more
convenient to generate the distinguishing formulae in Section~\ref{sec:df}.}

For bisimulation (\ensuremath{\Varid{bisim}\;\Gamma\;\Varid{p}\;\Varid{q}}), we consider both cases of either side
leading a step. Hence, the definition of \ensuremath{\Varid{bisim}} consists of four \ensuremath{\mathbf{do}}-blocks
where the first two have the same structure as \ensuremath{\Varid{sim}\hspace{-.2ex}\anonymous \,} lead by \ensuremath{\Varid{p}},
and the other two are the cases lead by \ensuremath{\Varid{q}}. Note that
bisimulation (\ensuremath{\Varid{bisim}\;\Gamma\;\Varid{p}\;\Varid{q}}) is not the same as mutual simulation
(\ensuremath{\Varid{sim}\;\Gamma\;\Varid{p}\;\Varid{q}\mathrel{\wedge}\Varid{sim}\;\Gamma\;\Varid{q}\;\Varid{p}}) in general. In bisimulation, the leading and 
following sides do not always alternate regularly. For instance, after
the leading step from \ensuremath{\Varid{p}} to \ensuremath{\Varid{p'}} followed by \ensuremath{\Varid{q}} to \ensuremath{\Varid{q'}}, both cases of \ensuremath{\Varid{p'}}s
lead and \ensuremath{\Varid{q'}}s lead are considered in bisimulation whereas only \ensuremath{\Varid{p'}} continues
to lead in simulation. Hence, bisimulation distinguishes more processes than
mutual simulation.

Both versions of transition steps are used here:
the symbolic version (Figure~\ref{fig:OpenLTS}) for the leading step and
the fixed-world version (Figure~\ref{fig:IdSubLTS}) for the following step.
It is possible to implement bisimulation only using the symbolic version
because the fixed-world version can be understood as a symbolic transition
restricted to the identity substitution. More precisely, the properties
in Figure~\ref{fig:prop} hold. The fixed-world version is more efficient
because it avoids generating possible worlds other than the current one.
In contrast, the equivalent implementation using the symbolic transition
generates substitutions of other possible worlds only to be discard by
failing to match the empty list pattern.
\begin{figure}\small
\begin{hscode}\SaveRestoreHook
\column{B}{@{}>{\hspre}l<{\hspost}@{}}%
\column{3}{@{}>{\hspre}c<{\hspost}@{}}%
\column{3E}{@{}l@{}}%
\column{7}{@{}>{\hspre}l<{\hspost}@{}}%
\column{E}{@{}>{\hspre}l<{\hspost}@{}}%
\>[7]{}(\Varid{runFreshMT}\mathop{\texttt{\$}}\Varid{\Conid{IdSubLTS}.one}\;\Varid{p}\mathbin{::}[\mskip1.5mu (\Conid{Act},\Conid{Pr})\mskip1.5mu]){}\<[E]%
\\
\>[3]{}\equiv {}\<[3E]%
\>[7]{}(\Varid{runFreshMT}\mathop{\texttt{\$}}\mathbf{do}\;\{\mskip1.5mu ([\mskip1.5mu \mskip1.5mu],\Varid{r})\leftarrow \Varid{\Conid{OpenLTS}.one}\;\Gamma\;\Varid{p}\hspace{.2ex};\hspace{.2ex}\Varid{return}\;\Varid{r}\mskip1.5mu\}){}\<[E]%
\ColumnHook
\end{hscode}\resethooks
\begin{hscode}\SaveRestoreHook
\column{B}{@{}>{\hspre}l<{\hspost}@{}}%
\column{3}{@{}>{\hspre}c<{\hspost}@{}}%
\column{3E}{@{}l@{}}%
\column{7}{@{}>{\hspre}l<{\hspost}@{}}%
\column{E}{@{}>{\hspre}l<{\hspost}@{}}%
\>[7]{}(\Varid{runFreshMT}\mathop{\texttt{\$}}\Conid{IdSubLTS}.\textit{one}_{\textsc{b}}\;\Varid{p}\mathbin{::}[\mskip1.5mu (\textit{Act}_{\textsc{b}},\textit{Pr}_{\textsc{b}})\mskip1.5mu]){}\<[E]%
\\
\>[3]{}\equiv {}\<[3E]%
\>[7]{}(\Varid{runFreshMT}\mathop{\texttt{\$}}\mathbf{do}\;\{\mskip1.5mu ([\mskip1.5mu \mskip1.5mu],\Varid{r})\leftarrow \Conid{OpenLTS}.\textit{one}_{\textsc{b}}\;\Gamma\;\Varid{p}\hspace{.2ex};\hspace{.2ex}\Varid{return}\;\Varid{r}\mskip1.5mu\}){}\<[E]%
\ColumnHook
\end{hscode}\resethooks
\vspace*{-4.5ex}%
\caption{Equational properties between fixed-world and symobilic transition steps
	where \ensuremath{\Gamma} is a closing context of \ensuremath{\Varid{p}}.}
\label{fig:prop}
\vspace*{-1.5ex}
\end{figure}

The amount of change from \ensuremath{\Varid{sim}\hspace{-.2ex}\anonymous \,} to \ensuremath{\Varid{sim'}} is small. The only differences are
that {\small\ensuremath{\Varid{return}\mathbin{\circ}(\Varid{and}\mathbin{::}[\mskip1.5mu \Conid{Bool}\mskip1.5mu]\to \Conid{Bool})}} and {\small\ensuremath{\Varid{return}\mathbin{\circ}(\Varid{or}\mathbin{::}[\mskip1.5mu \Conid{Bool}\mskip1.5mu]\to \Conid{Bool})}}
in each \ensuremath{\mathbf{do}}-block of \ensuremath{\Varid{sim}\hspace{-.2ex}\anonymous \,} are replaced by
{\small\ensuremath{\Varid{return}_L\mathbin{\circ}\Conid{One}\;(\mathbin{...})}} and {\small\ensuremath{\Varid{return}_R\mathbin{\circ}\Conid{One}\;(\mathbin{...})}} in the the free step case
and by
{\small\ensuremath{\Varid{return}_L\mathbin{\circ}\Conid{One}_{\textsc{b}}\;(\mathbin{...})}} and {\small\ensuremath{\Varid{return}_R\mathbin{\circ}\Conid{One}_{\textsc{b}}\;(\mathbin{...})}} in the bound step case
of \ensuremath{\Varid{sim'}}. The rest of the definition is exactly the same. Similarly, there are
twice amount of such differences between \ensuremath{\Varid{bisim}\hspace{-.2ex}\anonymous \,} and \ensuremath{\Varid{bisim'}} to prepare for
the distinguishing formulae generation.

\begin{figure}\small
\begin{hscode}\SaveRestoreHook
\column{B}{@{}>{\hspre}l<{\hspost}@{}}%
\column{15}{@{}>{\hspre}c<{\hspost}@{}}%
\column{15E}{@{}l@{}}%
\column{16}{@{}>{\hspre}c<{\hspost}@{}}%
\column{16E}{@{}l@{}}%
\column{17}{@{}>{\hspre}c<{\hspost}@{}}%
\column{17E}{@{}l@{}}%
\column{18}{@{}>{\hspre}l<{\hspost}@{}}%
\column{21}{@{}>{\hspre}l<{\hspost}@{}}%
\column{23}{@{}>{\hspre}l<{\hspost}@{}}%
\column{24}{@{}>{\hspre}l<{\hspost}@{}}%
\column{25}{@{}>{\hspre}l<{\hspost}@{}}%
\column{27}{@{}>{\hspre}l<{\hspost}@{}}%
\column{28}{@{}>{\hspre}l<{\hspost}@{}}%
\column{30}{@{}>{\hspre}l<{\hspost}@{}}%
\column{38}{@{}>{\hspre}l<{\hspost}@{}}%
\column{41}{@{}>{\hspre}l<{\hspost}@{}}%
\column{42}{@{}>{\hspre}l<{\hspost}@{}}%
\column{43}{@{}>{\hspre}l<{\hspost}@{}}%
\column{45}{@{}>{\hspre}l<{\hspost}@{}}%
\column{52}{@{}>{\hspre}l<{\hspost}@{}}%
\column{54}{@{}>{\hspre}l<{\hspost}@{}}%
\column{55}{@{}>{\hspre}l<{\hspost}@{}}%
\column{58}{@{}>{\hspre}l<{\hspost}@{}}%
\column{E}{@{}>{\hspre}l<{\hspost}@{}}%
\>[B]{}\mathbf{module}\;\Conid{OpenBisim}\;\mathbf{where}{}\<[E]%
\\
\>[B]{}\mathbf{import}\;\Conid{PiCalc}\hspace{.2ex};\hspace{.2ex}\mathbf{import}\;\Conid{\Conid{Control}.Applicative}\hspace{.2ex};\hspace{.2ex}\mathbf{import}\;\Conid{\Conid{Control}.Monad}{}\<[E]%
\\
\>[B]{}\mathbf{import}\;\Conid{OpenLTS}\hspace{.2ex};\hspace{.2ex}\mathbf{import}\;\mathbf{qualified}\;\Conid{IdSubLTS}\hspace{.2ex};\hspace{.2ex}\mathbf{import}\;\Conid{\Conid{Data}.Tree}{}\<[E]%
\\
\>[B]{}\mathbf{import}\;\Conid{\Conid{Unbound}.LocallyNameless}\;\mathbf{hiding}\;(\Varid{empty}){}\<[E]%
\\[\blanklineskip]%
\>[B]{}\mathbf{data}\;\Conid{StepLog}{}\<[15]%
\>[15]{}\mathrel{=}{}\<[15E]%
\>[18]{}\Conid{One}\;{}\<[24]%
\>[24]{}\Conid{Ctx}\;\Conid{EqC}\;\Conid{Act}\;{}\<[38]%
\>[38]{}\Conid{Pr}{}\<[E]%
\\
\>[15]{}\mid {}\<[15E]%
\>[18]{}\Conid{One}_{\textsc{b}}\;{}\<[24]%
\>[24]{}\Conid{Ctx}\;\Conid{EqC}\;\textit{Act}_{\textsc{b}}\;{}\<[38]%
\>[38]{}\textit{Pr}_{\textsc{b}}\;{}\<[43]%
\>[43]{}\mathbf{deriving}\;(\Conid{Eq},\Conid{Ord},\Conid{Show}){}\<[E]%
\\[\blanklineskip]%
\>[B]{}\Varid{return}_L\;\Varid{log}\mathrel{=}\Varid{return}\mathbin{\circ}\Conid{Node}\;(\Conid{Left}\;\Varid{log}){}\<[42]%
\>[42]{}\mbox{\onelinecomment  for the step on \ensuremath{\Varid{p}}'s side}{}\<[E]%
\\
\>[B]{}\Varid{return}_R\;\Varid{log}\mathrel{=}\Varid{return}\mathbin{\circ}\Conid{Node}\;(\Conid{Right}\;\Varid{log}){}\<[42]%
\>[42]{}\mbox{\onelinecomment  for the step on \ensuremath{\Varid{q}}'s side}{}\<[E]%
\\[\blanklineskip]%
\>[B]{}\Varid{sim}\;\Gamma\;\Varid{p}\;\Varid{q}\mathrel{=}\Varid{and}\mathop{\texttt{\$}}\Varid{sim}\hspace{-.2ex}\anonymous \,\;\Gamma\;\Varid{p}\;\Varid{q}{}\<[E]%
\\[\blanklineskip]%
\>[B]{}\Varid{sim}\hspace{-.2ex}\anonymous \,\mathbin{::}\Conid{Ctx}\to \Conid{Pr}\to \Conid{Pr}\to [\mskip1.5mu \Conid{Bool}\mskip1.5mu]{}\<[E]%
\\
\>[B]{}\Varid{sim}\hspace{-.2ex}\anonymous \,\;\Gamma\;\Varid{p}\;\Varid{q}{}\<[16]%
\>[16]{}\mathrel{=}{}\<[16E]%
\>[21]{}\mathbf{do}\;{}\<[25]%
\>[25]{}(\sigma,\Varid{r})\leftarrow \Varid{runFreshMT}\;(\Varid{one}\;\Gamma\;\Varid{p})\hspace{.2ex};\hspace{.2ex}\mathbf{let}\;{\hat{\sigma}}\mathrel{=}\Varid{subs}\;\Gamma\;\sigma{}\<[E]%
\\
\>[25]{}\mathbf{let}\;(l_p,\Varid{p'})\mathrel{=}{\hat{\sigma}}\;\Varid{r}{}\<[E]%
\\
\>[25]{}\Varid{return}\mathbin{\circ}(\Varid{or}\mathbin{::}[\mskip1.5mu \Conid{Bool}\mskip1.5mu]\to \Conid{Bool})\mathbin{\circ}\Varid{runFreshMT}\mathop{\texttt{\$}}\mathbf{do}{}\<[E]%
\\
\>[25]{}\hsindent{2}{}\<[27]%
\>[27]{}(l_q,\Varid{q'})\leftarrow \Varid{\Conid{IdSubLTS}.one}\;({\hat{\sigma}}\;\Varid{q}){}\<[E]%
\\
\>[25]{}\hsindent{2}{}\<[27]%
\>[27]{}\Varid{guard}\mathop{\texttt{\$}}l_p\equiv l_q{}\<[E]%
\\
\>[25]{}\hsindent{2}{}\<[27]%
\>[27]{}\Varid{return}\mathbin{\circ}(\Varid{and}\mathbin{::}[\mskip1.5mu \Conid{Bool}\mskip1.5mu]\to \Conid{Bool})\mathop{\texttt{\$}}\Varid{sim}\hspace{-.2ex}\anonymous \,\;\Gamma\;\Varid{p'}\;\Varid{q'}{}\<[E]%
\\
\>[16]{}\,\mathop{\scalebox{1}[.7]{\raisebox{.3ex}{$\langle\hspace{-.1ex}\vert\hspace{-.1ex}\rangle$}}}\,{}\<[16E]%
\>[21]{}\mathbf{do}\;{}\<[25]%
\>[25]{}(\sigma,\Varid{r})\leftarrow \Varid{runFreshMT}\;(\textit{one}_{\textsc{b}}\;\Gamma\;\Varid{p})\hspace{.2ex};\hspace{.2ex}\mathbf{let}\;{\hat{\sigma}}\mathrel{=}\Varid{subs}\;\Gamma\;\sigma{}\<[E]%
\\
\>[25]{}\mathbf{let}\;(l_p,b_{\!\Varid{p'}})\mathrel{=}{\hat{\sigma}}\;\Varid{r}{}\<[E]%
\\
\>[25]{}\mathbf{let}\;\Varid{x'}\mathrel{=}\Varid{runFreshM}\mathop{\texttt{\$}}\Varid{freshFrom}\;(\Varid{quan2nm}\mathop{\scalebox{1}[.7]{\raisebox{.3ex}{$\langle$}}\hspace{-.7ex}\mathop{\texttt{\$}}\hspace{-.7ex}\scalebox{1}[.7]{\raisebox{.3ex}{$\rangle$}}}\Gamma)\;b_{\!\Varid{p'}}{}\<[E]%
\\
\>[25]{}\Varid{return}\mathbin{\circ}(\Varid{or}\mathbin{::}[\mskip1.5mu \Conid{Bool}\mskip1.5mu]\to \Conid{Bool})\mathbin{\circ}\Varid{runFreshMT}\mathop{\texttt{\$}}\mathbf{do}{}\<[E]%
\\
\>[25]{}\hsindent{2}{}\<[27]%
\>[27]{}(l_q,b_{\!\Varid{q'}})\leftarrow \Conid{IdSubLTS}.\textit{one}_{\textsc{b}}\;({\hat{\sigma}}\;\Varid{q}){}\<[E]%
\\
\>[25]{}\hsindent{2}{}\<[27]%
\>[27]{}\Varid{guard}\mathop{\texttt{\$}}l_p\equiv l_q{}\<[E]%
\\
\>[25]{}\hsindent{2}{}\<[27]%
\>[27]{}(\Varid{x},\Varid{q}_{1},\Varid{p}_{1})\leftarrow \Varid{unbind2'}\;b_{\!\Varid{q'}}\;b_{\!\Varid{p'}}{}\<[E]%
\\
\>[25]{}\hsindent{2}{}\<[27]%
\>[27]{}\mathbf{let}\;(\Varid{p'},\Varid{q'}){}\<[41]%
\>[41]{}\mid \Varid{x}\equiv \Varid{x'}{}\<[54]%
\>[54]{}\mathrel{=}(\Varid{p}_{1},\Varid{q}_{1}){}\<[E]%
\\
\>[41]{}\mid \Varid{otherwise}{}\<[54]%
\>[54]{}\mathrel{=}\mathop{\sub{\Varid{x}}{(V\!\;\!\Varid{x'})}\!}\!\;(\Varid{p}_{1},\Varid{q}_{1}){}\<[E]%
\\
\>[25]{}\hsindent{2}{}\<[27]%
\>[27]{}\mathbf{let}\;\Gamma'\mathrel{=}\mathbf{case}\;l_p\;\mathbf{of}\;{}\<[52]%
\>[52]{}\hspace{-.15ex}\MVDnarrow^{\hspace{-.18ex}\textsc{b}\!}\;\anonymous \to \forall\hspace{-.75ex}\raisebox{.1ex}{\scalebox{1.1}[.9]{$/$}}\!\;\Varid{x'}\mathbin{:}\Gamma{}\<[E]%
\\
\>[52]{}\MVUparrow_{\!\textsc{b}\!}\;\anonymous \to \nabla\hspace{-.83ex}\raisebox{.1ex}{\scalebox{1.1}[.9]{$/$}}\!\;\Varid{x'}\mathbin{:}\Gamma{}\<[E]%
\\
\>[25]{}\hsindent{2}{}\<[27]%
\>[27]{}\Varid{return}\mathbin{\circ}(\Varid{and}\mathbin{::}[\mskip1.5mu \Conid{Bool}\mskip1.5mu]\to \Conid{Bool})\mathop{\texttt{\$}}\Varid{sim}\hspace{-.2ex}\anonymous \,\;\Gamma'\;\Varid{p'}\;\Varid{q'}{}\<[E]%
\\[\blanklineskip]%
\>[B]{}\Varid{sim'}\mathbin{::}\Conid{Ctx}\to \Conid{Pr}\to \Conid{Pr}\to [\mskip1.5mu \Conid{Tree}\;(\Conid{Either}\;\Conid{StepLog}\;\Conid{StepLog})\mskip1.5mu]{}\<[E]%
\\
\>[B]{}\Varid{sim'}\;\Gamma\;\Varid{p}\;\Varid{q}{}\<[17]%
\>[17]{}\mathrel{=}{}\<[17E]%
\>[23]{}\mathbf{do}\;{}\<[28]%
\>[28]{}(\sigma,\Varid{r})\leftarrow \Varid{runFreshMT}\;(\Varid{one}\;\Gamma\;\Varid{p})\hspace{.2ex};\hspace{.2ex}\mathbf{let}\;{\hat{\sigma}}\mathrel{=}\Varid{subs}\;\Gamma\;\sigma{}\<[E]%
\\
\>[28]{}\mathbf{let}\;(l_p,\Varid{p'})\mathrel{=}{\hat{\sigma}}\;\Varid{r}{}\<[E]%
\\
\>[28]{}\Varid{return}_L\;(\Conid{One}\;\Gamma\;\sigma\;l_p\;\Varid{p'})\mathbin{\circ}\Varid{runFreshMT}\mathop{\texttt{\$}}\mathbf{do}{}\<[E]%
\\
\>[28]{}\hsindent{2}{}\<[30]%
\>[30]{}(l_q,\Varid{q'})\leftarrow \Varid{\Conid{IdSubLTS}.one}\;({\hat{\sigma}}\;\Varid{q}){}\<[E]%
\\
\>[28]{}\hsindent{2}{}\<[30]%
\>[30]{}\Varid{guard}\mathop{\texttt{\$}}l_p\equiv l_q{}\<[E]%
\\
\>[28]{}\hsindent{2}{}\<[30]%
\>[30]{}\Varid{return}_R\;(\Conid{One}\;\Gamma\;\sigma\;l_q\;\Varid{q'})\mathop{\texttt{\$}}\Varid{sim'}\;\Gamma\;\Varid{p'}\;\Varid{q'}{}\<[E]%
\\
\>[17]{}\,\mathop{\scalebox{1}[.7]{\raisebox{.3ex}{$\langle\hspace{-.1ex}\vert\hspace{-.1ex}\rangle$}}}\,{}\<[17E]%
\>[23]{}\mathbf{do}\;{}\<[28]%
\>[28]{}(\sigma,\Varid{r})\leftarrow \Varid{runFreshMT}\;(\textit{one}_{\textsc{b}}\;\Gamma\;\Varid{p})\hspace{.2ex};\hspace{.2ex}\mathbf{let}\;{\hat{\sigma}}\mathrel{=}\Varid{subs}\;\Gamma\;\sigma{}\<[E]%
\\
\>[28]{}\mathbf{let}\;(l_p,b_{\!\Varid{p'}})\mathrel{=}{\hat{\sigma}}\;\Varid{r}{}\<[E]%
\\
\>[28]{}\mathbf{let}\;\Varid{x'}\mathrel{=}\Varid{runFreshM}\mathop{\texttt{\$}}\Varid{freshFrom}\;(\Varid{quan2nm}\mathop{\scalebox{1}[.7]{\raisebox{.3ex}{$\langle$}}\hspace{-.7ex}\mathop{\texttt{\$}}\hspace{-.7ex}\scalebox{1}[.7]{\raisebox{.3ex}{$\rangle$}}}\Gamma)\;b_{\!\Varid{p'}}{}\<[E]%
\\
\>[28]{}\Varid{return}_L\;(\Conid{One}_{\textsc{b}}\;\Gamma\;\sigma\;l_p\;b_{\!\Varid{p'}})\mathbin{\circ}\Varid{runFreshMT}\mathop{\texttt{\$}}\mathbf{do}{}\<[E]%
\\
\>[28]{}\hsindent{2}{}\<[30]%
\>[30]{}(l_q,b_{\!\Varid{q'}})\leftarrow \Conid{IdSubLTS}.\textit{one}_{\textsc{b}}\;({\hat{\sigma}}\;\Varid{q}){}\<[E]%
\\
\>[28]{}\hsindent{2}{}\<[30]%
\>[30]{}\Varid{guard}\mathop{\texttt{\$}}l_p\equiv l_q{}\<[E]%
\\
\>[28]{}\hsindent{2}{}\<[30]%
\>[30]{}(\Varid{x},\Varid{p}_{1},\Varid{q}_{1})\leftarrow \Varid{unbind2'}\;b_{\!\Varid{p'}}\;b_{\!\Varid{q'}}{}\<[E]%
\\
\>[28]{}\hsindent{2}{}\<[30]%
\>[30]{}\mathbf{let}\;(\Varid{p'},\Varid{q'}){}\<[45]%
\>[45]{}\mid \Varid{x}\equiv \Varid{x'}{}\<[58]%
\>[58]{}\mathrel{=}(\Varid{p}_{1},\Varid{q}_{1}){}\<[E]%
\\
\>[45]{}\mid \Varid{otherwise}{}\<[58]%
\>[58]{}\mathrel{=}\mathop{\sub{\Varid{x}}{(V\!\;\!\Varid{x'})}\!}\!\;(\Varid{p}_{1},\Varid{q}_{1}){}\<[E]%
\\
\>[28]{}\hsindent{2}{}\<[30]%
\>[30]{}\mathbf{let}\;\Gamma'\mathrel{=}\mathbf{case}\;l_p\;\mathbf{of}\;{}\<[55]%
\>[55]{}\hspace{-.15ex}\MVDnarrow^{\hspace{-.18ex}\textsc{b}\!}\;\anonymous \to \forall\hspace{-.75ex}\raisebox{.1ex}{\scalebox{1.1}[.9]{$/$}}\!\;\Varid{x'}\mathbin{:}\Gamma{}\<[E]%
\\
\>[55]{}\MVUparrow_{\!\textsc{b}\!}\;\anonymous \to \nabla\hspace{-.83ex}\raisebox{.1ex}{\scalebox{1.1}[.9]{$/$}}\!\;\Varid{x'}\mathbin{:}\Gamma{}\<[E]%
\\
\>[28]{}\hsindent{2}{}\<[30]%
\>[30]{}\Varid{return}_R\;(\Conid{One}_{\textsc{b}}\;\Gamma\;\sigma\;l_q\;b_{\!\Varid{q'}})\mathop{\texttt{\$}}\Varid{sim'}\;\Gamma'\;\Varid{p'}\;\Varid{q'}{}\<[E]%
\\[\blanklineskip]%
\>[B]{}\Varid{freshFrom}\mathbin{::}\Conid{Fresh}\;\Varid{m}\Rightarrow [\mskip1.5mu \Conid{Nm}\mskip1.5mu]\to \textit{Pr}_{\textsc{b}}\to \Varid{m}\;\Conid{Nm}{}\<[E]%
\\
\>[B]{}\Varid{freshFrom}\;\Varid{xs}\;\Varid{b}\mathrel{=}\mathbf{do}\;\{\mskip1.5mu \Varid{mapM}\hspace{-.2ex}\anonymous \,\;\Varid{fresh}\;\Varid{xs}\hspace{.2ex};\hspace{.2ex}(\Varid{y},\anonymous )\leftarrow (\hspace{.1ex}.\hspace{-.4ex}\backslash)^{{\text{-}\hspace{-.2ex}1\!}}\;\Varid{b}\hspace{.2ex};\hspace{.2ex}\Varid{return}\;\Varid{y}\mskip1.5mu\}{}\<[E]%
\ColumnHook
\end{hscode}\resethooks
\vspace*{-3.5ex}
\caption{An implementation of the open simulation (\ensuremath{\Varid{sim}}) and its variant (\ensuremath{\Varid{sim'}})
producing a forest.}
\label{fig:sim}
\end{figure}

\begin{figure}\small
\savecolumns
\begin{hscode}\SaveRestoreHook
\column{B}{@{}>{\hspre}l<{\hspost}@{}}%
\column{13}{@{}>{\hspre}c<{\hspost}@{}}%
\column{13E}{@{}l@{}}%
\column{18}{@{}>{\hspre}l<{\hspost}@{}}%
\column{22}{@{}>{\hspre}l<{\hspost}@{}}%
\column{E}{@{}>{\hspre}l<{\hspost}@{}}%
\>[B]{}\Varid{forest2df}\mathbin{::}[\mskip1.5mu \Conid{Tree}\;(\Conid{Either}\;\Conid{StepLog}\;\Conid{StepLog})\mskip1.5mu]\to [\mskip1.5mu (\Conid{Form},\Conid{Form})\mskip1.5mu]\hspace{10ex}{}\<[E]%
\\
\>[B]{}\Varid{forest2df}\;\Varid{rs}{}\<[E]%
\\
\>[B]{}\hsindent{13}{}\<[13]%
\>[13]{}\mathrel{=}{}\<[13E]%
\>[18]{}\mathbf{do}\;{}\<[22]%
\>[22]{}\Conid{Node}\;(\Conid{Left}\;(\Conid{One}\;\anonymous \;\sigma_{\!p}\;\Varid{a}\;\anonymous ))\;[\mskip1.5mu \mskip1.5mu]\leftarrow \Varid{rs}{}\<[E]%
\\
\>[22]{}\mathbf{let}\;\sigma_{\!q\!}\Varid{s}\mathrel{=}\Varid{subsMatchingAct}\;\Varid{a}\;(\Varid{right1s}\;\Varid{rs}){}\<[E]%
\\
\>[22]{}\Varid{return}\;(\Varid{prebase}\;\sigma_{\!p}\;\Varid{a}\,,\;\Varid{postbase}\;\sigma_{\!q\!}\Varid{s}\;\Varid{a}){}\<[E]%
\\
\>[B]{}\hsindent{13}{}\<[13]%
\>[13]{}\,\mathop{\scalebox{1}[.7]{\raisebox{.3ex}{$\langle\hspace{-.1ex}\vert\hspace{-.1ex}\rangle$}}}\,{}\<[13E]%
\>[18]{}\mathbf{do}~\ldots\mbox{\onelinecomment  do block symmetric to above omitted}{}\<[E]%
\ColumnHook
\end{hscode}\resethooks
\vspace*{-4.7ex}
\restorecolumns
\savecolumns
\begin{hscode}\SaveRestoreHook
\column{B}{@{}>{\hspre}l<{\hspost}@{}}%
\column{13}{@{}>{\hspre}c<{\hspost}@{}}%
\column{13E}{@{}l@{}}%
\column{18}{@{}>{\hspre}l<{\hspost}@{}}%
\column{22}{@{}>{\hspre}l<{\hspost}@{}}%
\column{E}{@{}>{\hspre}l<{\hspost}@{}}%
\>[13]{}\,\mathop{\scalebox{1}[.7]{\raisebox{.3ex}{$\langle\hspace{-.1ex}\vert\hspace{-.1ex}\rangle$}}}\,{}\<[13E]%
\>[18]{}\mathbf{do}\;{}\<[22]%
\>[22]{}\Conid{Node}\;(\Conid{Left}\;(\Conid{One}_{\textsc{b}}\;\anonymous \;\sigma_{\!p}\;\Varid{a}\;\anonymous ))\;[\mskip1.5mu \mskip1.5mu]\leftarrow \Varid{rs}{}\<[E]%
\\
\>[22]{}\mathbf{let}\;\sigma_{\!q\!}\Varid{s}\mathrel{=}\Varid{subsMatchingAct}_{\textsc{b}}\;\Varid{a}\;(\Varid{right1}{\scriptsize\textsc{b}}\Varid{s}\;\Varid{rs}){}\<[E]%
\\
\>[22]{}\Varid{return}\;(\Varid{preBbase}\;\sigma_{\!p}\;\Varid{a}\,,\;\Varid{postBbase}\;\sigma_{\!q\!}\Varid{s}\;\Varid{a}){}\<[E]%
\\
\>[13]{}\,\mathop{\scalebox{1}[.7]{\raisebox{.3ex}{$\langle\hspace{-.1ex}\vert\hspace{-.1ex}\rangle$}}}\,{}\<[13E]%
\>[18]{}\mathbf{do}~\ldots\mbox{\onelinecomment  do block symmetric to above omitted}{}\<[E]%
\ColumnHook
\end{hscode}\resethooks
\vspace*{-4.7ex}
\restorecolumns
\savecolumns
\begin{hscode}\SaveRestoreHook
\column{B}{@{}>{\hspre}l<{\hspost}@{}}%
\column{13}{@{}>{\hspre}c<{\hspost}@{}}%
\column{13E}{@{}l@{}}%
\column{18}{@{}>{\hspre}l<{\hspost}@{}}%
\column{22}{@{}>{\hspre}l<{\hspost}@{}}%
\column{E}{@{}>{\hspre}l<{\hspost}@{}}%
\>[13]{}\,\mathop{\scalebox{1}[.7]{\raisebox{.3ex}{$\langle\hspace{-.1ex}\vert\hspace{-.1ex}\rangle$}}}\,{}\<[13E]%
\>[18]{}\mathbf{do}\;{}\<[22]%
\>[22]{}\Conid{Node}\;(\Conid{Left}\;(\Conid{One}\;\anonymous \;\sigma_{\!p}\;\Varid{a}\;\anonymous ))\;\Varid{rs}_{\!R}\leftarrow \Varid{rs}{}\<[E]%
\\
\>[22]{}\mathbf{let}\;\Varid{rss'}\mathrel{=}[\mskip1.5mu \Varid{rs'}\mid \Conid{Node}\;\anonymous \;\Varid{rs'}\leftarrow \Varid{rs}_{\!R}\mskip1.5mu]{}\<[E]%
\\
\>[22]{}(\Varid{dfs}_{\!L},\Varid{dfs}_{\!R})\leftarrow \Varid{unzip}\mathop{\scalebox{1}[.7]{\raisebox{.3ex}{$\langle$}}\hspace{-.7ex}\mathop{\texttt{\$}}\hspace{-.7ex}\scalebox{1}[.7]{\raisebox{.3ex}{$\rangle$}}}\Varid{sequence}\;(\Varid{forest2df}\mathop{\scalebox{1}[.7]{\raisebox{.3ex}{$\langle$}}\hspace{-.7ex}\mathop{\texttt{\$}}\hspace{-.7ex}\scalebox{1}[.7]{\raisebox{.3ex}{$\rangle$}}}\Varid{rss'}){}\<[E]%
\\
\>[22]{}\Varid{guard}\mathbin{\circ}\neg \mathbin{\circ}\Varid{null}\mathop{\texttt{\$}}\Varid{dfs}_{\!L}{}\<[E]%
\\
\>[22]{}\mathbf{let}\;\sigma_{\!q\!}\Varid{s}\mathrel{=}\Varid{subsMatchingAct}\;\Varid{a}\;(\Varid{right1s}\;\Varid{rs}){}\<[E]%
\\
\>[22]{}\Varid{return}\;(\Varid{pre}\;\sigma_{\!p}\;\Varid{a}\;\Varid{dfs}_{\!L}\,,\;\Varid{post}\;\sigma_{\!q\!}\Varid{s}\;\Varid{a}\;\Varid{dfs}_{\!R}){}\<[E]%
\\
\>[13]{}\,\mathop{\scalebox{1}[.7]{\raisebox{.3ex}{$\langle\hspace{-.1ex}\vert\hspace{-.1ex}\rangle$}}}\,{}\<[13E]%
\>[18]{}\mathbf{do}~\ldots\mbox{\onelinecomment  do block symmetric to above omitted}{}\<[E]%
\ColumnHook
\end{hscode}\resethooks
\vspace*{-4.7ex}
\restorecolumns
\savecolumns
\begin{hscode}\SaveRestoreHook
\column{B}{@{}>{\hspre}l<{\hspost}@{}}%
\column{13}{@{}>{\hspre}c<{\hspost}@{}}%
\column{13E}{@{}l@{}}%
\column{18}{@{}>{\hspre}l<{\hspost}@{}}%
\column{22}{@{}>{\hspre}l<{\hspost}@{}}%
\column{27}{@{}>{\hspre}l<{\hspost}@{}}%
\column{29}{@{}>{\hspre}l<{\hspost}@{}}%
\column{E}{@{}>{\hspre}l<{\hspost}@{}}%
\>[13]{}\,\mathop{\scalebox{1}[.7]{\raisebox{.3ex}{$\langle\hspace{-.1ex}\vert\hspace{-.1ex}\rangle$}}}\,{}\<[13E]%
\>[18]{}\mathbf{do}\;{}\<[22]%
\>[22]{}\Conid{Node}\;(\Conid{Left}\;(\Conid{One}_{\textsc{b}}\;\Gamma\;\sigma_{\!p}\;\Varid{a}\;\anonymous ))\;\Varid{rs}_{\!R}\leftarrow \Varid{rs}{}\<[E]%
\\
\>[22]{}\mathbf{let}\;{}\<[27]%
\>[27]{}\Varid{rss'}\mathrel{=}[\mskip1.5mu \Varid{rs'}\mid \Conid{Node}\;\anonymous \;\Varid{rs'}\leftarrow \Varid{rs}_{\!R}\mskip1.5mu]{}\<[E]%
\\
\>[27]{}\Varid{x}\mathrel{=}\Varid{quan2nm}\mathbin{\circ}\Varid{head}\mathbin{\circ}\Varid{getCtx}\mathbin{\circ}\Varid{fromEither}{}\<[E]%
\\
\>[27]{}\hsindent{2}{}\<[29]%
\>[29]{}\mathbin{\circ}\Varid{rootLabel}\mathbin{\circ}\Varid{head}\mathop{\texttt{\$}}\Varid{head}\;\Varid{rss'}{}\<[E]%
\\
\>[22]{}(\Varid{dfs}_{\!L},\Varid{dfs}_{\!R})\leftarrow \Varid{unzip}\mathop{\scalebox{1}[.7]{\raisebox{.3ex}{$\langle$}}\hspace{-.7ex}\mathop{\texttt{\$}}\hspace{-.7ex}\scalebox{1}[.7]{\raisebox{.3ex}{$\rangle$}}}\Varid{sequence}\;(\Varid{forest2df}\mathop{\scalebox{1}[.7]{\raisebox{.3ex}{$\langle$}}\hspace{-.7ex}\mathop{\texttt{\$}}\hspace{-.7ex}\scalebox{1}[.7]{\raisebox{.3ex}{$\rangle$}}}\Varid{rss'}){}\<[E]%
\\
\>[22]{}\Varid{guard}\mathbin{\circ}\neg \mathbin{\circ}\Varid{null}\mathop{\texttt{\$}}\Varid{dfs}_{\!L}{}\<[E]%
\\
\>[22]{}\mathbf{let}\;\sigma_{\!q\!}\Varid{s}\mathrel{=}\Varid{subsMatchingAct}_{\textsc{b}}\;\Varid{a}\;(\Varid{right1}{\scriptsize\textsc{b}}\Varid{s}\;\Varid{rs}){}\<[E]%
\\
\>[22]{}\Varid{return}\;(\Varid{preB}\;\sigma_{\!p}\;\Varid{a}\;\Varid{x}\;\Varid{dfs}_{\!L}\,,\;\Varid{postB}\;\sigma_{\!q\!}\Varid{s}\;\Varid{a}\;\Varid{x}\;\Varid{dfs}_{\!R}){}\<[E]%
\\
\>[13]{}\,\mathop{\scalebox{1}[.7]{\raisebox{.3ex}{$\langle\hspace{-.1ex}\vert\hspace{-.1ex}\rangle$}}}\,{}\<[13E]%
\>[18]{}\mathbf{do}~\ldots\mbox{\onelinecomment  do block symmetric to above omitted}{}\<[E]%
\ColumnHook
\end{hscode}\resethooks
\vspace*{-4.7ex}
\restorecolumns
\savecolumns
\begin{hscode}\SaveRestoreHook
\column{B}{@{}>{\hspre}l<{\hspost}@{}}%
\column{3}{@{}>{\hspre}l<{\hspost}@{}}%
\column{5}{@{}>{\hspre}l<{\hspost}@{}}%
\column{7}{@{}>{\hspre}l<{\hspost}@{}}%
\column{13}{@{}>{\hspre}l<{\hspost}@{}}%
\column{14}{@{}>{\hspre}l<{\hspost}@{}}%
\column{15}{@{}>{\hspre}l<{\hspost}@{}}%
\column{19}{@{}>{\hspre}l<{\hspost}@{}}%
\column{24}{@{}>{\hspre}l<{\hspost}@{}}%
\column{30}{@{}>{\hspre}l<{\hspost}@{}}%
\column{32}{@{}>{\hspre}l<{\hspost}@{}}%
\column{39}{@{}>{\hspre}l<{\hspost}@{}}%
\column{40}{@{}>{\hspre}l<{\hspost}@{}}%
\column{44}{@{}>{\hspre}l<{\hspost}@{}}%
\column{51}{@{}>{\hspre}l<{\hspost}@{}}%
\column{54}{@{}>{\hspre}l<{\hspost}@{}}%
\column{62}{@{}>{\hspre}l<{\hspost}@{}}%
\column{E}{@{}>{\hspre}l<{\hspost}@{}}%
\>[3]{}\mathbf{where}{}\<[E]%
\\
\>[3]{}\hsindent{2}{}\<[5]%
\>[5]{}\Varid{prebase}\;\sigma\;\Varid{a}\mathrel{=}\Varid{pre}\;\sigma\;\Varid{a}\;[\mskip1.5mu \mskip1.5mu]{}\<[E]%
\\
\>[3]{}\hsindent{2}{}\<[5]%
\>[5]{}\Varid{postbase}\;\sigma{}\hspace{-.3ex}s\;\Varid{a}\mathrel{=}\Varid{post}\;\sigma{}\hspace{-.3ex}s\;\Varid{a}\;[\mskip1.5mu \mskip1.5mu]{}\<[E]%
\\
\>[3]{}\hsindent{2}{}\<[5]%
\>[5]{}\Varid{preBbase}\;\sigma\;\Varid{a}\mathrel{=}\Varid{preB}\;\sigma\;\Varid{a}\;(\Varid{s2n}\;\text{\tt \char34 ?\char34})\;[\mskip1.5mu \mskip1.5mu]{}\<[E]%
\\
\>[3]{}\hsindent{2}{}\<[5]%
\>[5]{}\Varid{postBbase}\;\sigma{}\hspace{-.3ex}s\;\Varid{a}\mathrel{=}\Varid{postB}\;\sigma{}\hspace{-.3ex}s\;\Varid{a}\;(\Varid{s2n}\;\text{\tt \char34 ?\char34})\;[\mskip1.5mu \mskip1.5mu]{}\<[E]%
\\
\>[3]{}\hsindent{2}{}\<[5]%
\>[5]{}\Varid{pre}\;\sigma\;\Varid{a}\mathrel{=}\Varid{boxMat}\;\sigma\mathbin{\circ}\Diamond\!\;\Varid{a}\mathbin{\circ}\Varid{conj}{}\<[E]%
\\
\>[3]{}\hsindent{2}{}\<[5]%
\>[5]{}\Varid{post}\;\sigma{}\hspace{-.3ex}s\;\Varid{a}\;\Varid{fs}\mathrel{=}\Box\!\;\Varid{a}\mathbin{\circ}\Varid{disj}\mathop{\texttt{\$}}\,(\Varid{diaMat}\mathop{\scalebox{1}[.7]{\raisebox{.3ex}{$\langle$}}\hspace{-.7ex}\mathop{\texttt{\$}}\hspace{-.7ex}\scalebox{1}[.7]{\raisebox{.3ex}{$\rangle$}}}\sigma{}\hspace{-.3ex}s)\plus \Varid{fs}{}\<[E]%
\\
\>[3]{}\hsindent{2}{}\<[5]%
\>[5]{}\Varid{preB}\;\sigma\;\Varid{a}\;\Varid{x}\mathrel{=}\Varid{boxMat}\;\sigma\mathbin{\circ}\Diamond_{^{\!}\textsc{b}}\!\;\Varid{a}\mathbin{\circ}\Varid{bind}\;\Varid{x}\mathbin{\circ}\Varid{conj}{}\<[E]%
\\
\>[3]{}\hsindent{2}{}\<[5]%
\>[5]{}\Varid{postB}\;\sigma{}\hspace{-.3ex}s\;\Varid{a}\;\Varid{x}\;\Varid{fs}\mathrel{=}\Box_{^{\!}\textsc{b}}\!\;\Varid{a}\mathbin{\circ}\Varid{bind}\;\Varid{x}\mathbin{\circ}\Varid{disj}\mathop{\texttt{\$}}\,(\Varid{diaMat}\mathop{\scalebox{1}[.7]{\raisebox{.3ex}{$\langle$}}\hspace{-.7ex}\mathop{\texttt{\$}}\hspace{-.7ex}\scalebox{1}[.7]{\raisebox{.3ex}{$\rangle$}}}\sigma{}\hspace{-.3ex}s)\plus \Varid{fs}{}\<[E]%
\\
\>[3]{}\hsindent{2}{}\<[5]%
\>[5]{}\Varid{boxMat}\;{}\<[13]%
\>[13]{}[\mskip1.5mu \mskip1.5mu]\mathrel{=}\Varid{id}\hspace{.2ex};\hspace{.2ex}\Varid{boxMat}\;{}\<[30]%
\>[30]{}\sigma\mathrel{=}\Box_{=}\!\;[\mskip1.5mu (V\!\;\Varid{x},V\!\;\Varid{y})\mid (\Varid{x},\Varid{y})\leftarrow \sigma\mskip1.5mu]{}\<[E]%
\\
\>[3]{}\hsindent{2}{}\<[5]%
\>[5]{}\Varid{diaMat}\;{}\<[13]%
\>[13]{}[\mskip1.5mu \mskip1.5mu]\mathrel{=}\bot\hspace{.2ex};\hspace{.2ex}\Varid{diaMat}\;{}\<[30]%
\>[30]{}\sigma\mathrel{=}\Diamond_{\!=}\!\;[\mskip1.5mu (V\!\;\Varid{x},V\!\;\Varid{y})\mid (\Varid{x},\Varid{y})\leftarrow \sigma\mskip1.5mu]\;\top{}\<[E]%
\\
\>[3]{}\hsindent{2}{}\<[5]%
\>[5]{}\Varid{right1s}\;{}\<[14]%
\>[14]{}\Varid{rs}\mathrel{=}[\mskip1.5mu \Varid{log}\mid \Conid{Node}\;(\Conid{Right}\;{}\<[39]%
\>[39]{}\Varid{log}\mathord{@}(\Conid{One}\;\{\mskip1.5mu \mskip1.5mu\}))\;\anonymous \leftarrow \Varid{rs}\mskip1.5mu]{}\<[E]%
\\
\>[3]{}\hsindent{2}{}\<[5]%
\>[5]{}\Varid{left1s}\;{}\<[14]%
\>[14]{}\Varid{rs}\mathrel{=}[\mskip1.5mu \Varid{log}\mid \Conid{Node}\;(\Conid{Left}\;{}\<[39]%
\>[39]{}\Varid{log}\mathord{@}(\Conid{One}\;\{\mskip1.5mu \mskip1.5mu\}))\;\anonymous \leftarrow \Varid{rs}\mskip1.5mu]{}\<[E]%
\\
\>[3]{}\hsindent{2}{}\<[5]%
\>[5]{}\Varid{right1}{\scriptsize\textsc{b}}\Varid{s}\;{}\<[15]%
\>[15]{}\Varid{rs}\mathrel{=}[\mskip1.5mu \Varid{log}\mid \Conid{Node}\;(\Conid{Right}\;{}\<[40]%
\>[40]{}\Varid{log}\mathord{@}(\Conid{One}_{\textsc{b}}\;\{\mskip1.5mu \mskip1.5mu\}))\;\anonymous \leftarrow \Varid{rs}\mskip1.5mu]{}\<[E]%
\\
\>[3]{}\hsindent{2}{}\<[5]%
\>[5]{}\Varid{left1}{\scriptsize\textsc{b}}\Varid{s}\;{}\<[15]%
\>[15]{}\Varid{rs}\mathrel{=}[\mskip1.5mu \Varid{log}\mid \Conid{Node}\;(\Conid{Left}\;{}\<[40]%
\>[40]{}\Varid{log}\mathord{@}(\Conid{One}_{\textsc{b}}\;\{\mskip1.5mu \mskip1.5mu\}))\;\anonymous \leftarrow \Varid{rs}\mskip1.5mu]{}\<[E]%
\\
\>[3]{}\hsindent{2}{}\<[5]%
\>[5]{}\Varid{getCtx}\;(\Conid{One}\;{}\<[19]%
\>[19]{}\Gamma\;\anonymous \;\anonymous \;\anonymous ){}\<[32]%
\>[32]{}\mathrel{=}\Gamma\hspace{.2ex};\hspace{.2ex}\Varid{getCtx}\;(\Conid{One}_{\textsc{b}}\;{}\<[54]%
\>[54]{}\Gamma\;\anonymous \;\anonymous \;\anonymous )\mathrel{=}\Gamma{}\<[E]%
\\
\>[3]{}\hsindent{2}{}\<[5]%
\>[5]{}\Varid{fromEither}\;(\Conid{Left}\;{}\<[24]%
\>[24]{}\Varid{t})\mathrel{=}\Varid{t}\hspace{.2ex};\hspace{.2ex}\Varid{fromEither}\;(\Conid{Right}\;{}\<[51]%
\>[51]{}\Varid{t})\mathrel{=}\Varid{t}{}\<[E]%
\\[\blanklineskip]%
\>[B]{}\Varid{subsMatchingAct}\mathbin{::}\Conid{Act}\to [\mskip1.5mu \Conid{StepLog}\mskip1.5mu]\to [\mskip1.5mu \Conid{EqC}\mskip1.5mu]{}\<[E]%
\\
\>[B]{}\Varid{subsMatchingAct}\;\Varid{a}\;\Varid{logs}\mathrel{=}{}\<[E]%
\\
\>[B]{}\hsindent{3}{}\<[3]%
\>[3]{}\mathbf{do}\;{}\<[7]%
\>[7]{}\Conid{One}\;\Gamma\;\sigma\;\Varid{a'}\;\anonymous \leftarrow \Varid{logs}{}\<[44]%
\>[44]{}\;\hspace{.2ex};\hspace{.2ex}~\;{}\<[62]%
\>[62]{}\mathbf{let}\;{\hat{\sigma}}\mathrel{=}\Varid{subs}\;\Gamma\;\sigma{}\<[E]%
\\
\>[7]{}\Varid{guard}\mathop{\texttt{\$}}{\hat{\sigma}}\;\Varid{a}\equiv {\hat{\sigma}}\;\Varid{a'}{}\<[44]%
\>[44]{}\;\hspace{.2ex};\hspace{.2ex}~{}\<[62]%
\>[62]{}\Varid{return}\;\sigma{}\<[E]%
\\[\blanklineskip]%
\>[B]{}\Varid{subsMatchingAct}_{\textsc{b}}\mathbin{::}\textit{Act}_{\textsc{b}}\to [\mskip1.5mu \Conid{StepLog}\mskip1.5mu]\to [\mskip1.5mu \Conid{EqC}\mskip1.5mu]{}\<[E]%
\\
\>[B]{}\Varid{subsMatchingAct}_{\textsc{b}}\;\Varid{a}\;\Varid{logs}\mathrel{=}{}\<[E]%
\\
\>[B]{}\hsindent{3}{}\<[3]%
\>[3]{}\mathbf{do}\;{}\<[7]%
\>[7]{}\Conid{One}_{\textsc{b}}\;\Gamma\;\sigma\;\Varid{a'}\;\anonymous \leftarrow \Varid{logs}{}\<[44]%
\>[44]{}\;\hspace{.2ex};\hspace{.2ex}~\;{}\<[62]%
\>[62]{}\mathbf{let}\;{\hat{\sigma}}\mathrel{=}\Varid{subs}\;\Gamma\;\sigma{}\<[E]%
\\
\>[7]{}\Varid{guard}\mathop{\texttt{\$}}{\hat{\sigma}}\;\Varid{a}\equiv {\hat{\sigma}}\;\Varid{a'}{}\<[44]%
\>[44]{}\;\hspace{.2ex};\hspace{.2ex}~{}\<[62]%
\>[62]{}\Varid{return}\;\sigma{}\<[E]%
\ColumnHook
\end{hscode}\resethooks
\vspace*{-3.5ex}
\caption{Generating distinguishing formulae from the forest produced by \ensuremath{\Varid{bisim'}}.}
\label{fig:df}
\end{figure}

\section{Distinguishing Formulae Generation}
\label{sec:df}
The distinguishing formulae generation is no more than a tree transformation.
(Figure~\ref{fig:df}), which generates a pair of
distinguishing formulae from the forest of rose trees produced
by \ensuremath{(\Varid{bisim'}\;\Gamma\;\Varid{p}\;\Varid{q})}. The first formula is satisfied by the left process (\ensuremath{\Varid{p}})
but fails to be satisfied by the other. Likewise, the second formula is
satisfied by the right process (\ensuremath{\Varid{q}}) but not by the other. The tree transformation
function \ensuremath{\Varid{forest2df}} returns a list (\ensuremath{[\mskip1.5mu (\Conid{Form},\Conid{Form})\mskip1.5mu]}) because there can be more
than one pair of such formulae for the given non-bisimilar processes.
For bisimilar processes, \ensuremath{\Varid{forest2df}} returns the empty list. The definition of
\ensuremath{\Varid{forest2df}} consists of eight \ensuremath{\mathbf{do}}-blocks where the first four are base cases
and the latter four are inductive cases. We only illustrate the cases lead by
the left side (\ensuremath{\Varid{p}}) while the cases lead by the right side (\ensuremath{\Varid{q}}) are omitted
in Figure~\ref{fig:df}.

It is a base case when the leading step has no matching following step.
That is, the children following the leading step specified by the root label of
the tree is an empty list, as you can observe from the beginning lines of
the first and third \ensuremath{\mathbf{do}}-blocks in Figure~\ref{fig:df}. 
The formula satisfied by the leading side is
\ensuremath{(\Diamond_{\!=}\!\;\sigma_{\!p}\;(\Diamond\!\;\Varid{a}\;\top))} or \ensuremath{(\Diamond_{\!=}\!\;\sigma_{\!p}\;(\Diamond_{^{\!}\textsc{b}}\!\;\Varid{a}\;(\Varid{w}\hspace{.1ex}.\hspace{-.4ex}\backslash\top))},
generated by \ensuremath{\Varid{prebase}} or \ensuremath{\Varid{preBbase}}, whose intuitive meaning is that
the process can make a step labeled with \ensuremath{\Varid{a}} in the world given by \ensuremath{\sigma_{\!p}}.
This formula clearly fails to be satisfied by the other side because there is
no following step (i.e., step labeled with \ensuremath{\Varid{a}} from \ensuremath{\Varid{q}} in the \ensuremath{\sigma_{\!p}}-world)
for the base case. If there were only one world to consider, the formula for
the other side would be \ensuremath{(\Box\!\;\Varid{a}\;\bot)} or \ensuremath{(\Box_{^{\!}\textsc{b}}\!\;\Varid{a}\;(\Varid{w}\hspace{.1ex}.\hspace{-.4ex}\backslash\bot))}, meaning that
the process cannot make a step labeled with \ensuremath{\Varid{a}}. However, we must consider
the possible worlds where such step exists for the following side.
Such worlds (\ensuremath{\sigma_{\!q\!}\Varid{s}}) are collected from the sibling nodes of the leading step
using the helper functions \ensuremath{\Varid{subsMatchingAct}} and \ensuremath{\Varid{subsMatchingAct}_{\textsc{b}}}.
The formula satisfied by the following side is
\ensuremath{(\Box\!\;(\Diamond_{\!=}\!\;\sigma_{\!q\!}\Varid{s}\;\top))} or \ensuremath{(\Box_{^{\!}\textsc{b}}\!\;(\Varid{w}\hspace{.1ex}.\hspace{-.4ex}\backslash\Diamond_{\!=}\!\;\sigma_{\!q\!}\Varid{s}\;\top)},
generated by \ensuremath{\Varid{postbase}} or \ensuremath{\Varid{postBbase}}.

In an inductive case where the leading step from \ensuremath{\Varid{p}} to \ensuremath{\Varid{p'}} is matched
by a following step \ensuremath{\Varid{q}} to \ensuremath{\Varid{q'}}, we find a pair of distinguishing formulae
for each pair of \ensuremath{\Varid{p'}} and \ensuremath{\Varid{q'}} at next step by recursively applying \ensuremath{\Varid{forst2df}}
to all the grandchildren following the steps lead by \ensuremath{\Varid{p}}, that is,
\ensuremath{(\Varid{sequence}\;(\Varid{forest2df}\,\mathop{\scalebox{1}[.7]{\raisebox{.3ex}{$\langle\hspace{-.1ex}\vert\hspace{-.1ex}\rangle$}}}\,\Varid{rss'}))\mathbin{::}[\mskip1.5mu (\Conid{Form},\Conid{Form})\mskip1.5mu]}.
The this list should not be empty; otherwise it had either been a base case
or it had been a forest generated from bisimilar processes.
The collected the left biased formulae (\ensuremath{\Varid{dfs}_{\!L}}) are used for constructing
the distinguishing formula satisfied by the leading side in the fifth and
seventh \ensuremath{\mathbf{do}}-blocks in Figure~\ref{fig:df}, which is
\ensuremath{(\Box_{=}\!\;\sigma_{\!p}\;(\Diamond\!\;\Varid{a}\;(\bigwedge\!\;\Varid{dfs}_{\!L})))} or
\ensuremath{(\Box_{=}\!\;\sigma_{\!p}\;(\Diamond_{^{\!}\textsc{b}}\!\;\Varid{a}\;(\Varid{w}\hspace{.1ex}.\hspace{-.4ex}\backslash\bigwedge\!\;\Varid{dfs}_{\!L})))} where \ensuremath{\Varid{w}} is fresh in \ensuremath{\Varid{dfs}_{\!L}}.
Similarly, the right biased formulae (\ensuremath{\Varid{dfs}_{\!R}}) are used for constructing
the formula satisfied by the other side, which is
\ensuremath{(\Box\!\;\Varid{a}\;(\bigvee\!\;(\Diamond_{\!=}\!\;\sigma_{\!q\!}\Varid{s}\;\top\plus \Varid{dfs}_{\!R})))} or
\ensuremath{(\Box_{^{\!}\textsc{b}}\!\;\Varid{a}\;(\Varid{x}\hspace{.1ex}.\hspace{-.4ex}\backslash\bigvee\!\;(\Diamond_{\!=}\!\;\sigma_{\!q\!}\Varid{s}\;\top\plus \Varid{dfs}_{\!R})))}.
Here, \ensuremath{\Varid{x}} corresponds to the \ensuremath{\Varid{x'}} in Figure~\ref{fig:sim}, which is
the fresh variable extending the context. Because we made sure that
the same variable is used to extend the context across
all the following bound steps from a leading step, we simply need to select
the first one, using some number of selector functions to go inside the list,
retrieve the context from the root, and grab the name
in the first quantifier of the context.

\section{Discussions}
\label{sec:discuss}
We point out three advantages of using Haskell for our problem of
generating distinguishing formulae (Section~\ref{sec:discuss:adv})
and discuss further optmizations and extensions to our current
implementation presented in this paper (Section~\ref{sec:discuss:fur}).%
\vspace*{-1ex}
\subsection{Advantages of using Haskell}
\label{sec:discuss:adv}
First, having a well-tailored generic name binding library such as
\textsf{unbound}~\cite{unbound11} saves a great amount of effort
on tedious boilerplate code for keeping track of freshness, 
collecting free variables, and capture-avoiding substitutions.
Due to value passing and name restriction in the $\pi$-calculus,
frequent management of name bindings is inevitable in implementations
involving the $\pi$-calculus.

Second, lazy evaluation and monadic encoding of nondeterminism in Haskell
makes it natural to view \emph{control flow} as \emph{data}.
Distinguishing formula generation can be defined as a tree transformation
(\ensuremath{\Varid{forest2df}}) over the forest of rose trees lazily produced from \ensuremath{\Varid{bisim'}}.
Only a small amount of change was needed to abstract the control flow of
computing a boolean by \ensuremath{\Varid{bisim}} into data production by \ensuremath{\Varid{bisim'}}.

The forest produced by \ensuremath{\Varid{bisim'}} is all possible traces of bisimulation steps.
The control flow of \ensuremath{\Varid{bisim}} for non-bisimilar processes
corresponds to a depth-first search traversal
until the return value is determined to be \ensuremath{\Conid{False}}.
For bisimilar processes, \ensuremath{\Varid{bisim}} returns \ensuremath{\Conid{True}} after the exhaustive traversal.

The traversal during the formulae generation
does not exactly match the pattern of traversal by \ensuremath{\Varid{bisim}}.
Alongside the depth-first search, there are traversals across the siblings of
the leading step to collect \ensuremath{\sigma_{\!q\!}\Varid{s}} in Figure~\ref{fig:df}.

For process calculi with less sophisticated semantics, it is possible to log
a run of bisimulation check and construct distinguishing formulae using
the information from those visited nodes only. In contrast, we need additional
information on other possible worlds, which come from the nodes not necessarily
visited by \ensuremath{\Varid{bisim}}.

Third, constraints are first-class values in constraint programming using Haskell.
We construct distinguishing formulae using substitutions (i.e., equality constraints)
as values (e.g., \ensuremath{\sigma_{\!p}} and \ensuremath{\sigma_{\!q\!}\Varid{s}} in Figure~\ref{fig:df}).
This is not quite well supported in (constraint) logic programming.
For example, consider a Prolog code fragment,
$\mathtt{ \cdots {\scriptsize\textcircled{1}}\,
            X=Y, {\scriptsize\textcircled{2}}\,
            Z=W, {\scriptsize\textcircled{3}}\, \cdots }$,
and let $\sigma_1$, $\sigma_2$, and $\sigma_3$ be the equality constraints at
the points marked by {\scriptsize\textcircled{1}}, {\scriptsize\textcircled{2}},
and {\scriptsize\textcircled{3}}. We understand that it should be
$\mathtt{\sigma_1\cup\{X=Y\}\equiv\sigma_2}$ and
$\mathtt{\sigma_2\cup\{Z=W\}\equiv\sigma_3}$.
However, $\sigma_1$, $\sigma_2$, and $\sigma_3$ are not values in a logic programming language.

The labeled transition semantics and open bisimulation can be elegantly
specified in higher-order logic programming systems~\cite{TiuMil10};
for those purposes, it fits better than functional programming. However, generating
certificates regarding open bisimulation requires the ability that amounts to
accessing meta-level properties of logic programs (e.g., substitutions) across
nondeterminisitc execution paths, where it is preferable to have constraints
as fist-class values.%
\vspace*{-1ex}
\subsection{Further Optimizations and Extensions}
\label{sec:discuss:fur}
One obvious optimization to our current implementation is to represent
the equality constraints as partitions instead of computing partitions
from the list of name pairs on the fly every time we need a substitution function.

We can enrich the term structure to model applied variants of $\pi$-calculi
by supporting unification in a more general setting \cite{Miler92uum} and
constraints other than the equalities solvable by unification. When the constraints
become more complex, we can no longer model them as integer set partitions. In addition,
it would be better to abstract constraint handling with another layer of monad
(e.g., state monad).  In this work, we did not bother to abstract the constraints
in a monad because they were very simple equalities over names only.

To handle infinite processes (or finite but quite large ones) effectively,
we should consider using more sophisticated search strategies. For this,
we would need to replace the list monad with a custom monad equipped with
better control over traversing the paths of nondeterministic computation.
Thanks to the monadic abstraction, the definitions could remain mostly the same
and only their type signatures would be modified to use the custom monad.

Memoization or tabling is a well known optimization technique to avoid
repetitive computation by storing results of computations associated with
their input arguments. When we have infinite processes, this is no longer
an optional optimization but a means to implement the coinductive definition of
bisimulation over possibly infinite transition paths.
Parallel computing may also help to improve scalability of traversing over
large space of possible transitions but memoization could raise additional
concurrency issues \cite{ZiaSivJag09,Bergstrom13phd}.

\section{Related Work}
\label{sec:relwork}
In this section, we discuss 
nondeterministic programming using monads (Section~\ref{sec:relwork:monad}),
bisimulation and its characterizing logic (Section~\ref{sec:relwork:logic}),
and related tools (Section~\ref{sec:relwork:tools}).
\vspace*{-1ex}
\subsection{Monadic encodings of Nondeterminism}
\label{sec:relwork:monad}
\citet{Wadler85listm} modeled nondeterminism with a list monad.
Monadic encodings of more sophisticated features involving nondeterminism
(e.g., \cite{FisOleSha09,Hinze00bmt,KisShaFriSab05logict}) have been developed
and applied to various domains (e.g., \cite{ChaGuoKohLoc98,Schrijvers09mcp}) afterwards.
\citet{FisOleSha09} developed a custom monadic datatype for lazy nondeterministic programming.
Their motivation was to find a way combine three desirable features found in
functional logic programming~\cite{Hanus10lea,LopHer99toy,TolSerNit04} and
probabilistic programming~\cite{ErwKol06pfp,Kiselyov16hakaru10}
-- lazyness, sharing (memoization), and nondeterminism, which are known to be tricky
to combine in functional programming. Having two versions of transitions
(Figures~\ref{fig:IdSubLTS} and \ref{fig:OpenLTS}) in our implementation
was to avoid an instance of undesirable side effects from this trickiness --
naive combination of laziness and nondeterminism causing needless traversals.
We expect our code duplication can be lifted by adopting
such a custom nondeterministic monad.
\vspace*{-1ex}
\subsection{Bisimulation and its Characterizing Logic}
\label{sec:relwork:logic}
Hennessy--Milner Logic (HML)~\cite{HenMil80hml} is a classical
characterizing logic for the Calculus of Communicating Systems (CCS)~\cite{Mil82ccs}.
The duality between diamond and box modalities related by negation 
(i.e., $[a]f \equiv \neg\langle a\rangle(\neg f)$ and $\langle a\rangle f \equiv \neg[a](\neg f)$)
holds in HML. This duality continues to hold in the characterizing logics for
early and late bisimulation for the $\pi$-calculus~\cite{MilParWal93lm}.
Presence of this duality makes it easy to obtain the distinguishing formula
for the opposite side by negation.
There have been attempts \cite{TiuMil10,ParBorEriGutWeb15} on developing
a characterizing logic for open bisimulation, but it has not been correctly
established until our recent development of \OM~\cite{AhnHorTiu17corr}.
Our logic \OM\ captures the intuitionistic nature of the open semantics, which
has a natural possible worlds interpretation typically found in Kripke-style model
of intuitionistic logic. The classical duality between diamond and box modalities
no longer hold in \OM. This is why we needed to keep track of pairs of formulae
for both sides 
during our distinguishing formulae generation in Section~\ref{sec:df}.%
\vspace*{-1ex}
\subsection{Tools for Checking Process Equivalence}
\label{sec:relwork:tools}
There are various existing tools that implement bisimulation or
other equivalence checking for variants and extensions of the $\pi$-calculus.
None of these tools generate distinguishing formulae for open bisimulation. 
The Mobility Workbench~\cite{VicMol94mwb} is a tool for the $\pi$-calculus
with features including open bisimulation checking.
It is developed using an old version of SML/NJ.
SPEC~\cite{TiuNamHor16spec} is security protocol verifier based on
open bisimulation checking \cite{TiuDaw10} for the spi-calculus~\cite{Abadi97ccs}.
The core of SPEC including open bisimulation checking is specified by
higher-order logic predicates in Bedwyr \cite{Bedwyr07} and the user interface
is implemented in OCaml.
ProVerif~\cite{BlaFou05} is another security protocol verifier based on
the applied $\pi$-calculus~\cite{AbaFou01appi}. It implements a sound
approximation of observational equivalence, but not bisimulation. 


There are few tools using Haskell for process equivalence.
Most relevant work to our knowledge is the symbolic (early) bisimulation for
LOTOS \cite{CalSha01lotos}, which is a message passing process algebra
similar to value-passing variant of CCS but with distinct features
including multi-way synchronization. Although not for equivalence checking,
\citet{Renzy14phi} implemented an interpreter that can be used as
a playground for executing applied $\pi$-calculus processes to
communicate with actual HTTP servers and clients over the internet.


\section{Conclusion}
\label{sec:concl}
We implemented automatic generation of modal logic formulae that witness
non-open bisimilarity of processes in the $\pi$-calculus.
These formulae can serve as certificates of process inequivalence,
which can be validated with an existing satisfaction checker for the modal logic \OM.
Our implementation enjoys the benefits of laziness, nondeterministic monad, and
first-class constraints; which are well known benefits of constraint programming
in Haskell. Laziness and monadic abstraction allows us to view all possible
control flow of nondeterminism as lazy generated trees, so that we can define
formula generation as a tree transformation. First-class constraints allows us
to manage information of possible worlds. Our problem setting particularly well
highlights these benefits because we needed additional information outside
the control flow of a usual bisimulation check. Our application of Haskell to
distinguishing formula generation demonstrates that Haskell and its ecosystem
are equipped with attractive features for analyzing equivalence properties of
labeled transition systems in an environment sensitive (or knowledge aware) setting.

\begin{acks}                            
  This material is based upon work supported by the
  \grantsponsor{MoEsg}{Ministry of Education, Singapore}{https://www.moe.gov.sg/}
  under Grant No.~\grantnum{MoEsg}{MOE2014-T2-2-076}.
\end{acks}



\end{document}